\documentclass{aa}
\usepackage{graphicx}
\usepackage{txfonts}
\usepackage{longtable}
\usepackage{natbib}
\usepackage{supertabular}
\usepackage{psfig}

\begin{document}
\title{Fundamental parameters of B Supergiants from the BCD System }\subtitle{I. Calibration of 
the ($\lambda_1$, D) parameters into $T_{\rm eff}$ \thanks {Data obtained at OHP, France}\,\thanks
{Data obtained at ESO La Silla, Chili}\, \thanks {Data obtained in CASLEO operated under agreement between the CONICET and the Universities of La Plata, C\'ordoba and San Juan, Argentina}}

\author{J. Zorec \inst{1}\and
L. Cidale \inst{2,3}\fnmsep\thanks{Member of the Carrera del Investigador Cient\'{\i}fico, 
CONICET, Argentina}      \and
M. L. Arias \inst{2,3}   \and
Y. Fr\'emat \inst{4}     \and
M. F. Muratore \inst{2}  \and
A. F. Torres \inst{2,3}     \and
C. Martayan \inst{4,5}}

\institute{Institut d'Astrophysique de Paris, UMR 7095 du CNRS, Universit\'e 
Pierre \& Marie Curie, 98bis bd. Arago, 75014 Paris, France                   \and
Facultad de Ciencias Astron\'omicas y Geof\'{i}sicas, Universidad Nacional de 
La Plata, Paseo del Bosque S/N, La Plata, Buenos Aires, Argentina             \and
Instituto de Astrof\'{i}sica de La Plata, (CCT La Plata - CONICET, UNLP), 
Paseo del Bosque S/N, La Plata, Buenos Aires, Argentina                       \and 
Royal Observatory of Belgium, 3 av. Circulaire, 1180 Brussels, Belgium        \and
Observatoire de Paris-Meudon, GEPI, UMR8111 du CNRS, 92195 Meudon Cedex, France}

\offprints{J. Zorec: \email{zorec@iap.fr}}
\date{Received ..., ; Accepted ...,}

\abstract
{Effective temperatures of early-type supergiants are important to test stellar atmosphere- and internal structure-models of massive and intermediate mass objects at different evolutionary phases. However, these $T_{\rm eff}$ values are more or less discrepant depending on the method used to determine them.} 
{We aim to obtain a new calibration of the $T_{\rm eff}$ parameter for early-type supergiants as a function of observational quantities that are: a) highly sensitive to the ionization balance in the photosphere and its gas pressure; b) independent of the interstellar extinction; c) as much as possible model-independent.}
{The observational quantities that best address our aims are the ($\lambda_1, D$) parameters of the BCD spectrophotometric system. They describe the energy distribution around the Balmer discontinuity, which is highly sensitive to $T_{\rm eff}$ and $\log g$. We perform a calibration of the ($\lambda_1, D$) parameters into $T_{\rm eff}$ using effective temperatures derived with the bolometric-flux method for 217 program stars, whose individual uncertainties are on average $|\Delta T_{\rm eff}|/T_{\rm eff}^f\!=\!0.05$.}
{We obtain a new and homogeneous calibration of the BCD ($\lambda_1, D$) parameters for OB supergiants and revisit the current calibration of the ($\lambda_1, D$) zone occupied by dwarfs and giants. The final comparison of calculated with obtained $T_{\rm eff}$ values in the $(\lambda_1,D)$ calibration show that the latter have total uncertainties, which on average are $\epsilon_{T_{\rm eff}}/T_{\rm eff}^f\!\simeq\pm0.05$ for all spectral types and luminosity classes.}
{The effective temperatures of OB supergiants derived in this work agree on average within some 2\,000 K with other determinations found in the literature, except those issued from wind-free non-LTE plane-parallel models of stellar atmospheres, which produce effective temperatures that can be overestimated by up to more than 5\,000 K near $T_{\rm eff}=25\,000$ K.\par
 Since the stellar spectra needed to obtain the ($\lambda_1, D$) parameters are of low resolution, a calibration based on the BCD system is useful to study stars and stellar systems like open clusters, associations or stars in galaxies observed with multi-object spectrographs and/or spectro-imaging devices.}


\maketitle

\titlerunning{Fundamental Parameters of B Supergiants} 
\authorrunning{Zorec et al.}

\section{Introduction}

 The effective temperatures of early-type stars of luminosity classes V to III are similar whatever the method used to determine them. Among the most commonly used methods are those based on: fits of the observed absolute spectral energy distributions (hereafter ASEDs) with \citet{kur79} line-blanketed stellar atmosphere models; line profile fittings; Str\"omgren and Geneva photometric color indices. Using Breger's spectrophotometric catalogue \citep{bre76a} for the ASED in the visual range, \citet{mor85} and \citet{mal90} obtained $T_{\rm eff}$ values with internal uncertainties below 5\%. The uncertainties are of the order of 10\%, either when the TD1 far-UV fluxes and IUE low resolution spectra are used in combination with fluxes from Breger's catalogue \citep{mal83,mal86,gul89}, or when H and He absorption line profiles are fitted with models \citep{mor80}. In general, $T_{\rm eff}$ values derived using calibrated Str\"omgren and Geneva photometric indices present low uncertainties \citep{bal84,moo85,cas91,ach93}. However \citet{nap93}, based on a critical comparison of several calibrations of Str\"omgren intermediate-band uvby$-\beta$ photometric indices, recommended the use of the calibration done by \citet{moo85}, corrected for gravity deviations.\par
 As regards B supergiants, \citet{cod76} and \citet{und79} derived effective temperatures based on spectrophotometric observations in the far-UV, visible and near-IR spectral regions. Later, effective temperature determinations for these stars were made using the silicon lines in the optical spectral region, either by fitting the line profiles or by measuring their equivalent widths \citep{bec90,mce99,tru04}. Nowadays, methods based on adjustments of \ion{He}{i} and \ion{He}{ii} line profiles, and/or the line intensity ratio \ion{Si}{iv}/\ion{Si}{iii} are preferred \citep{her02,rep04,mar05,cro06,ben07,mar08,sea08}. Recent estimates of the effective temperature of early B-type supergiants have shown that the values obtained with non-LTE blanketed models including winds (hereafter non-LTE BW models) are systematically lower than those derived with models without winds (hereafter wind-free models) \citep{lef07,cro06}. Differences range roughly from 0 to 6\,000 K and they tend to be lower the later the B-sub-spectral type \citep{mar08}. On the other hand, \citet{mor80} noted that the effective temperatures of B-supergiants derived with photometric methods are in general higher than those obtained spectroscopically. Physical characteristics and phenomena like activity and/or instabilities taking place in the extended atmospheric layers affect more significantly the spectral lines than the continuum spectrum. Thus, the genuine signature due to the $T_{\rm eff}$ carried by the lines could be somewhat blurred.\par
 The calibrations of effective temperatures for B supergiants are of great importance since these stars are in a significant phase of the evolutionary sequence of massive stars. They are also the main contributors to the chemical and dynamical evolution of galaxies. Accurate effective temperatures are then needed to construct HR diagrams and to test the theories of stellar structure and evolution, as well as to estimate the chemical content of the stellar environment. Effective temperatures are also necessary to study the physical processes in the atmosphere, such as non-radial pulsations or stellar winds (radiative forces; changes in ionization). Regarding the stellar winds, the effective temperatures are particularly useful in discussing terminal velocities, mass loss rates, the bi-stability jump, and the wind momentum luminosity relationship \citep{kud03,cro06,mar08}.\par 
 As a consequence of the large discrepancies found in the $T_{\rm eff}$ estimates of B supergiants, the current temperature scale is being revisited \citep{mar08, sea08}. In this context, we present an independent and homogeneous temperature calibration for B-type dwarfs to supergiants, based on the use of the BCD spectrophotometric system \citep{bar41,cha52}. This method has numerous advantages (see \S\ref{BCD}), mainly because it is based on : a) measurable quantities that are strongly sensitive to the ionization balance in the stellar atmosphere and to its gas pressure, thus being excellent indicators of $T_{\rm eff}$ and $\log g$; b) parameters that describe the visible continuum spectrum, whose atmospheric formation layers are on average deeper than those for spectral lines.\par
 Our first step will be to determine the effective temperatures of our sample of Galactic B-type supergiants. However, to perform a consistent calibration of the BCD $(\lambda_1,D)$ parameters for supergiants, we re-determine the calibration into effective temperature of the BCD $(\lambda_1,D)$ domain corresponding to B-type dwarfs and giants ($D\!=\!$ size of the Balmer jump; $\lambda_1\!=\!$ mean spectral position of the Balmer discontinuity (BD); see further explanations on these parameters in \S\ref{BCD} and Appendix \S\ref{bcd_expl}). We use a large and homogeneous sample of B stars observed in the BCD system, and newly-derived effective temperatures for all of them, based on the bolometric-flux method (hereafter BFM). We leave for another contribution the discussion of BCD calibrations related to $\log g$, visual and bolometric absolute magnitudes.\par 
 The present paper is organized as follows: In \S\ref{BCD} we briefly describe the BCD spectrophotometric system and the advantages of its use. In \S\ref{BFM} we present the BFM on which the determinations of the stellar effective temperature and angular diameter ($\theta$) are based.  Observations and the $T_{\rm eff}$ values determined with the BFM are presented in \S\ref{observations}. The uncertainties of the effective temperatures and angular diameters obtained with the BFM are discussed in \S\ref{comm}. Comparisons of our $T_{\rm eff}$ and $\theta$ determinations with those obtained by other authors are given in \S\ref{comp}. In \S\ref{bcdcalib}, we present the empirical temperature calibration curves and discuss the accuracy of the $T_{\rm eff}(\lambda_1, D)$ values obtained. A discussion and global conclusions are presented in \S\ref{disc} and \S\ref{concl}, respectively.\par

\section{The BCD system}\label{BCD}

 The Paris spectrophotometric classification system of stellar spectra, best known as BCD (Barbier-Chalonge-Divan), was defined by \citet{bar41} and \citet{cha52}. The original presentation of this system is given in French; explanations in English can be found in \citet{duf64,und66,und82,div92}. A short overview of the system is given in Appendix \S\ref{bcd_expl}. The BCD system is based on four measurable quantities in the continuum spectrum around the BD: D, the Balmer jump given in dex and determined at $\lambda$3700 \AA; $\lambda_1$, the mean spectral position of the BD, usually given as the difference $\lambda_1$-3700 \AA; $\Phi_{uv}$, the gradient of the Balmer energy distribution in the near-UV from $\lambda$3100 to $\lambda$3700 \AA, given in $\mu$m;  $\Phi_{rb}$, the gradient of the Paschen energy distribution in the wavelength interval $\lambda\lambda$4000-6200 \AA\ given in $\mu$m. The sole BCD parameters that are relevant to the present work are: $D$, which is a strong function of $T_{\rm eff}$, and $\lambda_1$ that is very sensitive to $\log g$.\par 
 The use of the ($\lambda_1$,$D$) pair to determine the spectral classification and the stellar fundamental parameters presents numerous advantages, not only because the BD is a well visible spectral characteristic for stars ranging from early O to late F spectral types but also because:\par
 a) The parameters ($\lambda_1$,$D$) are obtained from direct measurement on the stellar continuum energy distribution. This implies that, on average, they are relevant to the physical properties of photospheric layers which are deeper than those described by spectral lines;\par 
 b) Each MK (Morgan \& Keenan) spectral type-luminosity class (SpT/LC) is represented by wide intervals of $\lambda_1$ and $D$ values, which implies high SpT/LC classification resolution: $\lambda_1$ ranges from about 75 \AA\, for dwarfs to -5 \AA\, for supergiants, while D ranges from near 0.0 dex, for the hottest O stars and F9 stars, to about 0.5 dex, for the A3-4 stars; \par 
 c) Typical 1$\sigma$  measurement uncertainties affecting $D$ and $\lambda_1$ are: $\delta D\!\leq$ 0.02 dex and $\delta\lambda_1\!\leq$ 2 \AA, respectively. Thus, from b) and c) we find that hardly any other classification system has reached such a high resolution, especially concerning the luminosity class; \par 
 d) The parameter $\lambda_1$ is independent of the interstellar medium (ISM) extinction, while $D$ has a low E(B-V) color excess dependence, roughly $\delta D$ = 0.03 E(B-V) dex, which is almost insensitive to the selective absorption ratio $R_{\rm V}=A_V/E(B-V)$. The $\delta D$ difference is produced by ex\-tr\-apo\-la\-tion of the Paschen energy distribution from  $\lambda$4000 \AA\ to  $\lambda$3700 \AA, which carries the ISM reddening of the Paschen continuum. The low ISM extinction dependence is however of great interest for the study of early-type stars, since they are frequently distant and strongly reddened;\par
 e) Spectra needed to obtain the ($\lambda_1$,$D$) measurements are of low resolution (8 \AA\ at the BD). This means that numerous faint stars can be observed with short exposure times. Furthermore, the flux calibrations required to derive the BCD parameters rely on spectral reduction techniques which are easy-to-use and of common practice; \par 
 f) The BCD system is generally used for `normal' stars, i.e. objects whose atmospheres can be modeled in the framework of hydrostatic and radiative equilibrium approximations. However, since both the photospheric and the circumstellar components of the BD are spectroscopically well separated, it can also be used to study some `peculiar' objects, like: i) Be stars \citep{div83,zor86,zor91,cha01,zor05,vin06}; ii) objects with the B[e] phenomenon \citep{cid01}; iii) chemically peculiar stars (He-W group) \citep{cid07}.\par

\section{The bolometric-flux method (BFM)}\label{BFM}

  To obtain the calibration of the BCD ($\lambda_1$,$D$) parameters into effective temperature, we could simply adopt for each star the average of all the $T_{\rm eff}$ values found in the literature. Nevertheless, these quantities were obtained with several heterogeneous methods which carry more or less systematic differences on the estimates of $T_{\rm eff}$. Since most $T_{\rm eff}$ determinations are based on adjustment of the observed ASEDs with theoretical ASEDs, or on fittings of line profiles with models, the properties of the studied stars are implicitly assumed to be in accordance with the physical characteristics of the best fitted stellar model atmosphere. Instead, in the present contribution, we preferred to determine the effective temperature by using the total amount of radiated energy, so that the details of its distribution are of marginal importance and the dependence on models of stellar atmospheres are kept to a minimum. While models of stellar atmospheres produce Balmer jumps which are close to the observed ones, the theoretical $\lambda_1$ parameter may differ somewhat. In fact, $\lambda_1$ depends on the distribution of the emergent radiation fluxes near the limit of the Balmer line series, where the theoretical uncertainties concern the treatment of the non-ideal effects in the hydrogen upper level populations \citep{roh03}.\par
 We decided to adopt a single method for all program stars and to estimate their effective temperatures using its definition, where the incidence of the model-dependence is in principle strongly minimized.\par 

 By definition, the effective temperature of a star is:

\begin{equation} 
\label{teff}
T_{\rm eff} = \bigl[\frac{4f}{\sigma_R\theta^2}\bigr]^{1/4},
\end{equation}

\noindent where $\sigma_R$ is the \v{S}tefan-Boltzmann constant; $f$ is the stellar bolometric radiation flux received at the Earth, corrected for the ISM extinction; $\theta$ is the angular diameter of the star. If the radiation field coming from the stellar interior were the sole energy source in the atmosphere, the effective temperature deduced from the bolometric flux would be the same as that derived from the analysis of stellar spectra with model atmospheres in radiative and hydrostatic equilibrium. If the bolometric flux $f$ and the angular diameter $\theta$ were issued entirely from observations, the $T_{\rm eff}$ deduced from Eq. (\ref{teff}) could then also be considered a genuine observational parameter. The BFM was applied in this way by \citet{cod76} using stellar fluxes observed with the OAO-2 satellite and the angular diameters determined interferometrically \citep{hanb74}. Unfortunately, this could not happen in our case, as we do not have observed fluxes over the entire spectrum, nor do we have angular diameters measured for all program stars. In what follows the effective temperature and the angular diameter derived with the `bolometric flux method' are called $T_{\rm eff}^f$ and $\theta^f$, respectively.\par
 \citet{blck77} showed that a stellar angular diameter can be well reproduced with observed fluxes in near-IR and model atmospheres. The monochromatic stellar angular diameter is thus given by: 

\begin{equation}
\label{teta}
\theta_{\lambda} = 2\,\bigl[\frac{f^o_{\lambda}}{{\mathcal F}_{\lambda}}\bigr]^{1/2},
\end{equation}

\noindent  where $f^o_{\lambda}$ is the absolute monochromatic flux received at the Earth corrected for the ISM extinction, ${\mathcal F}_{\lambda} = \pi\,F_{\lambda}$ is the emitted monochromatic flux at the stellar surface, $F_{\lambda}$ is the so called `astrophysical' flux predicted by a model atmosphere. To represent $F_{\lambda}$ we have used the grids of ATLAS9 model atmospheres calculated by \citet{cas03}. In wavelengths lying in the Rayleigh-Jeans tail of the energy distribution, not only are the theoretical fluxes nearly independent of model characteristics, but the observed ASEDs also are mildly affected by the ISM extinction. However, they have the inconvenience of being frequently marred by infrared flux excesses of non stellar origin. We decided to calculate Eq. (\ref{teta}) in the red extreme of the Paschen continuum: $\lambda\lambda$\,$0.58-0.8\mu$m. In this spectral region, the theoretical radiative fluxes are still only slightly dependent on the particular model characteristics and the layers where this radiation field is formed have local electron temperatures close to $T_{\rm eff}$, so that the color temperature of the energy distribution of OB stars in this wavelength interval also approaches $T_{\rm eff}$ closely. While $T_{\rm eff}$ derived with Eq. (\ref{teff}) is insensitive to $\log g$ within the characteristic uncertainties of its determination, the parameter $\theta_{\lambda}$ does depend slightly on models. Therefore, in Eq.(\ref{teta}) model atmospheres giving $F_{\lambda}$ are chosen for gravity parameters $\log g$ = $\log g(\beta,T_{\rm eff}$), where $\beta$ is the H$\beta$-line index of the uvby$-\beta$ Str\"omgren photometry. The $\beta$ parameters used are from the \citet{haucmer75} compilation and the $\log g(\beta,T_{\rm eff})$ relations used are from \citet{caku06}. An extensive use of the BFM was made by \citet{und79}, where unfortunately Eqs. (\ref{teff}) and (\ref{teta}) were iterated only twice, which left the derived effective temperatures strongly correlated with their initial approximate values.\par

\subsection{Effective temperatures of B-type dwarfs to giants}
\label{tdwgi}

 The ASEDs of our program stars, taken from the literature to calculate $f$, are in most cases observed in the wavelength interval ranging from  $\lambda_a \sim$  1200-1300 \AA\ to some $\lambda_b$ in the far-IR. Many times they present intermediate gaps, according to the observational method or the instruments used. On the other hand, the angular diameter has been measured only for a few bright stars. Nevertheless, as in Eq. (\ref{teff}) $T_{\rm eff}$ depends on $f^{1/4}$, the effective temperature can still be reliably determined even though the unobserved spectral regions are represented using `modestly realistic' model atmospheres, as discussed in \S\ref{comm}. The calculation of the effective temperature for dwarfs to giants is based on the following iteration:\par
 1) Adopt initial values of $T_{\rm eff}^f$ and $\log g$;\par
 2) Interpolate the model fluxes in the far-UV and IR spectral regions for the adopted $(T_{\rm eff},\log g)$ values ;\par
 3) Calculate the angular diameter $\theta^f$ using relation (\ref{teta}) in the near IR spectral region as detailed in the explanation following relation (\ref{teta}), and assume $\theta^f$ independent of the wavelength;\par
 4) Calculate the bolometric flux $f$ as follows:

\begin{eqnarray}
\begin{array}{lcl}
          f & = &f_{\rm obs}\times[1+\delta] \ \ \ \ \\
f_{\rm obs}& = &\int_{\lambda_a}^{\lambda_b}\!f^o_{\lambda}{\rm d}\lambda \\
\delta & = & \delta_{\rm UV}+\delta_{IR} \ \ \ \ \\
       & = & \bigl\{\frac{\pi}{4}\theta^2\frac{\int_0^{\lambda_a}F_{\lambda}{\rm d}\lambda}{f_{\rm obs}}\bigr\}_{\rm UV} +\bigl\{\frac{\pi}{4}\theta^2\frac{\int_{\lambda_b}^{\infty}F_{\lambda}{\rm d}\lambda}{f_{\rm obs}}\bigr\}_{\rm IR}, \ \ \ \ \\
\end{array}
\label{fl}
\end{eqnarray}

\noindent where $f_{\rm obs}$ and $\delta$ represent, respectively, the contribution of the observed and unobserved spectral regions (extreme-UV, far-UV and IR) to the bolometric flux; $\lambda_a$ and $\lambda_b$ delimit the spectral range of ASEDs actually observed; $f^o_{\lambda}$ and ${\mathcal F}_{\lambda}$ have the same meaning as given in Eqs. (\ref{teff}) and (\ref{teta});\par 
 5) Introduce the bolometric flux given by (\ref{fl}) into relation (\ref{teff}) to obtain a new estimate of $T_{\rm eff}^f$ and accordingly, of $\log g= \log g(T_{\rm eff}^f,\beta)$;\par
 6) Use the new estimates of  ($T_{\rm eff}^f,\log g)$ to continue the iteration in step 2).\par 
 The iterations were performed until the difference between two consecutive $T_{\rm eff}^f$ values was smaller than 1 K. Depending on the star, this implies roughly 10 to 30 iterations. From (\ref{fl}) it is obvious that the estimates of the effective temperatures may in principle be more uncertain the higher the value of $\delta$; i.e. for the hottest stars. In Table \ref{deltas} are listed the fractions $\delta_{\rm UV}$ and $\delta_{IR}$ for different effective temperatures and gravities. As seen in this table, $\delta$ has a low dependence on $\log g$.\par

\begin{figure}
\centerline{\psfig{file=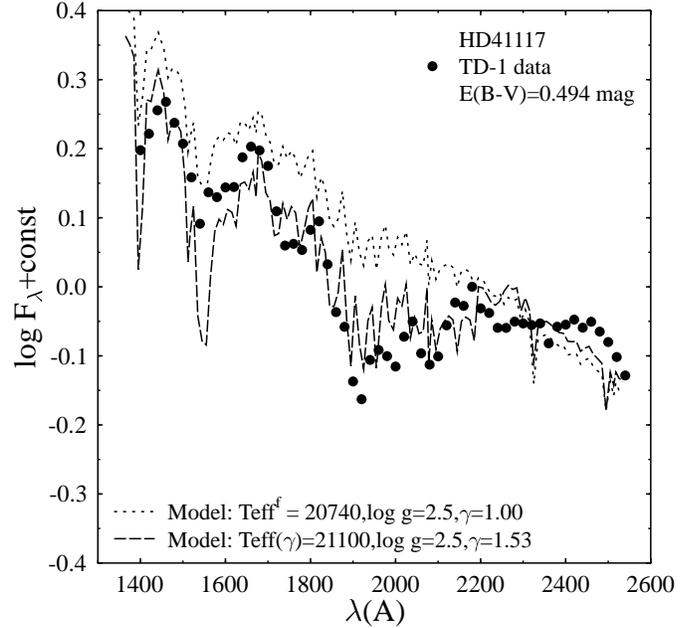}}
\caption{\label{compmod}Fitting of the far-UV observed energy distribution of \object{HD 41117} with $\gamma$-unmodified model fluxes for $T^f_{\rm eff}$ (pointed line), and with $\gamma$-modified model for $T_{\rm eff}(\gamma)$ (dashed line).}
\end{figure}

\begin{table*}
\caption{\label{deltas}Far-UV and IR fractions of the unobserved bolometric flux as a function of $T_{\rm eff}$ and $\log g$, the function $\Lambda(T_{\rm eff},\log g)$, and temperatures $T_{\rm eff}^f$ as a function of $T_{\rm eff}(\gamma)$.}
\tabcolsep 4.0pt
\begin{tabular}{c|cc|ccr||cc|ccr||cc|ccr||}
\hline
\noalign{\smallskip}
&\multicolumn{5}{c|}{$\log g = 4.0$} & \multicolumn{5}{c|}{$\log g = 3.0$} &\multicolumn{5}{c|}{$\log g = 2.5$}\\
\noalign{\smallskip}
\hline 
\noalign{\smallskip} 
$T_{\rm eff}(\gamma)$ & $\delta^*_{UV}$ &  $\delta^*_{IR}$ & $\Lambda[T_{\rm eff}(\gamma),g]$ & $\gamma$ & $T_{\rm eff}^f$ & $\delta^*_{UV}$ &  $\delta^*_{IR}$ & $\Lambda[T_{\rm eff}(\gamma),g]$ & $\gamma$ & $T_{\rm eff}^f$ &
$\delta^*_{UV}$ &  $\delta^*_{IR}$ & $\Lambda[T_{\rm eff}(\gamma),g]$ & $\gamma$ & $T_{\rm eff}^f$ \\
\noalign{\smallskip}
\hline 
10000&0.023&0.059&0.024&1.5& 9970& 0.024& 0.059& 0.028 & 1.5 &  9970 & 0.023 & 0.059 & 0.029 & 1.5 &  9960 \\
     &     &     &     &2.0& 9940&      &      &       & 2.0 &  9930 &       &       &       & 2.0 &  9930 \\
12500&0.105&0.034&0.088&1.5&12360& 0.099& 0.034& 0.086 & 1.5 & 12360 & 0.093 & 0.034 & 0.081 & 1.5 & 12370 \\
     &     &     &     &2.0&12220&      &      &       & 2.0 & 12220 &       &       &       & 2.0 & 12240 \\
15000&0.230&0.022&0.152&1.5&14710& 0.218& 0.022& 0.135 & 1.5 & 14740 & 0.202 & 0.022 & 0.122 & 1.5 & 14770 \\
     &     &     &     &2.0&14390&      &      &       & 2.0 & 14470 &       &       &       & 2.0 & 14520 \\
17500&0.386&0.016&0.182&1.5&17090& 0.353& 0.016& 0.155 & 1.5 & 17150 & 0.318 & 0.016 & 0.134 & 1.5 & 17200 \\
     &     &     &     &2.0&16640&      &      &       & 2.0 & 16780 &       &       &       & 2.0 & 16880 \\
20000&0.552&0.012&0.194&1.5&19450& 0.484& 0.012& 0.157 & 1.5 & 19600 & 0.450 & 0.012 & 0.135 & 1.5 & 19650 \\
     &     &     &     &2.0&18950&      &      &       & 2.0 & 19160 &       &       &       & 2.0 & 19290 \\
22500&0.706&0.009&0.191&1.5&21940& 0.596& 0.010& 0.139 & 1.5 & 22090 & 0.546 & 0.010 & 0.109 & 1.5 & 22190 \\
     &     &     &     &2.0&21340&      &      &       & 2.0 & 21670 &       &       &       & 2.0 & 21860 \\
25000&0.843&0.008&0.178&1.5&24420& 0.693& 0.008& 0.106 & 1.5 & 24670 & 0.635 & 0.008 & 0.071 & 1.5 & 24780 \\
     &     &     &     &2.0&23800&      &      &       & 2.0 & 24310 &       &       &       & 2.0 & 24540 \\
27500&0.955&0.007&0.153&1.5&26950& 0.823& 0.007& 0.093 & 1.5 & 27180 &       &       &       &     &       \\
     &     &     &     &2.0&26380&      &      &       & 2.0 & 26840 &       &       &       &     &       \\
30000&1.060&0.006&0.126&1.5&29520& 1.023& 0.006& 0.087 & 1.5 & 29670 &       &       &       &     &       \\
     &     &     &     &2.0&29010&      &      &       & 2.0 & 29330 &       &       &       &     &       \\
32500&1.241&0.005&0.109&1.5&32050&      &      &       &     &       &       &       &       &     &       \\
     &     &     &     &2.0&31580&      &      &       &     &       &       &       &       &     &       \\
35000&1.539&0.004&0.103&1.5&34540&      &      &       &     &       &       &       &       &     &       \\
     &     &     &     &2.0&34060&      &      &       &     &       &       &       &       &     &       \\
\noalign{\smallskip}
\hline
\noalign{\smallskip}
\multicolumn{16}{l}{For $\Lambda(T_{\rm eff})$ and $T_{\rm eff}$ the temperature $T_{\rm eff}(\gamma)$ is simply the effective temperature written in the first column of the table.}\\
\multicolumn{16}{l}{Parameters $\delta^*_{\rm UV}$ and $\delta^*_{\rm IR}$ are values of  $\delta_{\rm UV}$ and $\delta_{\rm IR}$ calculated with $T_{\rm eff}(\gamma)$ and fluxes $F_{\lambda}[T_{\rm eff}(\gamma),\log g]$, which do not undergo the transforma-}\\
\multicolumn{16}{l}{tion given by Eq.~\ref{gama}. To calculate them we have used $\lambda_a=1380$ \AA\ and $\lambda_b=11084$ \AA. Note that $T^f_{\rm eff}=$ $T_{\rm eff}(\gamma=1)$.}\\
\noalign{\smallskip}
\hline
\end{tabular}
\end{table*}

\subsection{Effective temperatures of B-type supergiants}\label{tspg} 

 The theoretical ASEDs predicted for supergiants by the wind-free plane-parallel model atmospheres with the effective temperatures issued directly from Eq. (\ref{teff}), do not always fit well to the observed ASEDs in the near- and far-UV. Generally, the predicted fluxes are higher than the observed ones in wavelengths $\lambda\!\la\!2200$ \AA, while they are lower in the near-UV. For the obtained $T_{\rm eff}$ they yield somewhat larger Balmer discontinuities than those observed. This might be partially due to the plane-parallel approximation of model atmospheres that probably is not suited to the extended atmospheric layers of these stars, to an incomplete treatment of the spectral line formation in such diluted atmospheres, to the omission of the effects produced on the photosphere by the stellar winds \citep{abb85,gab89,smi02,mori04}, and also to an insufficient line blocking in the model atmospheres, as discussed by \citet{rem81}. Thus, to obtain the $T_{\rm eff}^f$ parameter of supergiants, we have slightly modified  the use of relations (\ref{teff}) and (\ref{teta}) through the following iteration procedure:\par
 1) Adopt approximate values of $T_{\rm eff}$ and $\log g$;\par
 2) Interpolate the model fluxes in the far-UV and IR for the adopted ($T_{\rm eff},\log g$) parameters;\par
 3) Use a least square procedure to search for an enhancement parameter $\gamma$ of the line blocking in the $1400\!\leq\!\lambda\!\leq\!2150$ \AA\ wavelength interval, and correct the failure of the wind-free model fluxes used to fit the far-UV spectral region. The validity of $\gamma$ is then extended to the entire $0\!\leq\!\lambda\!\leq\!2150$ \AA\ spectral region. The empirical method used to calculate $\gamma$ and to modify the line-blocking is explained in \S\ref{mtf};\par
 4) Define a new effective temperature, $T_{\rm eff}(\gamma)$, to account for the lack of energy in the far-UV, produced by the increased absorption induced by $\gamma$ and for the redistribution of this energy in
the longer wavelengths. The calculation of $T_{\rm eff}(\gamma)$ is explained in \S\ref{efftg};\par
 5) Interpolate the model fluxes in the far-UV and IR for the $T_{\rm eff}(\gamma)$ and $\log g\!=\!\log
g[T_{\rm eff}(\gamma),\beta]$ parameters;\par
 6) Calculate the angular diameter $\theta^f$ using Eq.~(\ref{teta}) and the IR fluxes dependent on the [$T_{\rm eff}(\gamma),\log g(T_{\rm eff}(\gamma),\beta)$] pair of fundamental parameters. With the fluxes interpolated in step 5), calculate also the bolometric corrections $\delta_{\rm UV}$ and $\delta_{\rm IR}$ as indicated in Eq.~(\ref{fl}). The bolometric correction $\delta_{\rm UV}$ is calculated using the $\gamma$-modified far-UV fluxes $\widetilde{F}_{\lambda}$ defined in \S\ref{mtf} by Eq.~(\ref{gama}), where $T_{\rm eff}$ is replaced by $T_{\rm eff}(\gamma)$:

\begin{equation}
\label{deltauv}
\delta_{\rm UV} = \frac{\pi}{4}\theta^2\frac{\int_0^{\lambda_a}\widetilde{F_{\lambda}}[T_{\rm eff}(\gamma),
\log g(T_{\rm eff}(\gamma),\beta)]{\rm d}\lambda}{f_{\rm obs}}\ ;
\end{equation}

 7) Use $\delta_{\rm UV}$ and $\delta_{\rm IR}$, and $\theta^f$ derived in step 6) to calculate the bolometric flux $f$ with relation (\ref{fl}) and obtain from Eq.~(\ref{teff}) a new estimate of $T_{\rm eff}^f$;\par
 8) Continue the iteration in step 3) by searching for a new enhancement parameter $\gamma$ using the far-UV fluxes $F_{\lambda}[T_{\rm eff}(\gamma)]$ interpolated in step 5).\par
\medskip
\subsubsection{\it Modification of theoretical fluxes}\par
\label{mtf}
\medskip
  Since we are not interested here in reproducing detailed energy distributions to fit the observed ones, but rather in estimating the integrated amount of the emitted energy over the whole spectrum, we use an empirical method to minimize the disagreement between the predicted and the observed far-UV energy distributions. However, the method is not able to simulate the consequences due to the presence of winds which become conspicuous in the extreme-UV \citep{smi02}. Nevertheless, this fact cannot significantly change the estimate of $T_{\rm eff}^f$, as we shall see in \S\ref{commse}.\par 
 Thus, we have proceeded in a similar way as previously attempted by \citet{rem81} and \citet{zor87}, i.e. we have modified the blocking degree of spectral lines in the $\lambda\,\la\!2150$ \AA\ region by writing the radiative flux emitted by a star at a given $\lambda$ as:

\begin{equation}
\label{gama1}
F_{\lambda}(T_{\rm eff},\log g) = F_{c}(1-b_{\lambda}),
\end{equation}

\noindent where $F_c$ is the continuum flux for a given set of parameters $(T_{\rm eff},\log g)$, and $b_{\lambda}$ is the line blocking factor. The line blocking factor $b_{\lambda}$ easily can be  calculated using the $F_{\lambda}$ and $F_c$ fluxes listed by \citet{cas03} (http://kurucz.harvard.edu/grids.html). The enhanced line blocking factor was then calculated by multiplying $b_{\lambda}$ by a parameter $\gamma$, which we assumed constant over all the wavelength interval $0\!\leq\!\lambda\!\leq\!2150$ \AA. The modified theoretical flux at a given $\lambda$ is then:

\begin{equation}
\label{gama}
\widetilde{F}_{\lambda}(T_{\rm eff},\log g) = F_{\lambda}(T_{\rm eff},\log g)\bigl(\frac{1-\gamma b_{\lambda}}{1-b_{\lambda}}\bigl)\ . 
\end{equation}

 For each star and at each iteration step of its $T_{\rm eff}^f$, we looked for the value of $\gamma$ that produced the best possible fit between the observed and theoretical fluxes, in the $1400\,\la\,\lambda\,\la\!2150$ \AA\ wavelength interval. The model fluxes are calculated for a new effective temperature $T_{\rm eff}(\gamma)$, which is different from $T_{\rm eff}^f$, and for a $\log g$, which change as the iteration of the effective temperature continues.\par
\medskip
 \subsubsection{\it Effective temperature $T_{\rm eff}(\gamma)$ of models}\par
\label{efftg}
\medskip
 It can readily be understood that the estimate of $\delta_{\rm UV}$ with the modified fluxes given by Eq.~(\ref{gama}), where in most cases $\gamma\!>\!1$, leads to an effective temperature lower than the nominal value, $T_{\rm eff}$, of the model used. Thus, when using Eq. (\ref{gama}) in the wavelength range $0\!\leq\!\lambda\!\leq\!2150$ \AA, we can recover the bolometric flux initially represented by the model characterized by $(T_{\rm eff},\log g)$, by means of a new effective temperature, named hereafter $T_{\rm eff}(\gamma)$, which is larger than $T_{\rm eff}$ if $\gamma\!>\!1$. Mathematically, this is due to the lack of energy produced by an increased absorption in the far-UV. Physically, this accounts for a redistribution of the excess of absorbed energy in the far- and extreme-UV, towards the near-UV, visible and IR spectral regions produced by the back-warming induced by the enhanced line blocking. To find the relation between $T_{\rm eff}$ and  $T_{\rm eff}(\gamma)$, we write the same bolometric flux, $F_{\rm bol}$, in terms of both temperatures. On the one hand, it can be calculated using model fluxes dependent on $T_{\rm eff}$ as:

\begin{equation}
\label{tgama0}
F_{\rm bol} = \int_0^{\lambda_a}\!\!F_{\lambda}(T_{\rm eff}){\rm d}\lambda+ 
\int_{\lambda_a}^{\infty}\!\!F_{\lambda}(T_{\rm eff}){\rm d}\lambda,
\end{equation}

\noindent and, on the other hand, the same $F_{\rm bol}$ can be obtained with models that are a function of $T_{\rm eff}(\gamma)$, as follows:

\begin{equation}
\label{tgama1}
F_{\rm bol} = \int_0^{\lambda_a}\!\!\widetilde{F}_{\lambda}[T_{\rm eff}(\gamma)]{\rm d}\lambda+ 
\int_{\lambda_a}^{\infty}\!\!F_{\lambda}[T_{\rm eff}(\gamma)]{\rm d}\lambda,
\end{equation}

\noindent where $\widetilde{F}_{\lambda}[T_{\rm eff}(\gamma)]$ is given by the relation (\ref{gama}). Since

\begin{equation}
\label{tgama2}
F_{\rm bol} = \frac{\sigma_R}{\pi}T^4_{\rm eff}
\end{equation}

\noindent and

\begin{equation}
\label{tgama3}
\frac{\sigma_R}{\pi}T^4_{\rm eff}(\gamma) = \int_0^{\lambda_a}\!\!F_{\lambda}[T_{\rm eff}(\gamma)]{\rm d}\lambda+\int_{\lambda_a}^{\infty}\!\!F_{\lambda}[T_{\rm eff}(\gamma)]{\rm d}\lambda,
\end{equation}

\noindent by subtracting (\ref{tgama1}) from (\ref{tgama2}) and this result from (\ref{tgama3}), we obtain the sought relation:

\begin{equation}
\label{tgama}
T_{\rm eff}(\gamma)^4 = T_{\rm eff}^4+\bigl(\frac{\pi}{\sigma_R}\bigr)\!\int_{0}^{\lambda_a}\!\!\bigl\{F_{\lambda}[T_{\rm eff}(\gamma)]-\widetilde{F}_{\lambda}[T_{\rm eff}(\gamma)]\bigr\}{\rm d}\lambda\ , 
\end{equation}

\noindent where we have not made explicit the dependence on $\log g$, but have written that in the $0\!\la\!\lambda\!\la\!2150$ \AA\ wavelength interval, the unmodified fluxes $F_{\lambda}$ and modified fluxes $\widetilde{F}_{\lambda}$, are both calculated for the effective temperature $T_{\rm eff}(\gamma)$. It is obvious that to have the corresponding $T_{\rm eff}(\gamma)$ at each iteration step of the stellar effective
temperature we put $T_{\rm eff}\!=\!T^f_{\rm eff}$ and iterate the relation (\ref{tgama}). Then, replacing 
Eq.~(\ref{gama}) into $(\ref{tgama})$, we derive:

\begin{equation}
\label{tgamal}
T_{\rm eff}^f = T_{\rm eff}(\gamma)\bigl\{1-(\gamma-1)\Lambda[T_{\rm eff}(\gamma),\log g]\bigr\}^{1/4}\ ,
\end{equation}

\noindent where the function $\Lambda[T_{\rm eff}(\gamma),\log g]$ is given by:

\begin{equation}
\label{tgamal2}
\Lambda[T_{\rm eff}(\gamma),\log g] = \bigl(\frac{\pi}{\sigma_R}\bigr)\bigl[\frac{\int_0^{\lambda_a}F_{\lambda}[T_{\rm eff}(\gamma)]\bigl(\frac{b_{\lambda}}{1-b_{\lambda}}\bigr)}{T^4_{\rm eff}(\gamma)}{\rm d}\lambda\bigr]\ , 
\end{equation}

\noindent which can be calculated as a function of $(T_{\rm eff},\log g)$ and used to derive the required value of $\Lambda[T_{\rm eff}(\gamma)]$ by interpolation. The function $\Lambda[T_{\rm eff}(\gamma),\log g]$ is given in Table \ref{deltas}, where we also give the values of $T_{\rm eff}^f$ derived from (\ref{tgamal}). In this table $T_{\rm eff}(\gamma)$ appears as the entering temperature. Actually, in the calculation of the stellar effective temperatures, we enter relation (\ref{tgamal}) with $T_{\rm eff}^f$ and deduce the corresponding value of $T_{\rm eff}(\gamma)$ at each iteration step. At each iteration step of $T^f_{\rm eff}$, obtained by means of Eqs. (\ref{teff}) and (\ref{teta}), and $\delta_{\rm UV}$ calculated with (\ref{gama}), we changed the gravity parameter accordingly using the tables of $\log g\!=\!\log g(\beta,T_{\rm eff})$ given by \citet{caku06}.  For instance, in Fig. \ref{compmod} we show the results obtained at the final step of the analysis carried out for \object{HD 41117}. For this particular star we have obtained $T_{\rm eff}\!=\!20\,740$ K, $T_{\rm eff}(\gamma)\!=\!21\,100$ K, $\gamma\!=1.53$ and $\log g\!=\!2.5$. All fluxes in Fig.\ref{compmod}, observed and modelled, are normalized to a given flux in the visible wavelengths where the angular diameter is calculated. However, in the plot the logarithm of the fluxes is shifted by a constant value. We can also see that for $T_{\rm eff}^f$ and $\gamma\!=1.0$ the model fluxes in the far-UV are higher than the observed ones. In spite of the roughness of our approach, the fluxes from models computed with $T_{\rm eff}(\gamma)$ and $\gamma\!=1.53$ improve the fit of the observed energy distribution. It is also seen that the $\gamma$-modified flux is slightly higher in $\lambda\,\ge\!2200$ \AA\ than the fluxes calculated with $T_{\rm eff}^f$, which leads to a smaller Balmer discontinuity as desired, i.e. $D_{\rm obs}\!=0.050$ dex, $D[T_{\rm eff}(\gamma)]\!=0.065$ dex, while it is $D(T^f_{\rm eff})\!=0.074$ dex.\par
 It is important to note that $\gamma\!>\!1$ produces a lowering of fluxes in the far- and extreme-UV, while the effect carried by the stellar wind on the photosphere, neglected here, increases the emitted fluxes (see \citet{smi02}). As we shall see in \S\ref{commse}, this increase happens at global flux levels that may have an effect on the estimate of $\delta_{\rm UV}$ at effective temperatures higher than 20\,000 K, as shown in \S\ref{commse}. In general, the values of $\delta_{\rm UV}$ obtained with wind-free models can differ significantly from those derived with non-LTE BW models if $\gamma\!\gtrsim\!1.5$. The variation of $\delta_{\rm UV}$ with $\gamma$ can be obtained by noting that we can write:

\begin{equation}
\label{duvg1}
\delta_{\rm UV}(\gamma) = \frac{\int_0^{\lambda_a}\widetilde{F}_{\lambda}[T_{\rm eff}(\gamma)]{\rm d}\lambda}{\int_{\lambda_a}^{\lambda_b}F_{\lambda}[T_{\rm eff}(\gamma)]{\rm d}\lambda}\ ,
\end{equation}

\noindent where $\widetilde{F}_{\lambda}$ is given by Eq.~(\ref{gama}). We introduce the bolometric corrections $\delta_{\rm UV}^*$ and $\delta_{\rm IR}^*$ defined as:

\begin{equation}
\label{duvg2}
\delta_{\rm UV}^* = \frac{\int_0^{\lambda_a}F_{\lambda}[T_{\rm eff}(\gamma)]{\rm d}\lambda}
{\int_{\lambda_a}^{\lambda_b}F_{\lambda}[T_{\rm eff}(\gamma)]{\rm d}\lambda}\ ;\ 
\delta_{\rm IR}^* = \frac{\int_{\lambda_b}^{\infty}F_{\lambda}[T_{\rm eff}(\gamma)]{\rm d}\lambda}
{\int_{\lambda_a}^{\lambda_b}F_{\lambda}[T_{\rm eff}(\gamma)]{\rm d}\lambda}\ ,
\end{equation}

\noindent which are written in terms of fluxes calculated for $T_{\rm eff}(\gamma)$ that do not undergo
the transformation used for $\widetilde{F}_{\lambda}$ in the far- and extreme-UV. Subtracting $\delta_{\rm UV}^*$ from (\ref{duvg1}) and taking into account the definition of $T_{\rm eff}(\gamma)$ written as:

\begin{equation}
\label{duvg3}
\frac{\sigma_R}{\pi}T^4_{\rm eff}(\gamma) = \int_{\lambda_a}^{\lambda_b}F_{\lambda}[T_{\rm eff}(\gamma)]{\rm d}\lambda\times(1+\delta^*_{UV}+\delta^*_{IR})\ ,
\end{equation}

\noindent by making use of (\ref{tgamal2}), we readily obtain:

\begin{equation}
\label{duvg}
\delta_{\rm UV}(\gamma) = \delta^*_{\rm UV}\!-\!(\gamma\!-\!1)(1\!+\!\delta^*_{\rm UV}\!+\!\delta^*_{\rm IR})\times\Lambda[T_{\rm eff}(\gamma),\log g]\ ,
\end{equation}

\noindent where $\delta^*_{\rm UV}$, $\delta^*_{\rm IR}$ and the function $\Lambda$ are given in Table~\ref{deltas}.\par
\begin{figure}
\centerline{\psfig{file=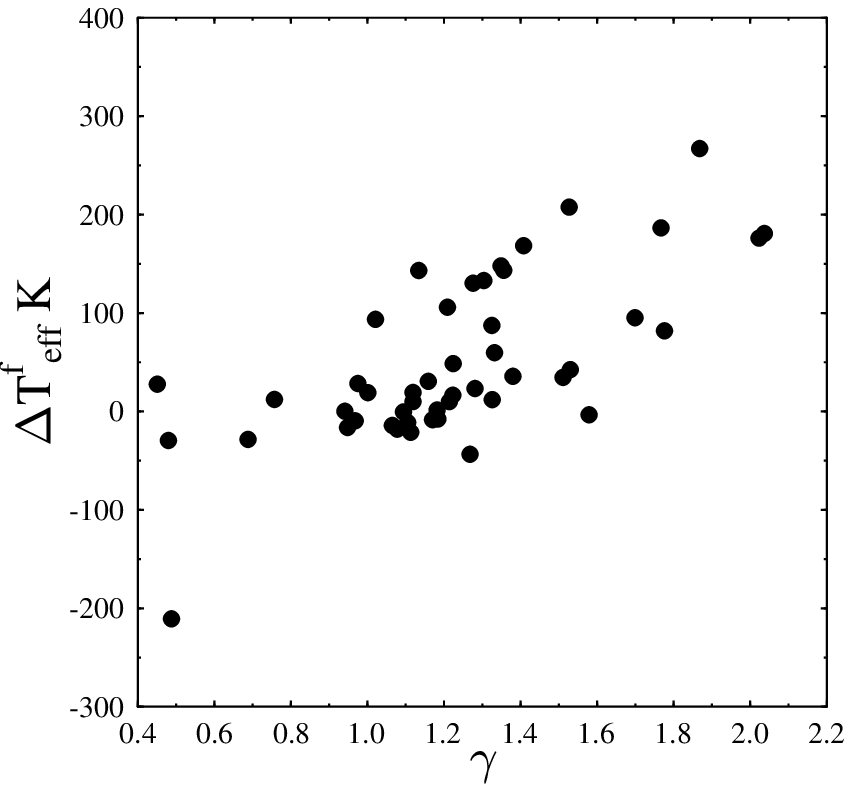,width=6.5truecm,height=6.5truecm}}
\caption{\label{gamma}Difference $\Delta T^f_{\rm eff}=T_{\rm eff}^f(\gamma\neq1)-T_{\rm 
eff}^f(\gamma=1)$ against $\gamma$.}
\end{figure}

 In the visible wavelengths used to calculate the stellar angular diameter, the blocking factor is $b_{\lambda}\,\sim0$ and the enhancement parameter is reduced to $\gamma\!=\!1$. However, the model fluxes employed to obtain $\theta_{\lambda}$ in Eq. (\ref{teta}) are for the effective temperature $T_{\rm eff}(\gamma)$, i.e. $F_{\lambda}[T_{\rm eff}(\gamma)]$. In general, when $\gamma\neq1$ the effective temperature $T_{\rm eff}^f$ issued from (\ref{teff}) is higher than for $\gamma=1$. Figure \ref{gamma} shows $\Delta T^f_{\rm eff}=T^f_{\rm eff}(\gamma\neq1)-T^f_{\rm eff}(\gamma=1)$ for our sample of supergiants. Thus, even though $\gamma$ can sometimes be as high as $\gamma\!\sim\!2$, we see that the differences in the final value of $T_{\rm eff}^f$ are less than 300 K, which is much lower than the average expected error affecting the determination of $T_{\rm eff}^f$ (see \S\ref{comm}). Finally, we note that for the stars in common with \citet{rem81}, our $\gamma$ values are systematically smaller by $\Delta\gamma\sim0.5$ than theirs, probably because in the models used here the line blocking effect is more realistic than in the Kurucz (1979) models.\par

\section{The program stars and the observed quantities}\label{observations}

\subsection{BCD parameters}

 The program B-type stars, dwarfs to supergiants, are simply those for which both the BCD
($\lambda_1, D$) parameters and calibrated fluxes from the far-UV to the IR, at least up to $1\mu$m were available. In this work we excluded stars with the Be phenomenon, but included some A0 to A2 type stars in the cold extreme of the hot fold of the BCD ($\lambda_1$,$D$) diagram. The list of the program stars and their ($\lambda_1$,$D$) are given in Table \ref{prog_stars}.\par
 Our sample contains 217 stars with MK luminosity classes from V to Ia. Observations in the BCD system were carried out over different periods. Most of them were observed with the Chalonge spectrograph \citep{bachadi73} attached to several telescopes at the Haute Provence Observatory from 1977 to 1987 and at the ESO (La Silla) from 1978 to 1988. More recently, on January 2006, low resolution spectra in the optical wavelength range $\lambda\lambda$\,3400-5400 \AA\ were obtained at CASLEO (Argentina), with the Boller \& Chivens spectrograph. These spectra were wavelength and flux calibrated. Since the parameter $\lambda_1$ depends on the spectral resolution, we selected a resolution of 7 \AA\ at $\lambda$ 3760 \AA, which is similar to the resolution required in the original BCD system. From the spectra obtained at CASLEO, the BCD ($\lambda_1$ and $D$) parameters were directly measured on the spectrograms. We have determined the parameter $D = \log(F_{+3700}/F_{-3700})$ following the original BCD prescriptions \citep{cha52}. We also controlled that $F_{-3700}$ corresponds to the flux level where the higher members of Balmer lines merge. Most stars have been observed many times, so that for each star the adopted ($\lambda_1, D$) set comes from 2 to about 50 determinations. The uncertainties that characterize these quantities are then $\sigma_D\simeq0.015$ dex and $\sigma_{\lambda_1}\simeq2$\AA, respectively.\par

\subsection{ASED data}\label{ased-data}

  The ASED data were collected in the CDS astronomical database. The far-UV fluxes used in this work were obtained with the IUE spectra in low resolution mode and the 59 narrow-band fluxes, between 
1380 and  2500 \AA\, measured during the S2/68 experiment from the TD1 satellite \citep{jam76,mac78}. We have compared the TD1 fluxes with the low resolution IUE spectra calibrated in absolute fluxes, but we have not detected any systematic deviation neither in the far-UV ASED nor in the values of $f$ obtained. The TD1 fluxes give a homogeneous far-UV flux data set, for they are laboratory based calibrations.\par 
  In the visible and near-IR spectral region, fluxes are from Breger's catalogue \citep{bre76a,bre76b} and the 13-color photometry calibrated in absolute fluxes \citep{joh75} (hereinafter JM). The normalized 13-color fluxes of the stars in common were compared to \citet{bre76a,bre76b} spectrophotometry and to the monochromatic fluxes observed by \citet{tug80}. The comparison of JM's absolute fluxes with those of known flux standards  $\alpha$ Lyr (\object{HD 172167}), 109 Vir (\object{HD 130109}) and $\eta$~UMa (\object{HD 120315}) given in the \citet{hay75} system revealed no noticeable differences in the $\lambda\lambda$0.58-0.80 $\mu$m spectral region, which was chosen in this work to obtain the stellar angular diameter $\theta^{f}$. Nevertheless, there are small differences near the BD that lie within the uncertainties of other calibrations, but they do not affect the estimate of $f$. We also noticed that JM's fluxes give consistent continuations to the IUE and ANS fluxes \citep{wes82}.\par 

\subsection{Correction for interstellar extinction}\label{ISM}

 The adopted ISM extinction law is from \citet{car89} and \citet{odo94}. The ratio $R_{\rm V}=\,A_V/E(B-V)$ was adopted in this work according to the galactic region, as specified in \citet{gul87} and \citet{gul89}. The adopted color excess $E(B-V)$ is the average of several more or less independent determinations based on the following methods: a) UBV photometry with the standard intrinsic colors given by \citet{lang92}; b) BCD system with the intrinsic gradients $\Phi^o_{rb}$ taken from the calibration done by  \citet{cha73} from where we obtain that $E(B-V)$ = 0.55\,[$\Phi_{rb}-\Phi^o_{rb}$]; c) depths of the 2200 \AA\, ISM absorption band \citep{Bee80,zor85}; d) profile parameters of the 2200 \AA\, band \citep{gur82,fri83,fri87} calibrated in $E(B-V)$ \citep{mou93}; e) diagrams of $E(B-V)$ vs. distance obtained with `normal' stars surrounding the program stars within less than 2$^o$. The UBV photometry used is from the CDS compilation and the distances are either from the {\sc hipparcos} satellite or spectroscopy.

\subsection{Results}

 The list of the program stars is given in Table \ref{prog_stars}: 1) HD number of the star; 2) the MK SpT/LC determined with the BCD system; 3) the parameter $\lambda_1$ in $\lambda_1-3700$ \AA; 4) the Balmer discontinuity $D$ in dex; 5) the adopted average color excess $E(B-V)$, in mag; 6) the effective temperature $T_{\rm eff}^f$ with its estimated uncertainty, given in K; 7) the angular diameter with its estimated uncertainty, given in mas (milliarcseconds); 8) the parameter $\gamma$ used to fit the observed far-UV ASED (see \S \ref{tspg}); 9) the $T_{\rm eff}(\lambda_1, D)$ read on the new calibration curves (see Fig. \ref{BCD-temp}a).\par

\begin{figure*}
\centerline{\psfig{file=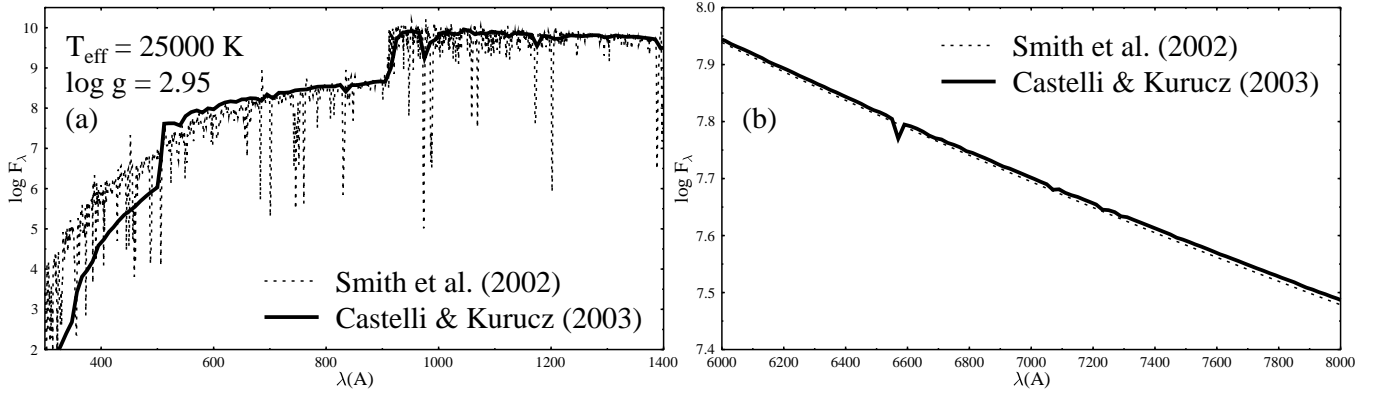}}
\caption{\label{fluxcomp}Comparison of theoretical ASEDs from non-LTE blanketed model atmospheres with stellar winds for $T_{\rm eff}\!=$ 25\,000 K and $\log g\!=$ 2.95 calculated by Smith et al. (2002) (doted line), with LTE wind-free models obtained by  Castelli \& Kurucz (2003) (full line).}
\end{figure*}

\section{Comments on the uncertainties affecting the $T_{\rm eff}^f$ and $\theta^f$ determinations}
\label{comm} 
\subsection{Systematic deviations}\label{commse}

 Hot supergiants have massive winds whose optical depths can be high enough to heat somewhat the photosphere back to the continuum formation region. Thus, both spectral lines and the emitted continuum energy distribution can in principle be modified \citep{abb85,gab89,smi02,mori04}. Since the models used in this work are wind-free, doubts can be raised about whether the use of these models introduces systematic effects on the estimate of $T^f_{\rm eff}$. As is seen in \S\ref{comp}, this question may be of particular interest at $T_{\rm eff}=$ 25\,000 K. We then compared the absolute energy distribution produced by a non-LTE BW model atmosphere for supergiants \citep{smi02} with that predicted by an LTE wind-free model atmosphere for $T_{\rm eff}=$ 25\,000 K and $\log g=$ 2.95, similar to those used in the present work \citep{cas03}. This comparison is shown in Fig. \ref{fluxcomp} where we see [Fig. \ref{fluxcomp}(a)] that the most significant differences appear only in the extreme-UV energy distribution, at $\lambda\la500$ \AA. In this spectral region the fluxes are between two and three orders of magnitude lower than in slightly longer wavelengths, where the most significant contribution to the value of the the filling factor $\delta_{\rm UV}$ in relation (\ref{fl}) arises. Although the differences seen in $\lambda\la500$ \AA\ can certainly be important for the excitation/ionization of the stellar environment, as we shall see below, they seem to have a limited incidence on the estimate of the stellar bolometric flux $f$ on which the estimate of $T_{\rm eff}^f$ relies.\par
\begin{figure}
\centerline{\psfig{file=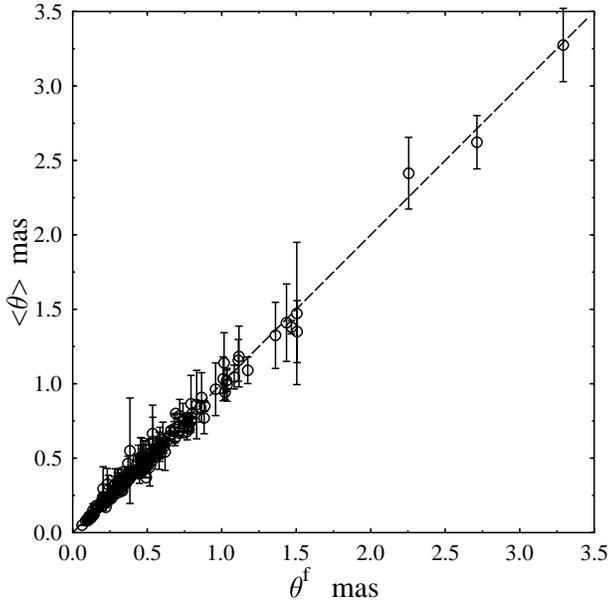,width=8.7truecm,height=8.7truecm}}
\caption{\label{theta}Average apparent angular diameters $\langle\theta\rangle$ in mas of our program stars determined by other authors (ordinates) against $\theta^f$ obtained in the present work (abscissa).}
\end{figure}

 In Fig. \ref{fluxcomp}(b) we also see that the fluxes produced by both types of models are similar in the visible wavelengths where the angular diameter $\theta^f$ is calculated. This ensures that $\theta^f$ does not suffer from the use of plane-parallel model atmospheres either. The same conclusion is reached from Fig. \ref{theta} by comparing the obtained $\theta^f$ values with the measured angular diameters.\par
\begin{figure}
\centerline{\psfig{file=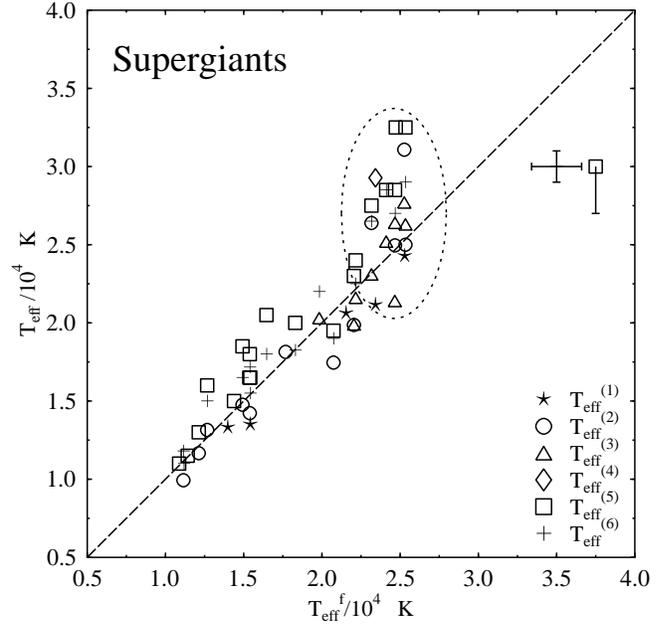,width=8.7truecm,height=8.7truecm}}
\caption{\label{spg_otros}Effective temperatures of supergiant stars determined by other authors (ordinates) against the $T_{\rm eff}^f$ estimates obtained in the present work (abscissa). The error bars correspond to temperatures inside the ellipse taken from \citet{cro06}(vertical) and in the present work (horizontal). The square with a downward error bar indicates the systematic average shift that the \citet{mce99} data might have.}
\end{figure}

 
  Let us now explore what systematic deviation can be expected in our $T_{\rm eff}^f$ estimates due to the use of fluxes predicted by wind-free models of stellar atmospheres. In what follows, we use the notation $T_{\rm eff}^{f,w}$ and $\delta^w_{\rm UV,IR}$ to designate the parameters that should be derived with non-LTE BW model atmospheres. In \S\ref{appad} and Fig.~\ref{theta} it is demonstrated that the angular diameters $\theta^f$ obtained agree well with the observed ones. This means that they cannot introduce a systematic deviation in the derivation of $T_{\rm eff}^f$ with Eq.~(\ref{teff}). Then, the ratio between $T_{\rm eff}^{f,w}$ and $T_{\rm eff}^f$ is given by:

\begin{equation}
\label{rteff}
\frac{T_{\rm eff}^{f,w}}{T_{\rm eff}^f} = \frac{1+\delta^w_{\rm UV}+\delta^w_{\rm IR}}{1+\delta_{\rm UV}(\gamma)+\delta_{\rm IR}(\gamma)}\ ,
\end{equation}

\noindent where $\delta_{\rm UV}(\gamma)$ and $\delta_{\rm IR}(\gamma)$ are the bolometric corrections calculated with wind-free models for different values of the enhancement parameter $\gamma$. The value of $\delta_{\rm UV}(\gamma)$ is obtained from Eq.~(\ref{duvg}), while it is reasonable to assume that $\delta^w_{\rm IR}\!\simeq\!\delta_{\rm IR}(\gamma)\!=\!\delta_{\rm IR}^*$. Concerning the non-LTE BW model $\delta^w_{\rm UV}$ bolometric correction, we ask what extreme-UV flux excess carried by the wind-related effects must exist to explain the underestimations suggested by the comparison shown in Fig.~\ref{spg_otros} that possibly affect our $T_{\rm eff}^f$ determinations for supergiants. To this end we simply write: 

\begin{equation}
\label{udvw}
\delta_{\rm UV}(w) = w\times\delta_{\rm UV}^*\ ,
\end{equation}

\noindent where the factor $w$ mimics the extreme-UV flux excess due to wind-related effects. Using for $\delta_{\rm UV,IR}^*$ and $\Lambda(T_{\rm eff},\log g)$ the values given in Table~\ref{deltas}, we readily obtain the estimates of $w$ for a series of imposed underestimations $\Delta T_{\rm eff}^f$ which are displayed in Table~\ref{estw}.\par

\setcounter{table}{2}
\begin{table}
\caption{\label{estw}Extreme-UV flux excess factors $w$ as a function of the effective temperature 
and $\gamma$, needed to explain the $\Delta T_{\rm eff}^f$ underestimations affecting the $T_{\rm eff}^f$ values for supergiants.}
\center
\begin{tabular}{cr|rrr}
\noalign{\smallskip}
 \hline
\noalign{\smallskip}
     &  & \multicolumn{3}{|c}{$w(\gamma)$} \\
\noalign{\smallskip}
\cline{3-5}      
\noalign{\smallskip}
 $T_{\rm eff}$ &  $\Delta T_{\rm eff}^f$ & $\gamma\!=\!1.0$ & $\gamma\!=\!1.5$ & $\gamma\!=\!2.0$ \\
        K      &             K \ \ \ \ \ &                  &                  &                  \\
\noalign{\smallskip}
\hline 
\noalign{\smallskip} 
10000 &  $-$500 & 10.73 &  9.96 &  9.19 \\
      & $-$1000 & 21.94 & 21.02 & 20.09 \\    
      & $-$1500 & 34.80 & 33.69 & 32.59 \\    
      & $-$2000 & 49.45 & 48.14 & 46.83 \\    
\noalign{\smallskip}
\hline 
\noalign{\smallskip} 
15000 &  $-$500 & 1.78 &  1.36 &  0.92 \\
      & $-$1000 & 2.68 &  2.18 &  1.68 \\    
      & $-$1500 & 3.64 &  3.08 &  2.52 \\    
      & $-$2000 & 4.70 &  4.06 &  3.43 \\    
\noalign{\smallskip}
\hline 
\noalign{\smallskip} 
20000 &  $-$500 & 1.32 &  1.05 &  0.79 \\
      & $-$1000 & 1.67 &  1.37 &  1.08 \\    
      & $-$1500 & 2.04 &  1.71 &  1.39 \\    
      & $-$2000 & 2.43 &  2.08 &  1.72 \\    
\noalign{\smallskip}
\hline 
\noalign{\smallskip} 
25000 &  $-$500 & 1.20 &  1.06 &  0.92 \\
      & $-$1000 & 1.41 &  1.27 &  1.11 \\    
      & $-$1500 & 1.64 &  1.48 &  1.32 \\    
      & $-$2000 & 1.89 &  1.71 &  1.53 \\    
\noalign{\smallskip}
\hline 
\noalign{\smallskip} 
30000 &  $-$500 & 1.14 &  1.04 &  0.95 \\
      & $-$1000 & 1.28 &  1.18 &  1.08 \\    
      & $-$1500 & 1.43 &  1.32 &  1.22 \\    
      & $-$2000 & 1.59 &  1.47 &  1.36 \\    
\noalign{\smallskip}
\hline
\noalign{\smallskip}
\multicolumn{5}{l}{The estimates are done for $\log g=3.0$}\\
\noalign{\smallskip}
\hline
\end{tabular}
\end{table}

 From the non-LTE BW models published by \citet{smi02} we obtain,

\begin{eqnarray}
\label{ktg}
\begin{array}{ccc}
w(25\,000,2.95) = 1.05 \\
w(30\,200,3.14) \simeq 1.08 \\
\end{array}
\end{eqnarray}

\noindent for $T_{\rm eff}\!=$ 25\,000 K, $\log g\!=$ 2.95 and $T_{\rm eff}\!=$ 32\,200 K, $\log g\!=$ 3.14,respectively, where $T_{\rm eff}\!=$ 25\,000 K corresponds to the effective temperature where there are strong deviations in the diagram of Fig.~\ref{spg_otros}. Unfortunately \citet{smi02} have not made available fluxes for $T_{\rm eff}<$ 25\,000 K. Then, by extrapolation and approximate calculation with our codes for extended spherical atmospheres \citep{cruz07}, we get the value $w(20\,000,3.0)\simeq1.04$. The values of $w$ derived here are close to those in Table~\ref{estw} for $\gamma\!=\!1.5$. Since for the superginats that are in common with those in the $T_{\rm eff}^{(6)}$ category
(see \S\ref{efftemp}), we have $\gamma\lesssim1.5$, it means that we may expect our $T_{\rm eff}^f$ to be systematically smaller by less than 470 K, 480 K and 640 K, for objects whose temperatures are 20\,000 K, 25\,000 K and 30\,000 K, respectively. In addition, for later B sub-spectral types, the underestimates can be even smaller, as it is difficult to believe that the $w$ parameters are larger than those quoted above when $T_{\rm eff}\!<\!20\,000$ K. For $\gamma\!\simeq\!2$ (see Table~\ref{estw}) differences easily can approach 1\,000 K. Nevertheless, there is only one common late type supergiant with $\gamma\!>\!1.5$ (HD 202850, $\gamma\!=\!1.78$), for which curiously the difference between our estimate and that in the $T_{\rm eff}^{6)}$ group is $+170$ K. We note that to obtain the imposed underestimation $\Delta T_{\rm eff}^f$ when $w\!<\!1$ for a given factor $\gamma$, the non-LTE BW models should predict lower extreme-UV fluxes than wind-free models do.\par
\begin{figure}
\centerline{\psfig{file=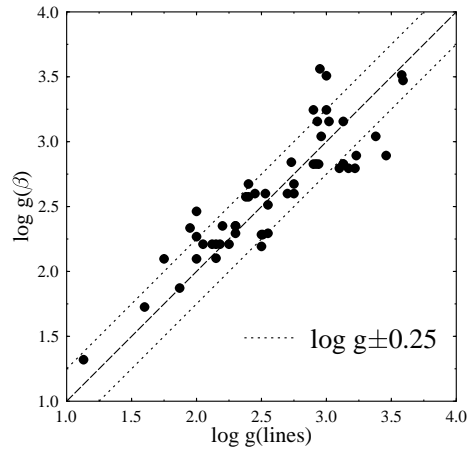,width=6.5truecm,height=6.5truecm}}
\caption{\label{complogg}Comparison of $\log g(\beta)$ parameters used in this work and derived with the uvby$-\beta$ photometry, with those for the common supergiant stars estimated by other authors through spectral line fitting with model atmospheres.}
\end{figure}

  The model atmospheres used in this work to obtain $T_{\rm eff}^f$ are interpolated here for $\log g$ parameters that were estimated using the uvby$-\beta$ photometry. As the index $\beta$ is calibrated mainly for dwarf and giant stars, to see whether its extrapolation induces significant systematic effect for supergiants, we compare in Fig. \ref{complogg} the $\log g(\beta)$ adopted in the present work, with the $\log g$(lines) parameters derived in the literature by detailed fitting of spectral lines with model atmospheres, mainly hydrogen H$\gamma$ and H$\delta$  lines \citep{mce99,rep04,mar05,
cro06,ben07,mar08, sea08}. From this comparison we can conclude that our estimates of $\log g$ do not deviate either strongly or in a systematic way from those based on spectral lines. Although in the cited works the authors claim that their $\log g$(lines) are determined with errors ranging from 0.10 to 0.15 dex, they differ among them by 0.10 to 0.29 dex. The comparison in Fig.~\ref{complogg} of our $\log g(\beta)$ estimates with $\log g$(lines) shows that most of the points deviate randomly from the first diagonal within $0.25$ dex. This implies that we do not expect that the $\log g(\beta)$ values used in the present work will induce systematic errors in the $T_{\rm eff}^f$ and $\theta^f$ parameters.\par
 In conclusion, we do not see how we can attribute to our BFM method systematic effective temperature deviations attaining 2\,000 K, or more than 5\,000 K, as observed in Fig. \ref{spg_otros}.\par

\subsection{Random errors}
\label{rderr}

 The $T_{\rm eff}^f$ values issued from relation (\ref{teff}) and the $\theta^f$ parameters derived with Eq. (\ref{teta}) have the following sources of error: a) the ISM color excess $E(B-V)$; b) the line blocking enhancement parameter $\gamma$; c) the $\log g$ parameter on which depend the model fluxes used to estimate the angular diameter and the filling factor due to the unobserved spectral region; d) the filling factor $\delta$ used to calculate the bolometric flux in relation (\ref{fl}).\par
 In Appendix \S\ref{err_calc} we explain in detail how we have calculated the uncertainties of $T_{\rm eff}^f$ and $\theta^f$ by taking into account the combined effect of all error sources mentioned above. Since the most important uncertainty on $T_{\rm eff}^f$ and $\theta^f$ is produced by the error on the ISM color excess estimate $E(B-V)$, in this section we consider only the effect caused by this parameter as if it were the sole error source. Let us recall that the $E(B-V)$ affects the bolometric flux estimate through $f_{\rm obs}$ in (\ref{fl}) of \S\ref{tdwgi}, and the monochromatic fluxes on which the calculation of the stellar angular diameter depends. However, from (\ref{teff}) and (\ref{teta}) we find that:

\begin{equation}
\label{ertt}
T_{\rm eff}^f \propto \bigl(f/f_{\lambda}\bigr)^{1/4}
\end{equation}

\noindent which implies that the effect on the estimate of $T_{\rm eff}^f$ by an error on $E(B-V)$ is reduced by the error of the monochromatic fluxes entering the calculation of $\theta^f$. In Table \ref{erttt} are given the estimates of the errors produced on $T_{\rm eff}^f$ and $\theta^f$ by the uncertainty $\Delta E(B-V)$ in the ISM color excess, assuming that it is the only source of error. We note the asymmetric propagation of uncertainties in $T_{\rm eff}^f$ and $\theta^f$ due to those in $E(B-V)$. Since for us $\Delta E(B-V)\la0.03$, from Table \ref{erttt} we find that on average the corresponding errors are of 5\% in $T_{\rm eff}^f$ and 2\% in $\theta^f$.\par

\setcounter{table}{3}
\begin{table}
\caption{\label{erttt}Uncertainties on $T_{\rm eff}^f$ and $\theta^f$ as a function of $T_{\rm eff}$ and $\log g$ due to errors in the estimate of the ISM color excess $E(B-V)$.}
\tabcolsep 2.5pt
\begin{tabular}{c|cccc|cccc}
\noalign{\smallskip}
 \hline
\noalign{\smallskip}
     & \multicolumn{4}{c|}{$\Delta T_{\rm eff}/T_{\rm eff}^f$} & \multicolumn{4}{c}{$\Delta\theta/\theta^f$} \\
\noalign{\smallskip}
\cline{2-9}      
\noalign{\smallskip}
 $T_{\rm eff}$    & \multicolumn{4}{c|}{$\Delta E(B-V)$ mag} & \multicolumn{4}{c}{$\Delta E(B-V)$ mag} \\
      K           & $-$0.07 & $-$0.03 & 0.03 & 0.07 & $-$0.07 & $-$0.03 & 0.03 & 0.07 \\
\noalign{\smallskip}
\hline 
\noalign{\smallskip} 
     & \multicolumn{8}{c}{$\log g = 4.5$} \\
\noalign{\smallskip} 
10000 & $-$0.054 &  $-$0.025 & 0.027 &  0.067 &  $-$0.054 &  $-$0.023 &  0.023 &  0.054 \\
15000 & $-$0.087 &  $-$0.040 & 0.043 &  0.106 &  $-$0.041 &  $-$0.017 &  0.017 &  0.039 \\
20000 & $-$0.100 &  $-$0.045 & 0.049 &  0.120 &  $-$0.036 &  $-$0.015 &  0.015 &  0.034 \\
25000 & $-$0.107 &  $-$0.048 & 0.052 &  0.127 &  $-$0.033 &  $-$0.014 &  0.014 &  0.032 \\
30000 & $-$0.111 &  $-$0.050 & 0.054 &  0.132 &  $-$0.031 &  $-$0.013 &  0.013 &  0.030 \\
\noalign{\smallskip}
     & \multicolumn{8}{c}{$\log g = 3.5$} \\
\noalign{\smallskip}
10000 & $-$0.050 &  $-$0.023 & 0.025 &  0.062 &  $-$0.052 &  $-$0.022 &  0.022 &  0.051 \\
15000 & $-$0.081 &  $-$0.037 & 0.040 &  0.098 &  $-$0.038 &  $-$0.016 &  0.015 &  0.035 \\
20000 & $-$0.093 &  $-$0.042 & 0.045 &  0.111 &  $-$0.032 &  $-$0.013 &  0.013 &  0.029 \\
25000 & $-$0.100 &  $-$0.045 & 0.048 &  0.117 &  $-$0.028 &  $-$0.012 &  0.012 &  0.026 \\
30000 & $-$0.104 &  $-$0.046 & 0.050 &  0.121 &  $-$0.027 &  $-$0.011 &  0.011 &  0.025 \\
\noalign{\smallskip}
     & \multicolumn{8}{c}{$\log g = 2.5$} \\
\noalign{\smallskip}   
10000 & $-$0.059 &  $-$0.027 & 0.030 &  0.073 &  $-$0.056 &  $-$0.024 &  0.024 &  0.055 \\
15000 & $-$0.094 &  $-$0.043 & 0.047 &  0.116 &  $-$0.044 &  $-$0.018 &  0.018 &  0.042 \\
20000 & $-$0.108 &  $-$0.049 & 0.053 &  0.131 &  $-$0.039 &  $-$0.016 &  0.016 &  0.037 \\
25000 & $-$0.116 &  $-$0.052 & 0.057 &  0.139 &  $-$0.036 &  $-$0.015 &  0.015 &  0.035 \\
30000 & $-$0.120 &  $-$0.054 & 0.059 &  0.144 &  $-$0.034 &  $-$0.015 &  0.014 &  0.034 \\
\noalign{\smallskip}
\hline
\end{tabular}
\end{table}
 
\begin{figure}
\centerline{\psfig{file=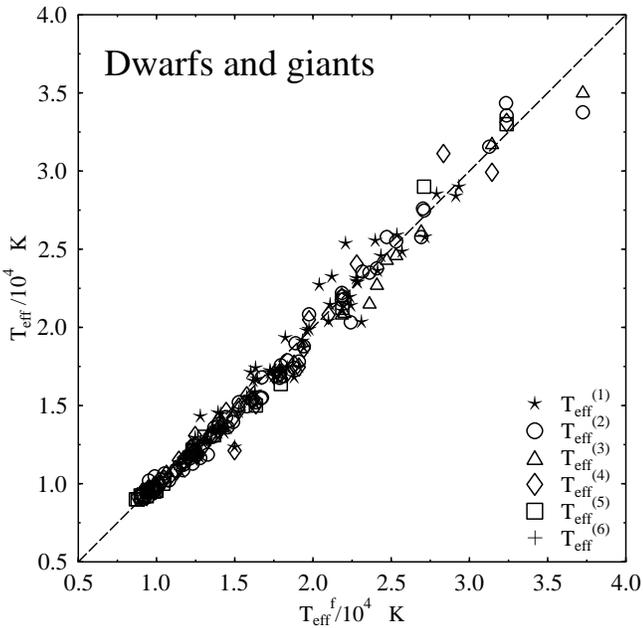,width=8.7truecm,height=8.7truecm}}
\caption{\label{engig_otros}Effective temperatures of dwarfs and giants determined by other authors (ordinates) against the $T_{\rm eff}^f$ estimates obtained in the present work (abscissa).}
\end{figure}

\section{Comparison with other $\theta$ and $T_{\rm eff}$ determinations}\label{comp}

\subsection{Apparent angular diameters}\label{appad}

 In spite of the fact that for the wavelengths where $\theta^f$ is calculated, all models predict roughly the same flux level, the angular diameter is in principle the quantity among those estimated in this work that can be the most strongly model-dependent. To test our angular diameter determinations, we have compared them with those found in the literature. We used the data collected in \citet{pasi01}, where we discarded the very old determinations and privileged those determined by interferometry. The comparison is shown in Fig. \ref{theta}, where we put on the ordinate axis the average values from the data in \citet{pasi01} and the respective $1\sigma$ error-bars. In this figure we can see that there is no systematic deviation of points from the first bisecting line and that they distribute around it with a fairly uniform dispersion: $\sigma_{\theta}=0.045$ mas. The uncertainty that can affect the $T_{\rm eff}^f$ values due to errors in the angular diameter estimates ranges then from $\Delta T_{\rm eff}^f/T_{\rm eff}^f=$ 1.5\% to 9\% as the angular diameter goes from $\theta=3.0$ mas to $\theta=0.5$ mas.\par

\subsection{Effective temperatures}\label{efftemp}
 
 In order to test the accuracy of the effective temperatures obtained with the BFM method, we compared our values to those found in the literature. The $T_{\rm eff}$ values were gathered according to the method used to determine them:

\begin{itemize}  
\item[1)] The temperature is derived empirically using either the reddening-free color index QUV, as in \citet{gul89}, or the c$_0$ and $\beta$ indices of the Str\"omgren uvby$-\beta$ photometry \citep{cas91}.

\item[2)] The temperature is determined from integration of fluxes over a large interval of wavelengths, as was done by \citet{cod76, und79, mal86}.

\item[3)] The temperature is obtained by means of the BFM by \citet{rem81}.

\item[4)] The temperature is evaluated by comparing the observed visual or UV fluxes with predictions from line-blanketed LTE model atmospheres, as in \citet{mal83,mor85,mal90}.

\item[5)] The temperature is computed by means of: a) NLTE line-blanketed model analysis of ionization balances due to \ion{He}{i}/\ion{}{ii}, \ion{Si}{iii}/\ion{}{iv }and \ion{Si}{ii}/\ion{}{iii} on moderate resolution spectra, as done by \citet{mce99} using the code TLUSTY \citep{hub88}, or b) LTE line-blanketed model fitting of the optical region and H$\gamma$ profile using ATLAS9 model atmospheres as in Adelman et al. (2002). 

\item[6)] The temperature is determined from synthetic fits to the optical spectral range of B supergiants employing unified NLTE line- and non-LTE BW extended model atmosphere codes, such as FASTWIND \citep{pul05} or CMFGEN \citep{hil03}, as in \citet{sea08,mar08,cro06}.

\end{itemize}

\begin{figure}
\centerline{\psfig{file=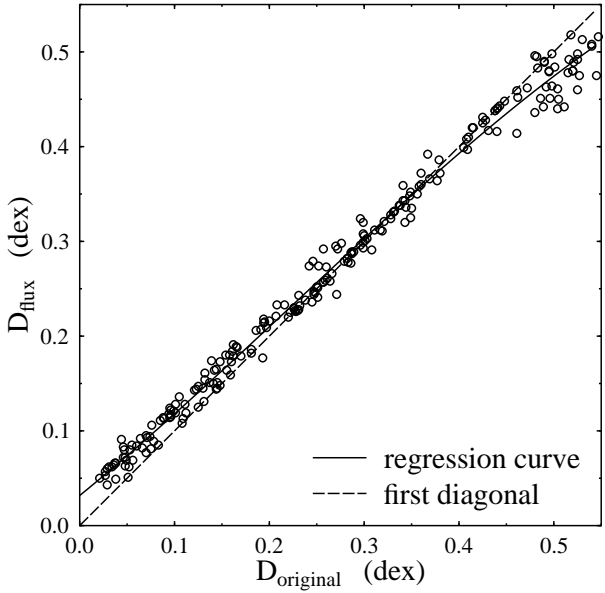,width=8.7truecm,height=8.7truecm}}
\caption{\label{compdd}Comparison between the original BCD Balmer discontinuities of the program stars (abscissa) and those determined with absolute flux-calibrated energy distributions (ordinates).}
\end{figure}

 For each program star, Table \ref{tab_comp} lists our determinations of $T^f_{\rm eff}$ together with the corresponding uncertainties and, in the following columns, the $T^{(i)}_{\rm eff}$ ($i=$ 1,...,6) obtained by other authors with the methods described in items: 1) to 6). When more than one determination of $T_{\rm eff}$ exists for the same star, obtained using similar techniques, we adopted the mean value.\par 
 The difficulties in determining $T_{\rm eff}$ can be better shown by separating dwarfs and giants from supergiants. Figure \ref{engig_otros} displays the comparison of $T^{(1)}_{\rm eff}$ (stars), $T^{(2)}_{\rm eff}$ (open circles), $T^{(3)}_{\rm eff}$ (open triangles), $T^{(4)}_{\rm eff}$ (diamonds), $T^{(5)}_{\rm eff}$ (open squares), and $T^{(6)}_{\rm eff}$ (plus signs) with our $T^f_{\rm eff}$ values (abscissa) for dwarfs and giants. From this figure, we can see that the differences in the $T_{\rm eff}$ estimates seem to be related to the absolute value of temperatures rather than to the adopted techniques. For $T_{\rm eff}\la20\,000$ K, all methods produce temperatures within $\sigma_{T_{\rm eff}}= 750$ K, while for $T_{\rm eff}\gtrsim20\,000$ K, $\sigma_{T_{\rm eff}}= 1\,300$ K. We notice, however, that the $T_{\rm eff}$ derived with the BFM agree best with those obtained using line-blanketed LTE model atmospheres \citep{mal83,mor85,mal90}.\par
 The same type of comparison, but for supergiants, is shown in Fig. \ref{spg_otros} which appears less ordered than the plot for dwarfs and giants. However, the $T_{\rm eff}^{(5)}$ values do deviate strongly and in a systematic way. The $T_{\rm eff}^{(5)}$ values for supergiants are taken only from \citet{mce99}, and they were obtained with non-LTE wind-free model atmospheres. The observed deviations form a kind of `temperature-step' rising at $T_{\rm eff}\sim15\,000$ K and $T_{\rm eff}\sim25\,000$ K, i.e. at temperatures identified as specific to the bi-stability phenomenon \citep{lam95,vink99,cro06,mar08}. This fact may reveal that there can be effects related to the presence of stellar winds which have not been taken into account in their models. Moreover, the noted deviations also reveal a dependence on the method used to determine the effective temperature. The $T_{\rm eff}$ values from \citet{mce99} determined by \ion{He}{ii} line profile fits deviate on average from our temperatures by $\langle\Delta T_{\rm eff}\rangle\!=5\,500\!\pm\!1\,800$ K at abscissa $\langle T_{\rm eff}\rangle\!=24\,400\!\pm\!800$ K; the temperatures determined by the \ion{Si}{iii}/\ion{Si}{iv} ionization balance deviate by $\langle\Delta T_{\rm eff}\rangle\!=1\,300\!\pm\!380$ K at $\langle T_{\rm eff} \rangle\!=20\,400\!\pm\!1\,900$ K; those determined from the \ion{Si}{ii}/\ion{Si}{iii} ioniation balance deviate by $\langle\Delta T_{\rm eff}\rangle\!=2\,200\!\pm\!1\,400$ K at $\langle T_{\rm eff}\rangle\!=1\,4600\!\pm\!1\,500$ K. The $T_{\rm eff}$ values from \citet{mce99} have random errors $\pm1\,000$ K. \citet{mce99} pointed out that their effective temperatures could in fact be 10\% lower, which means that for the stars lying in the ellipse of Fig. \ref{spg_otros} the temperatures are on average 3\,000 K too high. Thus, excluding the $T_{\rm eff}^{(5)}$ values, the average dispersions with the remaining sources cited in Fig. \ref{spg_otros} are $\sigma_{T_{\rm eff}}= 1\,250$ K for $T_{\rm eff}\lesssim20\,000$ K, and $\sigma_{T_{\rm eff}}= 2\,800$ K for $T_{\rm eff}\gtrsim20\,000$ K.\par
 Stars in the $T_{\rm eff}^{(6)}$ group are from several sources, but their temperatures were derived using the same models, although corresponding to different implementation generations. \citet{cro06} note that the current non-LTE BW model atmospheres lead to lower effective temperatures by 1\,000 to 2\,000 K, as compared to some earlier determinations in \citet{kud99,cro02,rep04}, and that their latest effective temperature estimates have uncertainties of about 1\,000 K. However, three of the stars lying in the ellipse of Fig. \ref{spg_otros}: \object{HD 30614}, \object{HD 37128} and \object{HD 38771}, were assigned temperatures that are from 2\,300 K to 3\,700 K higher than those obtained in the present work. From the discussion in Sect. \ref{commse} it appears that our method of determining the effective temperature does not introduce systematic deviations larger than 500 K to 700 K as the effective temperature goes from 20\,000 K to 30\,000 K. We suspect then that the spectral lines used to derive some $T_{\rm eff}^{(6)}$ may still undergo a perturbation that the existing non-LTE BW model atmospheres do not account for entirely. Among the phenomena that may explain the perturbed lines are: non-isothermal multicomponent plasma effects \citep{spr92}, shocks and/or the presence of exo-photospheric density clumps. Regarding the latter, it is worth noting that the use of an average continuous stellar wind to mask a possible clumpy environment, where a mass-loaded wind must appear \citep{hart86}, can lead to misleading conclusions. The radiation transfer effects expected from the resulting average opacity are quite different from those the clumpy environment is able to produce \citep{boiss90}. Furthermore, the actual opacity of the medium is higher than that expected from line diagnostic and the back warming on the photospheric layers could have different characteristics than those expected from winds with regular density and temperature structures.\par

\begin{figure*}
\centerline{\psfig{file=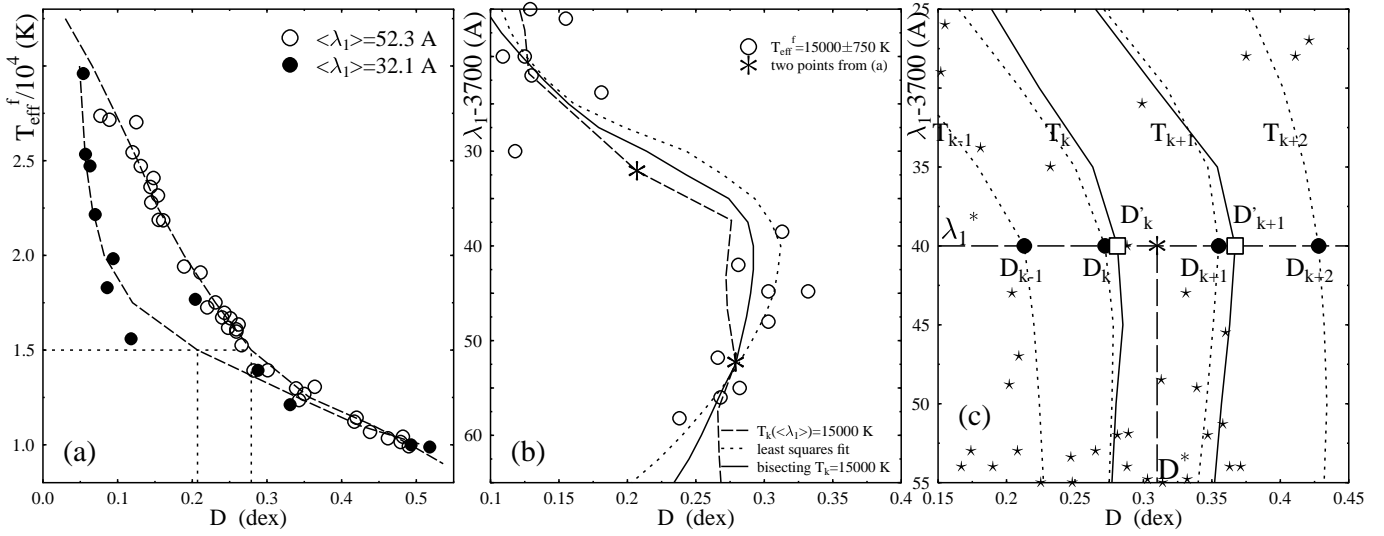}}
\caption{\label{calib_iter}The iteration of $T_k\!=$ constant curves in the plane ($\lambda_1,D$). a) $T^f_{\rm eff}\!=\!T^f_{\rm eff}(D)$ relations for stars in strips of total width $\Delta\lambda_1\!=\!10$ \AA\ and average $\langle\lambda_1\!-\!3700\rangle\!=32.1$ and 52.3 \AA. b) Bisecting $\lambda_1\!=\!\lambda_1(D)$ curve for $\overline{T_k}\!=\!15\,000$ K (full line), between the $\lambda_1\!=\!\lambda_1(D|\overline{T_k})$ obtained from panel a) (dashed curve) and the similar one derived by a least squares fit of a polynomial passing through stars in the plane ($\lambda_1,D)$ having effective temperatures $\overline{T_k}\!\pm\!\Delta T_k$ (dotted curve). c) Iterated correction of the $T_{\rm eff}\!=\!T_{\rm eff}(\lambda_1,D)\!=const.$ layout. ($\lambda^*_1,D^*)=$ coordinates of a test star. $\bullet=$ abscissa of intersections between the $\lambda_1^*=$ constant line and the $T_k\!=$ constant curves at iteration step ``n";  $\square=$ abscissa of intersections between the $\lambda_1^*=$ constant line and the new $T_k\!=$ constant curves at iteration step ``n+1"; $\star=$ program stars.}
\end{figure*}

\section{Calibration of the $(\lambda_1, D)$ parameters into $T_{\rm eff}$}\label{bcdcalib}

\subsection{The scale of the Balmer discontinuities}\label{sbd}

 In the original BCD system, the Balmer discontinuities were obtained by comparing newly observed stars with stars observed simultaneously for which $D$ was known. The `zero' of these $D$ values was known within an uncertainty of some 0.012 dex (L. Divan, private communication). Conversely, for our program stars, we have derived the BD using absolute calibrated fluxes, and compared them with those determined in the original BCD system. The result is shown in Fig. \ref{compdd}. In this figure we can see that, on average, the deviation between both types of BD determinations amounts to the expected 0.012 dex in the region of early sub-spectral types, but that the difference is actually a function of $D$. The least-square fitted relation between the `old' and `new' BDs valid for the $0.03<D<0.55$ dex interval is given by :

\begin{equation}
D_{\rm new} = 0.032+0.817\times D_{\rm old}+0.524\times D_{\rm old}^2-0.775\times D_{\rm old}^3,
\label{dd}
\end{equation}

\noindent where the $D$ are given in dex. Therefore, in this work we used the scale of BDs determined from absolute fluxes, which enable the measured Balmer discontinuities to be directly compared with models. All diagrams in the present work are also given in the flux-calibrated $D-$scale.\par

\begin{figure*}
\centerline{\psfig{file=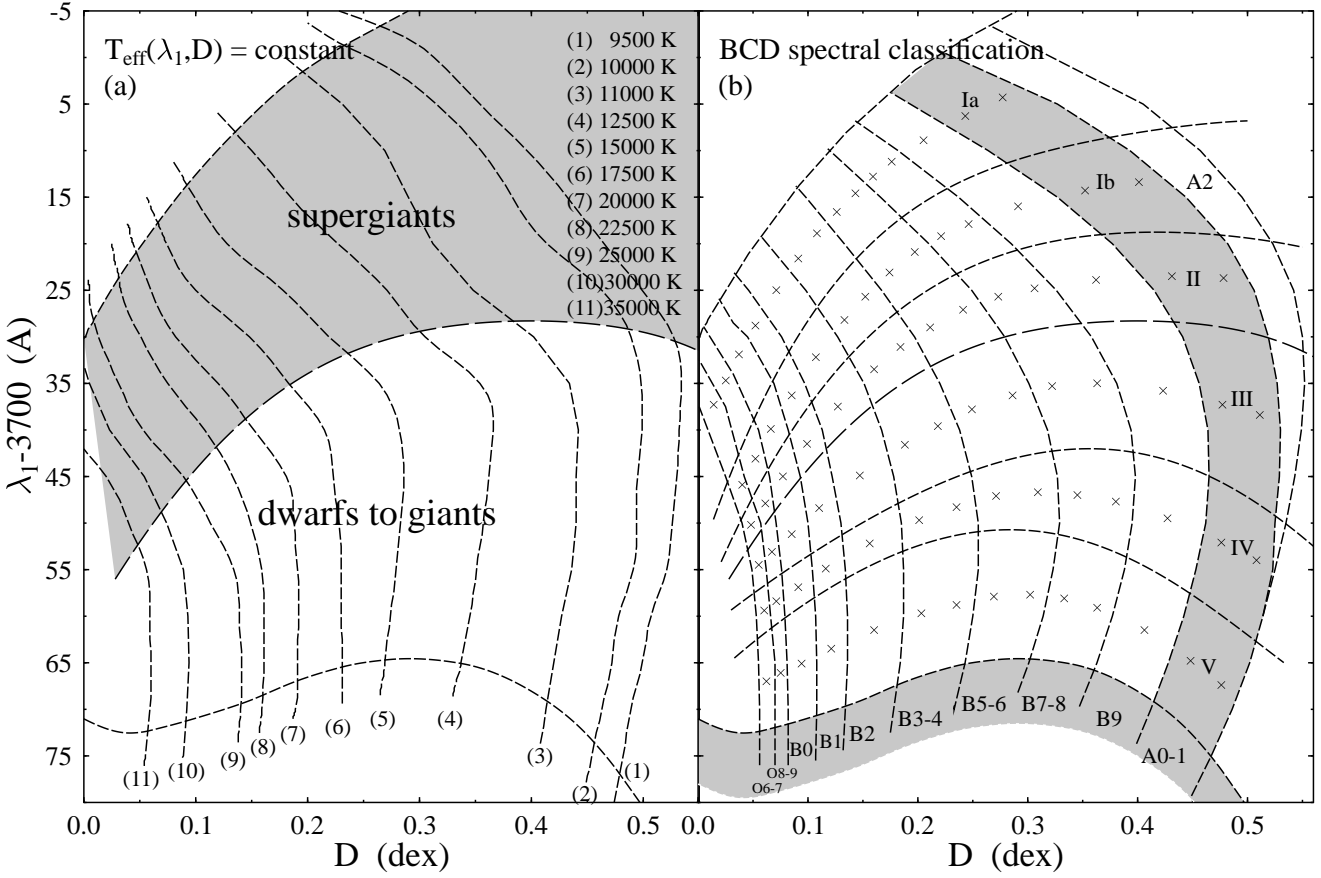,width=18truecm,height=12truecm}}
\caption{\label{BCD-temp}(a): Spline-smoothed $T_{\rm eff}(\lambda_1, D)=const$ curves for the values of effective temperature given in the box at right-top; (b) The curvilinear-quadrilateral BCD $(\lambda_1, D)$ spectral classification diagram, where the corresponding 2D MK spectral types (bottom shaded strip) and the luminosity classes (left shaded strip) are indicated. The cross-marks indicate the mid-point of each MK spectral type-luminosity class box where the $T_{\rm eff}(\lambda_1, D)$ values given in Table 3 were obtained.}
\end{figure*}

\subsection{The calibration}\label{thecal}

 A calibration of the $(\lambda_1, D)$ parameters into effective temperatures for dwarf to giant B-type
stars has been presented in \citet{div83}. In the present work, we are mainly interested in extending the BCD calibration for B supergiants. To ensure a better consistency of the layout of iso-effective temperature curves, we need to calibrate the entire region in the bi-folded BCD ($\lambda_1, D$) surface that corresponds to OB-type stars of all luminosity classes. So, the iso-effective temperature curves for stars earlier than A2 of all luminosity classes are obtained in two successive steps:  1) First, we obtain an approximate layout of curves $T_{\rm eff}(\lambda_1,D)\!=\!T_k\!=\!\!const.$; 2) The shape of the approximate curves $T_{\rm k}\!=\!const.$ is then corrected by iteration. In what follows, a detailed explanation of the procedure used is presented.\par
 1) {\it Approximate system of curves $T_{\rm eff}(\lambda_1,D)\!=\!T_{\rm k}\!=\!\!const.$}:\par
 The system is established as the mean regression curve between the variables $x$ and $y$ from a set of measured pairs of points $(x,y)$. The regression $x\!=\!x(y)$ does not necessarily represent the same locus of points as the one calculated in the form $y\!=\!y(x)$, nonetheless issued from the same $(x,y)$ data set. The bisecting curve between the direct $x\!=\!x(y)$ and the inverse function $x\!=\!y^{-1}(x)$ can then be used to represent the sought relation. Thus, in this work we calculate two series of functions to obtain the approximate average/bisecting system of $\lambda_1\!=\!\lambda_1(D|T_k)$ curves\footnote{Here, the notation $F\!=\!F(u|v)$ means that $F$ is a function of $u$, so that $v\!=\!v(F,u)\!=\!constant$ over the entire space of variables ($F,u$).}, where $T_k\!=\!const.$ with $k\!=\!1,2,...N_{\rm K}$ are constant values of the effective temperature:\par
 $i$) A series of $N_{\rm K}$ regression polynomials $\lambda_1\!=\!\lambda_1(D)$ is calculated using stars whose effective temperatures are in intervals $(T_k-\Delta T_k,T_k+\Delta T_k)$, where $\Delta T_k/T_k\!=\!0.05$. For a given $T_k$ the obtained regression polynomial is assigned to the average $\overline{T_k}$ temperature calculated from those entering the interval $(T_k-\Delta T_k,T_k+\Delta T_k)$;\par
 $ii$) With the $T_{\rm eff}^f$ values of stars lying in successive strips of constant $\lambda_1$ parameters and total width $\Delta\lambda_1\!=\!10$ \AA\ in the $(\lambda_1, D)$ plane, we obtained a series of curves $T_{\rm eff}^f\!=\!T_{\rm eff}^f(D|\langle\lambda_1\rangle)$, where $\langle\lambda_1\rangle$ is the average of $\lambda_1$ parameters of the stars entering a given strip, as shown in Fig. \ref{calib_iter}a for two $\langle\lambda_1\rangle$ values. Then, for a given $\overline{T_k}$ temperature obtained in step $i$), we read the abscissae determined by all the $\langle\lambda_1\rangle = \langle\lambda_1\rangle(D)$ curves, as in the example shown in Fig. \ref{calib_iter}a for the particular case $\overline{T_k}\!=\!15\,000$ K, to draw the corresponding $\langle\lambda_1\rangle = \langle\lambda_1\rangle(D|\overline{T_k})$ (see Fig. \ref{calib_iter}b, dashed curve).\par
 The final curve adopted as the first approach to $\overline{T_k}(\lambda_1,D)\!=\!const.$, here called $\widehat{\lambda_1} = \widehat{\lambda_1}(D|\overline{T_k})$, is the smoothed bisecting locus of points (full line in Fig. \ref{calib_iter}b) between the regression polynomial $\lambda_1\!=\!\lambda_1(D|\overline{T_k})$ obtained in $i$) (dotted line in Fig. \ref{calib_iter}b) and the homologous one $\langle\lambda_1\rangle = \langle\lambda_1\rangle(D|\overline{T_k})$ constructed in $ii$). Using a second order interpolation, the set of  $\widehat{\lambda_1} = \widehat{\lambda_1}(D|\overline{T_k})$ curves with $k\!=\!1,2,...N_{\rm K}$ is then transformed into $\widehat{\lambda_1} = \widehat{\lambda_1}(D|T_k)$ curves, where $T_k\!=\!9\,500, 10\,000,...,35\,000$ K are rounded integer values used in the following step. Hereafter the curves $\widehat{\lambda_1}(D|\overline{T_k})$ are called $\lambda_1(D|T_k)$. \par
 2) {\it Corrected $T_{\rm eff}(\lambda_1,D)\!=\!\!const$ curves}:\par
 The layout of $T_k(\lambda_1,D)\!=\!const$ curves obtained in 1) in principle might be adopted as the sought calibration. We have noticed, however, that we can reduce even more the residuals $\epsilon_{T_{\rm eff}}\!=\!|T_{\rm eff}(\lambda_1^*,D^*)-T_{\rm eff}^f|$, where $T_{\rm eff}(\lambda_1^*,D^*)$ represents the temperature read in the calibrated $(\lambda_1,D)$ plane for the star whose parameters are $(\lambda_1^*,D^*,T_{\rm eff}^f)$. To reduce $\epsilon_{T_{\rm eff}}$ somewhat, we proceed as follows. We consider a star whose parameters are $(\lambda_1^*,D^*,T_{\rm eff}^f)$. We call $D_{k-1}$, $D_k$, $D_{k+1}$ and $D_{k+2}$ the abscissae of the intersections of the $\lambda_1\!=\!\lambda_1^*$ line with the curves of constant effective temperature $\lambda_1\!=\!\lambda_1(D|T_k)$, ordered so that $T_{k-1}\!>\!T_k\!>\!T_{\rm eff}^f\!>\!T_{k+1}\!>\!T_{k+2}$. We calculated the `corrected' abscissae $D^{'}_k$, $D^{'}_{k+1}$ of the curves $T_k$ and $T_{k+1}$ at
$\lambda_1\!=\!\lambda_1^*$ with:

\begin{eqnarray}
\begin{array}{lcl}
D'_k & = & D^*+\bigl(\frac{1/T_k-1/T_{\rm eff}^f}{1/T_{k-1}-1/T_{\rm eff}^f}\bigr)(D_{k-1}-D^*) \\
D'_{k+1} & = & D^*+\bigl(\frac{1/T_{k+1}-1/T_{\rm eff}^f}{1/T_{k+2}-1/T_{\rm eff}^f}\bigr)(D_{k+2}-D^*). \\
\end{array}
\label{corrd}
\end{eqnarray}

\noindent For a given $(\lambda_1^*,D^*)$, the abscissae $D_{k-1}$ and $D_{k+2}$ are maintained unchanged (see Fig. \ref{calib_iter}c). This operation is done for all program stars, so that each of them contributes with its corresponding correcting displacement $D'\!-\!D$. In this way the position of each curve $T_k\!=\!const.$ depends also on the nearest ones. The smoothed curve passing through the new $(\lambda_1,D')$ points relative to the same $T_k\!=\!const.$ is considered as the new $\lambda_1 = \lambda_1(D|T_k)$ curve. Thus, at each iteration the whole system of $T_k$ curves is somewhat modified. The iteration is continued until two conditions are obeyed: 1) the correlation of the read effective temperatures of the program stars in the newly obtained layout of $T_{\rm eff}\!=$ constant curves as a function of the respective $T_{\rm eff}^f$ values does not show any systematic deviation from the first diagonal; 2) the dispersion of points around this line attains its lowest possible value.\par 
 The $T_{\rm eff}(\lambda_1, D)$ diagram thus obtained is shown in Fig. \ref{BCD-temp}a. The upper limiting curve in the diagram of Fig. \ref{BCD-temp}a corresponds to the lowest $\lambda_1$ parameters ever observed in the BCD system for OB supergiants. Whilst $\lambda_1$ values higher than the bottom limiting curve of this diagram have never been measured for normal dwarf OB stars, they are normally observed for O sub-dwarfs and/or white dwarfs.\par  
 The method used to obtain the curves of constant $T_{\rm eff}(\lambda_1, D)$ makes the position of each $T_{\rm eff}\!=\!const$ curve depend on the global $T_{\rm eff}(\lambda_1, D)$ pattern; i.e. a change in one of these curves has an incidence on the shape of the surrounding curves. We also notice that the layout of the $T_{\rm eff}\!=\!const$ curves in the sector of supergiants is strongly constrained by the way the curves are arranged in the $(\lambda_1, D)$ region of dwarfs and giants which are traced by a large sample of stars, and by the natural limiting condition which imposes that $D\!\to\!0$ as $\lambda_1$ becomes negative. Actually, before the limit $D\!\to\!0$ is attained, for supergiants with $T_{\rm eff}\gtrsim 25\,000$ K, there is a luminosity class interval, roughly from $\lambda_1\!-\!3700\!\sim\!55$ \AA\ to $\lambda_1\!-\!3700\!\!\sim\!5$ \AA, the extended low density atmosphere of these stars behaves like the circumstellar envelope of Be stars, producing $D\!<\!0$, i.e. an emission-like Balmer discontinuity. The black body-like behavior $D\!=\!0$ is reached for the theoretical limit $\lambda_1\!-\!3700\!\sim\!-47$~\AA.\par
 To examine the correspondence between the $T_{\rm eff}$ values and the MK spectral classes, we show in Fig. \ref{BCD-temp}b the calibration of the BCD $(\lambda_1, D)$ surface into B-spectral subtypes and luminosity classes. We can see that the pattern of $T_{\rm eff}\!=\!const$ curves mirrors the empirical MK spectral type calibration. The delimiting strips of constant sub-spectral types are determined with color gradient curves $\Phi_{rb}\!=\!const$. This system of curves resembles each other because the color temperature in the wavelength region over which $\Phi_{rb}$ is defined is close to the stellar effective temperature. In Table \ref{midt} we give the values of $T_{\rm eff}$ for the points marked in Fig. \ref{BCD-temp}b with crosses. These are mid-points of more or less large curvilinear boxes corresponding to a given MK SpT/LC. We note that it may not be straightforward to compare the effective temperatures for the nominal MK SpT/LCs given in Table \ref{midt} with the average values determined in the literature for the same MK spectral classes. This is because the baricenters of the $(\lambda_1,D$) parameters in these averages may not correspond to the mid-points of the curvilinear quadrilaterals given here (see crosses in Fig. \ref{BCD-temp}b).\par
\begin{figure}
\centerline{\psfig{file=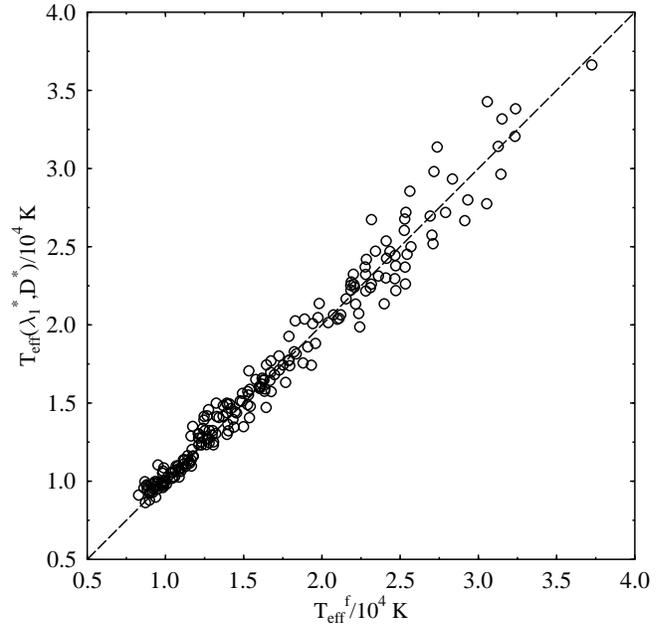,width=8.7truecm,height=8.7truecm}}
\caption{Comparison between effective temperatures read in the ($\lambda_1, D$) plane (ordinates) and those derived with bolometric fluxes of the program stars (abscissa).}
\label{plano}
\end{figure}

 Finally, to verify the reliability of the new BCD $T_{\rm eff}$ calibration, in Fig. \ref{plano} we compare the effective temperatures of the program stars read in the ($\lambda_1,D$) calibration to those derived with bolometric fluxes. The comparison reveals a good correlation, where the dispersion of points around the first diagonal remains within a constant limit $\Delta T_{\rm eff}/T_{\rm eff}\!=\!\pm0.17$. Considering that the $T^f_{\rm eff}$ values represent the reference for our $T_{\rm eff}(\lambda_1,D)$ calibration, the read $T_{\rm eff}$ values in the ($\lambda_1,D$) plane can be considered as affected by two kinds of uncertainties, one of them inherent to the measurement of the Balmer discontinuity, and the other related to the obtained layout of the $T_{\rm eff}\!=$ constant curves. Since the deviation of points around the first diagonal in Fig. \ref{plano} has a Gaussian distribution, which lies in the whole $3\sigma/T_{\rm eff}\!=\!\pm0.17$ interval, the expected total average error affecting the read $T_{\rm eff}(\lambda_1,D)$ quantities is thus \citep{sma58}:

\begin{equation} 
\frac{\overline{\epsilon}}{T_{\rm eff}} = \sqrt{\frac{2}{\pi}}\frac{\sigma}{T_{\rm eff}} \simeq\pm0.05,
\label{er}
\end{equation}

\noindent which is valid for the entire calibrated $(\lambda_1,D)$ plane.\par 

\setcounter{table}{5} 
\begin{table*}
\caption{\label{midt}Effective temperatures at the mid-point of each curvilinear quadrilateral representing a MK spectral type-luminosity class in the BCD ($\lambda_1, D)$ plane for early type stars.}
\tabcolsep 3.5pt
\begin{tabular}{l|ccr|ccr|ccr|ccr|ccr|ccr}
\noalign{\smallskip}
\hline
\noalign{\smallskip}
\multicolumn{1}{c}{LC :} & \multicolumn{3}{c|}{V} & \multicolumn{3}{c|}{IV} & \multicolumn{3}{c|}{III} & \multicolumn{3}{c|}{II} & \multicolumn{3}{c|}{Ib} & \multicolumn{3}{c}{Ia}  \ \ \ \ \\
\noalign{\smallskip}
\hline
\noalign{\smallskip}
SpT & $\lambda_1$ & $D$ & $T_{\rm eff}$ & $\lambda_1$ & $D$ & $T_{\rm eff}$ & $\lambda_1$ & $D$ & $T_{\rm eff}$ & $\lambda_1$ & $D$ & $T_{\rm eff}$ & 
$\lambda_1$ & $D$ & $T_{\rm eff}$ & $\lambda_1$ & $D$ & $T_{\rm eff}$   \ \ \ \ \\          
\noalign{\smallskip}
\hline 
\noalign{\smallskip}
 O6-7 & 67.0 & 0.062 & 34640 &  59.4 & 0.060 & 34800 &  54.5 & 0.055 & 35060 &  50.2 & 0.048 & 33970 &  45.9 & 0.040 & 32590 &  37.3 & 0.014 & 29670  \ \ \ \ \\
 O8-9 & 66.1 & 0.075 & 32750 &  58.4 & 0.071 & 33070 &  53.1 & 0.067 & 32460 &  47.9 & 0.061 & 30690 &  43.1 & 0.052 & 28350 &  34.7 & 0.025 & 26040  \ \ \ \ \\
 B0   & 65.1 & 0.094 & 30000 &  56.9 & 0.091 & 29970 &  51.2 & 0.085 & 29070 &  45.0 & 0.077 & 26710 &  39.9 & 0.066 & 24380 &  31.9 & 0.037 & 23890  \ \ \ \ \\
 B1   & 63.5 & 0.121 & 27080 &  54.9 & 0.116 & 26960 &  48.4 & 0.110 & 24880 &  41.5 & 0.099 & 23370 &  36.3 & 0.085 & 22360 &  28.8 & 0.052 & 22020  \ \ \ \ \\
 B2   & 61.5 & 0.160 & 22620 &  52.2 & 0.156 & 22380 &  44.9 & 0.147 & 21740 &  37.5 & 0.127 & 20570 &  32.2 & 0.107 & 19790 &  25.0 & 0.071 & 19310  \ \ \ \ \\
 B3   & 59.7 & 0.203 & 19270 &  49.7 & 0.201 & 19160 &  41.6 & 0.188 & 18900 &  33.5 & 0.160 & 18010 &  28.2 & 0.133 & 17520 &  21.6 & 0.091 & 17140  \ \ \ \ \\
 B4   & 58.8 & 0.235 & 17220 &  48.3 & 0.235 & 17130 &  39.6 & 0.218 & 17310 &  31.1 & 0.184 & 16500 &  25.7 & 0.152 & 16180 &  18.9 & 0.108 & 15660  \ \ \ \ \\
 B5   & 57.9 & 0.269 & 15380 &  47.1 & 0.271 & 15570 &  37.8 & 0.249 & 15890 &  29.0 & 0.211 & 15080 &  23.1 & 0.174 & 14770 &  16.6 & 0.126 & 14460  \ \ \ \ \\
 B6   & 57.7 & 0.302 & 14100 &  46.7 & 0.309 & 14200 &  36.3 & 0.286 & 14530 &  27.1 & 0.241 & 13790 &  20.9 & 0.197 & 13510 &  14.6 & 0.143 & 13560  \ \ \ \ \\
 B7   & 58.1 & 0.333 & 13000 &  47.0 & 0.345 & 13000 &  35.3 & 0.322 & 13460 &  25.7 & 0.273 & 12600 &  19.2 & 0.221 & 12540 &  12.8 & 0.159 & 12800  \ \ \ \ \\
 B8   & 59.1 & 0.363 & 12190 &  47.7 & 0.380 & 12120 &  35.0 & 0.363 & 12380 &  24.8 & 0.306 & 11790 &  17.9 & 0.246 & 11930 &  11.2 & 0.176 & 12280  \ \ \ \ \\
 B9   & 61.5 & 0.406 & 11340 &  49.5 & 0.427 & 11240 &  35.8 & 0.423 & 11240 &  23.9 & 0.362 & 10800 &  16.0 & 0.291 & 11070 &   8.9 & 0.205 & 11720  \ \ \ \ \\
 A0   & 64.8 & 0.448 & 10470 &  52.1 & 0.476 & 10310 &  37.3 & 0.477 & 10350 &  23.5 & 0.431 & 10140 &  14.3 & 0.352 & 10310 &   6.3 & 0.243 & 11010  \ \ \ \ \\
 A1   & 67.4 & 0.476 &  9860 &  54.0 & 0.508 &  9730 &  38.4 & 0.511 &  9820 &  23.7 & 0.478 &  9680 &  13.4 & 0.401 &  9690 &   4.3 & 0.277 & 10440  \ \ \ \ \\
\noalign{\smallskip}
\hline
\noalign{\smallskip}
\multicolumn{19}{l}{LC = Luminosity class; SpT = spectral type}  \ \ \ \ \\
\multicolumn{19}{l}{[$\lambda_1]=\lambda_1-3700$ \AA; $[D] =$ dex}  \ \ \ \ \\
\noalign{\smallskip}
\hline
\end{tabular}
\end{table*}

\section{Discussion}\label{disc}   

 In this work we have obtained a new  $T_{\rm eff}(\lambda_1, D)$ calibration which is homogeneous over the entire BCD plane of ($\lambda_1, D$) parameters corresponding to early-type stars. For dwarfs and giants, this calibration yields $T_{\rm eff}$ values which are consistent with those derived using photometric and spectroscopic techniques, and with model atmospheres. However, for supergiants our $T_{\rm eff}$ estimates show rather high systematic discrepancies with the $T_{\rm eff}$ values derived by \citet{mce99} with wind-free models of stellar atmospheres (see Fig. \ref{spg_otros}). As seen in Fig. \ref{spg_otros}, the wind-free model-dependent $T_{\rm eff}$ values are systematically higher than ours. Our $T_{\rm eff}$ estimates are also lower than those obtained with non-LTE BW models around $T_{\rm eff}\sim$25\,000 K, but the discrepancies are smaller. From the discussion in \S\ref{commse}, it appears that our $T_{\rm eff}^f$ estimates for supergiants could have possible systematic underestimates ranging from 500 K to about 700 K as the effective temperature goes from 20\,000 K to 30\,000 K. Although this can partially account for the deviation of points belonging to the $T_{\rm eff}^{(6)}$ group (crosses), it cannot explain the points above the the one-to-one line in Fig.~\ref{spg_otros} from other groups, because the systematic underestimations of our $T_{\rm eff}^f$ values are lower than 1\,000 K. We suggest that the $T_{\rm eff}$ values determined through line profile fitting with synthetic spectra obtained with non-LTE BW model atmospheres, which in most cases are isothermal, could be partially responsible for the observed effective temperature discrepancies. It should not be neglected either that some discrepancies could arise due to problematic convergences of models in the diluted external atmospheric layers of supergiants \citep{sim91,sim93,sim94a,sim94b}.\par

\subsection{Effects due to metallicity}

 \begin{figure*}
\centerline{\psfig{file=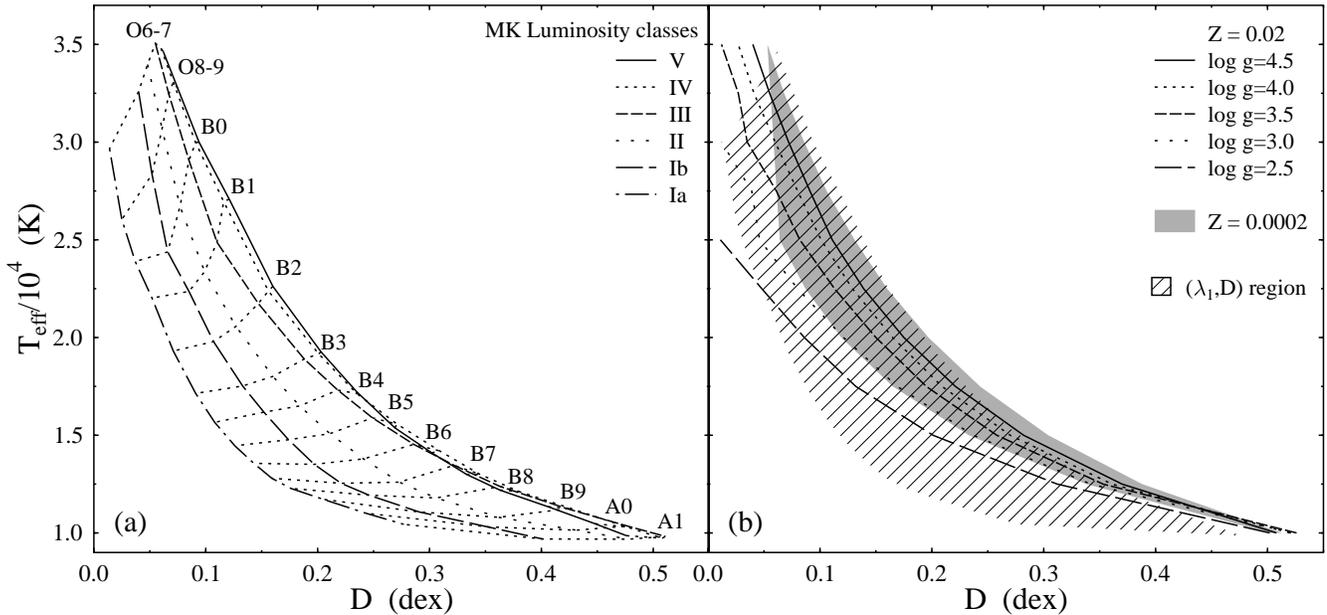,width=18truecm,height=8.7truecm}}
\caption{(a): Effective temperature $T_{\rm eff}(\lambda_1,D)$ against the Balmer discontinuity for different average MK luminosity classes and spectral types. The letters OBA head the lines of constant spectral type; (b) $T_{\rm eff}\!=\!T_{\rm eff}(D,\log g)$ relations from wind-free model atmospheres with metallicity Z = 0.02. The gray region corresponds to $T_{\rm eff}\!=\!T_{\rm eff}(D,\log g)$ relations for metallicity Z = 0.0002 and the same $\log g$ parameters. The shaded zone corresponds to the empirical $T_{\rm eff}(\lambda_1,D)$ curves from panel (a) of this figure.}
\label{teff_d}
\end{figure*}

The present BCD $T_{\rm eff}$ calibration was obtained using a sample of stars near the Sun, which are characterized by an average metallicity $Z=0.02$. We might then expect systematic differences between our $T_{\rm eff}(\lambda_1, D)$ estimates and those for stars having lower metallicities, as in the Magellanic Clouds. Since stars of the same mass, but with lower $Z$, have smaller radii, we cannot ensure that a given $T_{\rm eff}^f$ can correspond to the same pair $(\lambda_1, D)$ of a Magellanic Cloud-star. We can then wonder whether the $T_{\rm eff}^f$-based calibration of the $(\lambda_1, D)$ parameters obtained here can be used for stars with $Z=0.004$ and lower. To assess possible systematic differences in the $T_{\rm eff}$ estimates caused by the use of the present $T_{\rm eff}(\lambda_1, D)$ calibration for stars with an initial metallicity other than $Z=0.02$, we calculated the BD using model atmospheres for $Z=0.002$, 0.0002, and  $Z=0.2$, using the grids of ATLAS9 model atmospheres calculated by \citet{cas03}. The results are displayed in Table \ref{dmetal}. We have not extrapolated the fluxes to lower $\log g$ values because of the difficulties of model convergence that cannot ensure reliable $D$ values . In this table we see that for given values of $\log g$ and $T_{\rm eff}$, the $D$ parameter is larger the lower the metallicity. Although the differences in the values of $D$ depend on $T_{\rm eff}$ and $\log g$, on average we have $\overline{D(Z\!=\!0.002)\!-\!D(Z\!=\!0.02)}=$ $0.014\pm0.003$, $\overline{D(Z\!=\!0.0002)\!-\!D(Z\!=\!0.02)}=$ $0.023\pm0.005$; $\overline{D(Z\!=\!0.2)\!-\!D(Z\!=\!0.02)}=$ $-0.019\pm0.008$. These differences are slightly larger than those expected from the intrinsic uncertainties of the $D$ values. The present BCD calibration of effective temperatures would then produce slightly underestimated $T_{\rm eff}$ values for ($\lambda_1, D$) pairs of stars in environments with $Z\!<\!0.02$, or overestimated if $Z\!>\!0.02$. In both cases the $T_{\rm eff}(\lambda_1, D)$ of stars in environments with $Z\neq0.02$ determined with the present calibration can be easily corrected using Table \ref{dmetal}. The present $T_{\rm eff}^f$-based calibration of the $(\lambda_1, D)$ may be interesting also for early-type stars in the Magellanic Clouds, since there are no systematic measurements of far-UV fluxes that would enable us to use the BFM to obtain their $T_{\rm eff}^f$.\par
 In Fig. \ref{teff_d} we show the relation between the effective temperature and the Balmer discontinuity. Figure \ref{teff_d}a refers to the relations issued from the calibration obtained in this work, where the values are for the mid-points of the spectral type-luminosity class curvilinear quadrilaterals in the BCD plane of $(\lambda_1, D)$ parameters (c.f. Table \ref{midt}). In this figure we indicate the lines of constant mean MK spectral type of stars in different average luminosity classes near the Sun. In Fig. \ref{teff_d}b we show the ($T_{\rm eff},D$) curves calculated with wind-free models \citep{cas03} for several $\log g$ values and the standard metallicity $Z=0.02$. The gray-shaded region corresponds to ($T_{\rm eff},D$) curves for $Z=0.0002$ and for the same set of $\log g$ parameters as for $Z=0.02$. For comparison sake, the hatched zone demarcates the region occupied by the empirical curves displayed in Fig. \ref{teff_d}a. This shows that while wind-free model atmospheres can account reasonably for the ($T_{\rm eff},D$) relations from dwarfs to giants, they fail to do so for supergiants, and that the problem is particularly acute for supergiants with $T_{\rm eff}\!\gtrsim\!20000$ K. It is expected that non-LTE BW models of hot supergiants can explain more easily the observed ($T_{\rm eff},D$) relations, since such atmospheres are heated by the backward wind radiation, and the corresponding $D$ values should then become smaller.

\setcounter{table}{6}
\begin{table}[h!]
\caption{\label{dmetal}Balmer discontinuities $D$ as a function of $T_{\rm eff}$ and $\log g$ for different metallicities $Z$.}
\tabcolsep 4.0pt
\begin{tabular}{c|cccccc}
\noalign{\smallskip}
 \hline
\noalign{\smallskip}
      & \multicolumn{6}{c}{$\log g$} \\
      & 4.5 & 4.0 & 3.5 & 3.0 & 2.5 & 2.0 \\
\noalign{\smallskip}
\hline 
\noalign{\smallskip} 
$T_{\rm eff}$ & \multicolumn{6}{c}{$Z = 0.02$} \\
\noalign{\smallskip}
\hline 
\noalign{\smallskip} 
  10000 & 0.508 & 0.520 & 0.526 & 0.521 & 0.502 & 0.452\\
  12500 & 0.370 & 0.361 & 0.353 & 0.342 & 0.312 & 0.281\\
  15000 & 0.282 & 0.270 & 0.258 & 0.237 & 0.201 & 0.168\\
  17500 & 0.221 & 0.209 & 0.195 & 0.169 & 0.132 & 0.097\\
  20000 & 0.175 & 0.165 & 0.150 & 0.121 & 0.086 &      \\
  22500 & 0.139 & 0.129 & 0.113 & 0.080 & 0.048 &      \\
  25000 & 0.111 & 0.101 & 0.082 & 0.052 & 0.011 &      \\
  27500 & 0.091 & 0.080 & 0.061 & 0.030 &       & \\
  30000 & 0.072 & 0.060 & 0.035 & 0.012 &       & \\
  32500 & 0.055 & 0.041 & 0.027 &       &       & \\
  35000 & 0.040 & 0.027 & 0.012 &       &       & \\
\noalign{\smallskip}
\hline 
\noalign{\smallskip} 
$T_{\rm eff}$ & \multicolumn{6}{c}{$Z = 0.002$} \\
\noalign{\smallskip}
\hline 
\noalign{\smallskip} 
  10000 & 0.511 & 0.523 & 0.526 & 0.518 & 0.500 & 0.457\\
  12500 & 0.380 & 0.376 & 0.368 & 0.357 & 0.326 & 0.295\\
  15000 & 0.295 & 0.288 & 0.276 & 0.257 & 0.222 & 0.190\\
  17500 & 0.234 & 0.227 & 0.213 & 0.188 & 0.155 & 0.122\\
  20000 & 0.189 & 0.181 & 0.166 & 0.139 & 0.108 & \\
  22500 & 0.153 & 0.145 & 0.129 & 0.102 & 0.074 & \\
  25000 & 0.125 & 0.115 & 0.099 & 0.073 & 0.048 & \\
  27500 & 0.101 & 0.091 & 0.074 & 0.050 & & \\
  30000 & 0.082 & 0.070 & 0.052 & 0.032 & & \\
  32500 & 0.065 & 0.052 & 0.034 &  & & \\
  35000 & 0.051 & 0.036 & 0.018 &  & & \\
\noalign{\smallskip}
\hline 
\noalign{\smallskip} 
$T_{\rm eff}$ & \multicolumn{6}{c}{$Z = 0.0002$} \\
\noalign{\smallskip}
\hline 
\noalign{\smallskip} 
  10000 & 0.513 & 0.524 & 0.528 & 0.521 & 0.503 & 0.462\\
  12500 & 0.388 & 0.385 & 0.379 & 0.366 & 0.338 & 0.308\\
  15000 & 0.304 & 0.298 & 0.287 & 0.266 & 0.233 & 0.201\\
  17500 & 0.243 & 0.236 & 0.223 & 0.199 & 0.165 & 0.133\\
  20000 & 0.197 & 0.189 & 0.175 & 0.150 & 0.119 & \\
  22500 & 0.160 & 0.152 & 0.138 & 0.115 & 0.087 & \\
  25000 & 0.131 & 0.122 & 0.109 & 0.088 & 0.064 & \\
  27500 & 0.106 & 0.097 & 0.084 & 0.067 &  & \\
  30000 & 0.086 & 0.076 & 0.064 & 0.050 &  & \\
  32500 & 0.068 & 0.058 & 0.047 &  &  & \\
  35000 & 0.053 & 0.042 & 0.032 &  &  & \\
\noalign{\smallskip}
\hline 
\noalign{\smallskip} 
$T_{\rm eff}$ & \multicolumn{6}{c}{$Z = 0.2$} \\
\noalign{\smallskip}
\hline 
\noalign{\smallskip} 
  10000 & 0.499 & 0.515 & 0.528 & 0.525 & 0.507 & 0.458\\
  12500 & 0.361 & 0.357 & 0.346 & 0.334 & 0.317 & 0.299\\
  15000 & 0.265 & 0.258 & 0.248 & 0.222 & 0.200 & 0.181\\
  17500 & 0.199 & 0.190 & 0.179 & 0.151 & 0.127 & 0.104\\
  20000 & 0.151 & 0.142 & 0.127 & 0.103 & 0.079 & \\
  22500 & 0.115 & 0.106 & 0.089 & 0.070 & 0.048 & \\
  25000 & 0.088 & 0.079 & 0.061 & 0.046 & 0.027 & \\
  27500 & 0.067 & 0.058 & 0.041 & 0.028 &  & \\
  30000 & 0.050 & 0.040 & 0.027 & 0.014 &  & \\
  32500 & 0.037 & 0.026 & 0.017 &  &  & \\
  35000 & 0.024 & 0.014 & 0.011 &  &  & \\
\noalign{\smallskip}
\hline
\end{tabular}
\end{table}

\subsection{Effects related to the rotation}

  The new calibration of the ($\lambda_1, D$) plane as a function of $T_{\rm eff}$ was done assuming that all stars can be characterized by two parameters: $T_{\rm eff}$ and $\log g$. It was shown by \citet{fre05} that also the apparent visual spectral region emitted by the non-uniform and geometrically deformed surface of fast rotating early type stars can be well represented with parent non-rotating $T_{\rm eff}$ and $\log g$ parametric counterparts. The present calibration can then provide a first step for the interpretation of spectra emitted by rotating early-type objects, as already shown in \citet{lev03,nei03,zor05,vin06,fre06,
mar06,mar07,flo00,flo02}.\par 
 Rotation can also induce internal mixing of chemical elements. The mixing and transport of chemical elements throughout the star, produced in the stellar core, can be stronger with higher stellar rotation and it can be enhanced if the metallicity is low \citep{mey00,mey02}. In particular, this may concern helium, whose abundance could then be larger in the atmospheres of evolved stars. Since this excess of He contributes to the absorption in the stellar surface layers, the value of the Balmer discontinuity can be affected. It was shown by \citet{cid07} that in hot dwarf stars the larger the He abundance the smaller the value of $D$.\par
 In order to see in more detail the influence of the He/H abundance ratio on the emitted visual energy distribution, a grid of synthetic spectra for a range of effective temperatures was computed in non-LTE using the TLUSTY and SYNSPEC computing codes \citep{hub95} and the references therein, assuming model atmospheres with He/H ratios of 0.1, 0.2, 0.5 and 1.0, and $Z\!=\!0.02$. The atomic models we used are basically those provided on the TLUSTY website for \ion{H}{i} (9 levels), \ion{He}{i}  (20 individual levels) and \ion{He}{ii} (20 levels). In all cases, the micro-turbulence velocity was supposed to be 2 km~s$^{-1}$. We have reduced the synthetic spectra to the BCD resolution and calculated the BDs following the procedure used for the empirical ones. In Table~\ref{dheh} we give the obtained differences $\delta D\!=\!D\!-\!D(0.1)$ as a function of the model ($T_{\rm eff},\log g$, He/H) parameters. Here $D(0.1)$ indicates that $D$ is for the He/H = 0.1 abundance ratio. In this table we see that on average $\delta D<0$ and that $|\delta D|$ is larger the higher the He/H abundance ratio and for $19000\lesssim T_{\rm eff}\lesssim23000$ K. In recent studies of supergiants \citet{rep05,cro06,sea08,mar08} have found that an abundance He/H = 0.2 fits their atmospheres. From Table \ref{dheh} we see that a ratio He/H = 0.2 effects the value of $D$ by less than the error of measurement of $D$. Nevertheless, these differences are systematic and can be important for the hottest supergiants. In table (Table~\ref{ddddt}) we give the increments $\delta T_{\rm eff}$ that can be produced on the $T_{\rm eff}$ values for $\log g\!=\!3.0$ by $\delta D(0.2)\!=\!D({\rm He/H\!=\!0.2})\!-\!\!D({\rm He/H\!=\!0.1})$ with respect to the $T_{\rm eff}(D)$ scale at He/H = 0.1 given in Table~\ref{dmetal}.

\setcounter{table}{7}
\begin{table}[h!]
\caption{\label{ddddt}Increments $\delta T_{\rm eff}$ carried by $\delta D(0.2)$.}
\begin{center}
\begin{tabular}{ccccccc}
\hline
\noalign{\smallskip}
 & & & $T_{\rm eff}$ (K) & & & \\
\noalign{\smallskip}
17000 & 19000 & 21000 & 23000 & 25000 & 27000 & 30000 \\
\noalign{\smallskip}
 & & & $\delta T_{\rm eff}$ (K) & & & \\
\noalign{\smallskip}
$+$140 & $+170$ & $+270$ & $+590$ & $+600$ & $+610$ & $+600$ \\ 
\noalign{\smallskip}
\hline
\end{tabular}
\end{center}
\end{table}
 
 We note that: 1) Only those stars that were fast rotators during their Main Sequence evolutionary phase are likely to show an atmospheric increase of the He/H ratio in their blue supergiant phase; 2) The energy distributions in the near-IR where the angular diameter $\theta^f$ is calculated, using models for He/H = 0.1 and He/H = 0.2, give the same flux levels. Moreover, in the very few cases with $T_{\rm eff}\!>\!20\,000$ K where we have calculated the energy distributions in the far- and extreme-UV, no changes were obtained in the value of the bolometric correction $\delta_{\rm UV}$ between He/H = 0.1 and He/H = 0.2, so that the use of models for He/H = 0.2 in our BFM would not in principle change our $T_{\rm eff}^f$ estimates; 3) Even though the calculations of $\delta D$ given in Table~\ref{dheh} concern wind-free model atmospheres, the low dependence of deviations $\delta D$ with $T_{\rm eff}$ and $\log g$ ensures that the inclusion of the non-LTE BW will not change the results noticeably.\par

\setcounter{table}{8}
\begin{table}[h]
\caption{\label{dheh}Differences of Balmer discontinuities $\delta D\!=\!D({\rm He/H})\!-\!D(0.1)$ at metallicity $Z\!=\!0.02$, as a function of $T_{\rm eff}$, $\log g$ and for different He/H abundance 
ratios. $D(0.1)$ is for the He/H$\!=\!0.1$ ratio.}
\begin{center}
\begin{tabular}{cc|ccc}
\hline\hline
\noalign{\smallskip}
    &   & He/H=0.2 & 0.5 & 1.0 \\ 
\noalign{\smallskip}
\cline{3-5}
\noalign{\smallskip}
 $T_{\rm eff}$   & $\log g$  & \multicolumn{3}{c}{$\delta D$ (dex)} \\ 
\noalign{\smallskip}
\cline{1-2}
\noalign{\smallskip}
 12500  & 3.0   & $+$0.011  & $+$0.007 & $+$0.000 \\
        & 3.5   & $-$0.005  & $-$0.006 & $-$0.014 \\
        & 4.0   & $-$0.001  & $-$0.004 & $-$0.008 \\
\noalign{\smallskip}
 15000  & 3.0   & $+$0.000  & $-$0.014 & $-$0.020  \\
        & 3.5   & $-$0.005  & $-$0.014 & $-$0.029 \\
        & 4.0   & $-$0.005  & $-$0.014 & $-$0.029 \\
\noalign{\smallskip}
 17000  & 3.0   & $-$0.003  & $-$0.019 & $-$0.025 \\
        & 3.5   & $-$0.005  & $-$0.016 & $-$0.032 \\
        & 4.0   & $-$0.006  & $-$0.016 & $-$0.035 \\
\noalign{\smallskip}
 19000  & 3.0   & $-$0.003  & $-$0.020 & $-$0.026 \\
        & 3.5   & $-$0.004  & $-$0.016 & $-$0.032 \\
        & 4.0   & $-$0.006  & $-$0.017 & $-$0.035 \\
\noalign{\smallskip}
 21000  & 3.0   & $-$0.006  & $-$0.019 & $-$0.025 \\
        & 3.5   & $-$0.004  & $-$0.014 & $-$0.029 \\
        & 4.0   & $-$0.006  & $-$0.016 & $-$0.033 \\
\noalign{\smallskip}
 23000  & 3.0   & $-$0.006  & $-$0.017 & $-$0.022 \\
        & 3.5   & $-$0.003  & $-$0.013 & $-$0.026 \\
        & 4.0   & $-$0.005  & $-$0.015 & $-$0.029 \\
\noalign{\smallskip}
 25000  & 3.0   & $-$0.006  & $-$0.014 & $-$0.019 \\
        & 3.5   & $-$0.003  & $-$0.011 & $-$0.023 \\
        & 4.0   & $-$0.004  & $-$0.013 & $-$0.025 \\
\noalign{\smallskip}
 27000  & 3.0   & $-$0.004  & $-$0.010 & $-$0.016 \\
        & 3.5   & $-$0.002  & $-$0.009 & $-$0.019 \\
        & 4.0   & $-$0.003  & $-$0.011 & $-$0.021 \\
\noalign{\smallskip}
 30000  & 3.0   & $-$0.003  & $-$0.005 & $-$0.011 \\
        & 3.5   & $-$0.002  & $-$0.007 & $-$0.013 \\
        & 4.0   & $-$0.002  & $-$0.008 & $-$0.014 \\
\noalign{\smallskip}
\hline
\noalign{\smallskip}
\multicolumn{5}{l}{$\delta D=0$ for all He/H abundance ratios at $T_{\rm eff}=10000$ K}\\
\noalign{\smallskip}
\hline
\end{tabular}
\end{center}
\end{table}

\section{Conclusions}\label{concl}

  In this work we have presented a new and homogeneous calibration of the BCD plane ($\lambda_1, D$) as a function of $T_{\rm eff}$ for early-type stars of all luminosity classes, in particular supergiants, which complete a similar one made earlier only for dwarf to giant B-type stars \citep{div83}. The present calibration is based on effective temperatures calculated with the bolometric-flux method for all program stars, whose individual uncertainties are on average $\epsilon/T_{\rm eff}^f\!=\!0.05$. The average error of the obtained $T_{\rm eff}$ values on the $(\lambda_1, D)$ plane is the same in all early spectral types and luminosity classes, and they are of the same order as for the individual $T_{\rm eff}^f$ values. The effective temperatures of OB supergiants derived in this work agree within some 2000 K with other determinations found in the literature, except with those issued from wind-free non-LTE plane-parallel models of stellar atmospheres, which produce overestimates of up to more than 5000 K near $T_{\rm eff}=25000$ K.\par
 The $T_{\rm eff}(\lambda_1, D)$ calibration has the advantage of using measurable parameters of the continuum spectrum around the Balmer discontinuity, which are strongly sensitive to the ionization balance of the photosphere and to its gas pressure. The BCD parameters ($\lambda_1, D$) can be easily measured in spectra of low resolution and, even though $D$ is slightly affected by the ISM extinction, the correction can be precisely and easily accounted for. This makes the ($\lambda_1, D$) quantities useful for distant stars. Moreover, since their determination easily can be programmed for automatic measurements, they are convenient for studies of stars in clusters or other stellar systems observed with multi-object spectrographs and/or spectro-imaging devices. The BCD system can also be of interest in analyzing stars in far away galaxies that will be observed with the Extremely Large Telescopes in the near future. In this context, the present calibration of $T_{\rm eff}(\lambda_1, D)$, together with those in preparation for $\log g(\lambda_1, D)$ and $M_{\rm bol}(\lambda_1, D)$, can be helpful in translating the data in terms of physical quantities such as $T_{\rm eff}$ and $\log g$. Furthermore, introducing small corrections, these calibrations can be applied to stars in environments with different initial metallicity.\par
 The contamination of atmospheres in supergiants by He due to rotational mixing does reduce the value of the Balmer discontinuity. However, this change can exceed the average measurement uncertainty of this parameter only when the abundance ratio becomes He/H$\gtrsim$0.05, which is twice as large as the value assumed today in the model atmospheres of supergiants. Nevertheless, differences of up to 600 K can be produced by the estimated values of $T_{\rm eff}$ if the $D$ values affected by the abundance He/H$\simeq0.2$ are used in $T_{\rm eff}(D)$ relations calculated for He/H$\simeq0.1$ at temperatures ranging form 23\,000 K to 30\,000 K. \par
 On average, due to the measurement errors of $D$, the obtained $T_{\rm eff}(\lambda_1, D)$ parameters are obtained within uncertainties that for all studied early spectral types and luminosity classes are of the order of $\Delta T_{\rm eff}/T_{\rm eff}=0.05$, which is roughly the same as for the effective temperatures determined with the bolometric flux method. Since the temperatures obtained with the bolometric flux method are not strongly sensitive to the characteristics of model atmospheres, they may be taken as a reference to study the properties of stellar atmospheres using theoretical spectroscopy.\par
 The ($\lambda_1, D$) parameters are less easily obtained for very hot stars. However, for active stars, like Be, B[e] and He-peculiar OB-types, the determination of BCD parameters might be as straightforward as it is for OB-type stars without emission. This is due to the fact that the photospheric component of BD is separated from that originating in the circumstellar environment.\par

\begin{acknowledgements}
We thank the referee whose careful reading, remarks and criticisms allowed a significant improvement in the presentation of our results. We thank Mrs Martine Usdin for the language editing of this paper. LC acknowledges financial support from the Agencia de Promoci\'on Cient\'{\i}fica y Tecnol\'ogica (BID 1728 OC/AR PICT 111) and the Programa de Incentivos G11/089 of the National University of La Plata, Argentina.  
\end{acknowledgements}

\bibliographystyle{aa}
\bibliography{11147}

\begin{appendix}
\section{About the BCD system}
\label{bcd_expl}

\begin{figure}[ht]
\centerline{\psfig{file=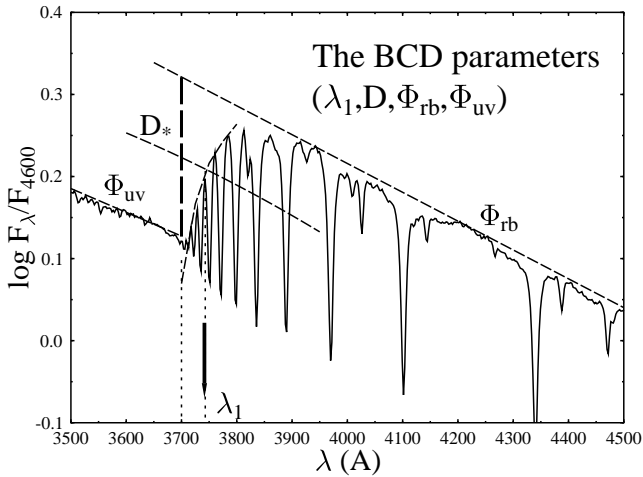}}
\caption{Graphical explanation of the BCD ($\lambda_1,D,\Phi_{\rm rb},\Phi_{\rm uv}$)
parameters.}
\label{bcd_ldphi}
\end{figure}

 Initially, the BCD (Barbier-Chalonge-Divan) system has been thought as a two-parameter system of stellar classification, based on the spectrophotometric study of the continuum spectrum around the Balmer discontinuity. It is used for O, B, A and F-type stars. The two parameters are the size of the Balmer jump, $D$ given in dex, and its mean spectral position, $\lambda_1$ given as the difference $\lambda_1\!-\!3700$ in \AA.\par
 The value of $D$ is calculated at $\lambda\!=\!3700$ \AA, as $D\!=\!\log_{10}F_{3700^+}/F_{3700^-}$, where $F_{3700^+}$ is the Paschen side of the flux and $F_{3700^-}$ is the flux in the Balmer continuum. The value of $F_{3700^+}$ is obtained by extrapolating the rectified Paschen continuum to $\lambda\!=\!3700$, for which a relation such as $\log F_{\lambda}/B_{\lambda}\!=\!p\times(1/\lambda)+q$ is used. $B_{\lambda}$ can be the flux of a comparison star or simply the Planck function of a higher effective temperature than that expected for the studied star. In the present work we adopt the Planck function, so that the expression for $D$ is:

\begin{equation}
D = \log\bigl[\frac{F_{3700^+}/B_{3700}}{F_{3700^-}/B_{3700}}\bigr] \ \ {\rm dex}\ .
\end{equation}

 For the empirical determination of $D$, spectra of low resolution are used: $\Delta\lambda=8$ \AA\ at $\lambda\!=\!3700$.\par
  The average spectral position of the Balmer discontinuity is given by the point of intersection between the curve that passes over the maxima of fluxes in the limit of the Balmer line series, and the flux curve determined by the points $\log F_{\lambda}-D/2$ on the Paschen side and $\log F_{\lambda}+D/2$ on the Balmer side. In Fig. \ref{bcd_ldphi} we show the method of determining $D$ and $\lambda_1$. Since the wavelength scale to determine the intersection of flux curves is based on the intrinsic wavelengths of the identified spectral lines, the parameter $\lambda_1$ is not affected by any displacements due to the radial velocity of stars. \par
  When carring out the spectrophotometric study of the energy distribution near the Balmer discontinuity, two other parameters are obtained: the color gradient $\Phi_{\rm uv}$, given in $\mu$m and defined for the $3200\!-\!3700$ \AA\ spectral region, and the Paschen gradient defined in two versions: $\Phi_{\rm b}$ or $\Phi_{\rm rb}$ for the spectral regions $4000\!-\!4800$ \AA\ and $4000\!-\!6700$ \AA, respectively, both given in $\mu$m. A color gradient is defined as \citep{all76}:

\begin{equation}
\Phi = 5\lambda - \frac{d\ln F_{\lambda}}{d(1/\lambda)}\ ,
\label{firgen}
\end{equation}

\noindent which for a black body at temperature $T$, becomes :

\begin{equation}
\Phi(T) = (C_2/T)\bigl(1-e^{-C_2/\lambda T}\bigr)\ ,
\label{firbb}
\end{equation}

\noindent where $C_2\!=\!hc/k\!=\!1.4388$ cm~deg is the radiation constant. Assuming that for a given stellar energy distribution $F_{\lambda}$ it is $\Phi\!=\!const.$ between wavelengths $\lambda_a$ and $\lambda_b$, the expression for $\Phi$ is:

\begin{equation}
\Phi = \ln\bigl[\frac{\lambda_a^5F_{\lambda_a}}{\lambda_b^5F_{\lambda_b}}\bigr]/(1/\lambda_a-1/\lambda_b)\ .
\label{firbf}
\end{equation}

   The color gradient $\Phi_{\rm b}$ was introduced in 1955 in the BCD system to distinguish F-type stars from B-type stars having the same ($\lambda_1,D$) pairs. This parameter also has been used as a third BCD quantity related to the metal abundance of late type stars, in particular to the abundance ratio [Fe/H] \citep{div77}.\par
   As the local temperature of the formation region of the Paschen continuum is close to the effective temperature, from (\ref{firbf}) and (\ref{firbb}) we note that stars with the same effective temperature but different surface gravity define a common region in the plane ($\lambda_1,D$). This fact was used by \citet{bar41} and \citet{cha52} to determine the curvilinear quadrilaterals that characterize the BCD classification system. To this end, the authors used the MK classification of stars made by Morgan and Keenan \citep{kee51} themselves. They simply delimited the common region occupied by stars of the same MK spectral type with curves of intrinsic constant $\Phi_{\rm rb}$ parameters. The same technique was used to draw the `horizontal' lines that separate the MK luminosty classes. The BCD authors attempted to keep inside a common strip stars of all spectral types, but having the same MK luminosity class label assigned by Morgan and Keenan.\par
  The color gradients $\Phi_{\rm b}$ and $\Phi_{\rm rb}$ can be written as a function of the $(B-V)$ color of the UBV photometric system \citep{muj98}. The relation between the color excesses in the BCD and UBV systems due to the ISM reddening is then \citep{div73}:

\begin{equation}
A_{\rm V}=3.1E(B-V)=1.7(\Phi_{\rm rb}\!-\!\Phi_{\rm rb}^o)=1.9(\Phi_{\rm b}\!-\!\Phi_{\rm b}^o)\ {\rm mag}\ ,
\end{equation}
 
\noindent where $\Phi_{\rm rb}^o$ is the stellar intrinsic color gradient.\par
  {\it One of the greatest advantages of the BCD method is that low resolution spectra are used. Since the BCD parameters can be easily obtained through automatic treatment of data, they are convenient for studies of stars in clusters or other stellar systems observed with multi-object spectrographs and/or spectro-imaging devices. The BCD system can also be of interest to analyze stars in far away galaxies that will be observed with the Extremely Large Telescopes.}\par

\section{Calculation of the random errors affecting $T_{\rm eff}^f$ and $\theta^f$}
\label{err_calc}

\begin{figure}
\centerline{\psfig{file=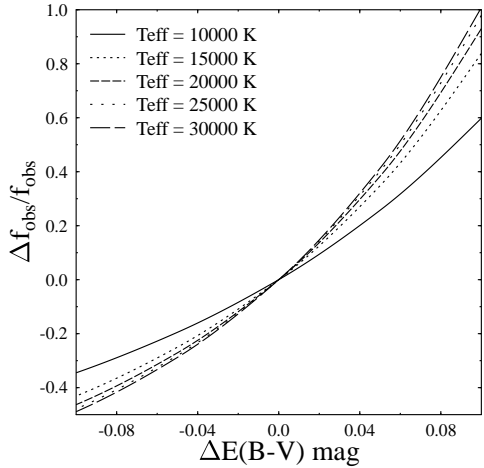,width=6.5truecm,height=6.5truecm}}
\caption{Error committed on the estimate of the observed fraction of the bolometric flux as a function of the uncertainty on the ISM color excess $E(B-V)$.}
\label{errdf}
\end{figure}

 As quoted in \S\ref{rderr}, the  $T_{\rm eff}^f$ and $\theta^f$ parameters derived in the present work have the following sources of error: a) the ISM color excess $E(B-V)$; b) the line blocking enhancement parameter $\gamma$; c) the $\log g$ parameter on which the model fluxes used to estimate the angular diameter depend and the filling factor due to the unobserved spectral region; d) the filling factor $\delta$ used in relation (\ref{fl}) to calculate the bolometric flux.\par

 a) {\it The ISM color excess $E(B-V)$.} The correction of the observed energy distributions for the ISM extinction is probably the source of error that has the heaviest consequences on the determination of $T_{\rm eff}^f$. Both the value of the bolometric flux $f$ in Eq. (\ref{fl}) and the flux ratio $f^o_{\lambda}/F_{\lambda}$ entering Eq. (\ref{teta}) depend on this correction.\par
  We write the error on the estimate of the color excess $E(B-V)$ as $\Delta E\!=\!E(B-V)-E_o(B-V)$, where $E_o(B-V)$ is the `unknown' correct value of this excess. The effect of $\Delta E$ on the integrated fluxes coming from the observed spectral region, $f_{\rm obs}$, can be estimated with model atmospheres, by calculating the ratio $\Delta~f_{\rm obs}/f_{\rm obs}\!=\!$ $[f_{\rm obs}(\Delta E)-f_{\rm obs}]/f_{\rm obs}$. The behavior of $\Delta~f_{\rm obs}/f_{\rm obs}$ against $\Delta E$
for several effective temperatures is shown in Fig. \ref{errdf}. This calculation is made only for $\log g\!=\!3.0$ since the effect of $\log g$ on $\Delta~f_{\rm obs}/f_{\rm obs}$ is negligible.\par 
 The uncertainty $\Delta E$ also affects the monochromatic fluxes $f^o_{\lambda}$ used to derive the angular diameter. The angular diameter affects $T_{\rm eff}^f$, which, in turn, determines $T_{\rm eff}(\gamma)$ and the model flux entering relation (\ref{teta}). For an estimate of the ratio $\Delta\theta^f/\theta^f$ we can simply use $\lambda\!=0.7\mu$m, the middle wavelength of the interval over which $\theta^f$ is calculated. To make these calculations easier, the flux $F_{\lambda0.7}$ is represented as a function of $T_{\rm eff}(\gamma)$ and $\log g$, using the following interpolation expression:

\begin{eqnarray}
\begin{array}{rcl}
F_{\lambda0.7}[T_{\rm eff}(\gamma),g] & = & A(\log g)T_{\rm eff}(\gamma)^{B(\log g)} \\
A(\log g) & = & 5.383-2.351\log g +0.32073\log^2 g \\
B(\log g) & = & 1.405+0.191\log g-0.027\log^2 g\ . \\ 
\end{array}
\label{interp}
\end{eqnarray}

 For the observed stellar flux at $\lambda\!=0.7\mu$m corrected for the ISM extinction is $f^o_{0.7}\!=\!f^{obs}_{0.7}10^{0.4k_{0.7}E(B-V)]}$, with $k_{0.7}\!=\!2.39$, using (\ref{interp}) the total uncertainty affecting the angular diameter can be written to the first order as:

\begin{eqnarray}
\begin{array}{lcl}
\label{envgamal}
\frac{\Delta\theta^f}{\theta^f} & = & 1.1\Delta E(B-V)-\frac{1}{2}\bigl\{B(\log g)\frac{\Delta T_{\rm eff}(\gamma)}{T_{\rm eff}(\gamma)}+\bigl[\frac{\partial\ln A(\log g)}{\partial\log g}+ \\
                         &   & \ln T_{\rm eff}(\gamma)\frac{\partial B(\log g)}{\partial\log g}\bigr]\Delta\log g\bigr\}\ ,
\end{array}
\end{eqnarray}

\noindent where it also includes the dependence on $\Delta\log g$ and the uncertainty $\delta\gamma$ in the factor $\Delta~T_{\rm eff}(\gamma)/T_{\rm eff}(\gamma)$, whose dependence on $\delta\gamma$ we shall discuss below. \par
 In this work, the uncertainty inherent to $E(B-V)$ was characterized with the $1\sigma$ dispersion of the independent estimates of $E(B-V)$ for a given star, i.e. $\Delta E(B-V)\!=\!\pm\sigma_{\rm E(B-V)}$. For the studied stars, it is on average $\sigma_{\rm E(B-V)}\simeq$ 0.03 mag. In (\ref{envgamal}) the direct dependence of $\Delta\theta^f$ on $\Delta E(B-V)$ is noted. However, it can be shown that the change of $f^o_{0.7}$ with $\Delta E(B-V)$ is larger than that of $F_{0.7}$ induced by the corresponding $\Delta T_{\rm eff}^f$. This means that if $\Delta E(B-V)$ carries an increase of $f^o_{0.7}$ and an overestimation of $T_{\rm eff}^f$ through a greater bolometric flux, the corresponding increase of $F_{0.7}$ will compensate for the increase of $f^o_{0.7}$, which maintains $\theta^f$ almost unchanged.\par

 b) {\it The line blocking enhancement parameter $\gamma$.} To some degree, the estimate of $\gamma$ is also dependent on the ISM extinction, whose effects in the far-UV can be confused with those produced by the line blanketing. The effect of the error on the enhancing factor $\gamma$ can be important as it enters the determination of the temperature $T_{\rm eff}(\gamma)$ used to calculate both $\theta$ and $\delta$. Nevertheless, the total effect of $\gamma$ on the value of $T_{\rm eff}^f$ does not seem to carry larger deviations than $\Delta T_{\rm eff}\sim300$ K, as shown in Fig. \ref{gamma}.\par 
 The relative error $\Delta T_{\rm eff}(\gamma)/T_{\rm eff}(\gamma)$ as a function of the uncertainty on $\gamma$ can be estimated using the relation (\ref{tgamal}). Since an interpolation expression of the function $\Lambda$ as a function of $T_{\rm eff}(\gamma)$ and $\log g$ is not easy to derive, to a good first order we can attempt an estimate of the uncertainty on $T_{\rm eff}(\gamma)$ produced by $\Delta\gamma$, with:

\begin{equation}
\label{ergamal}
\frac{\Delta T_{\rm eff}(\gamma)}{T_{\rm eff}(\gamma)} \simeq \frac{1}{4}\Lambda[T_{\rm eff}(\gamma),\log g]\Delta\gamma\ .
\end{equation}

 The model fitting to the observed energy distributions carries uncertainties on the $\gamma$ of the order of $\Delta\gamma\!\simeq\!0.25$. The consequences of this uncertainty on the value of $T_{\rm eff}(\gamma)$ for a given $T_{\rm eff}^f$ can be derived using (\ref{ergamal}) or by interpolation in Table \ref{deltas}.\par

 c) {\it The $\log g$ parameter.} We assumed that the adopted model energy distributions correspond to gravities determined within an error $\Delta~{\log g}\!=\!0.5$ dex, which is twice the typical dispersion seen in Fig. \ref{complogg}.\par 

 d) {\it The filling factor $\delta$}. The bolometric flux fraction $\delta$ would be realistic, if the actual stellar atmosphere of the observed star obeyed the basic assumptions that rule the models used. Although the non-LTE BW models produce somewhat different energy distributions in the far- and extreme-UV spectral regions, they cannot carry any significant change to the estimate of $\delta$, as compared to the values derived with wind-free models. Having no definite knowledge of the phenomena that models might still be missing, we assumed that we could have uncertainties affecting $\delta$ of the order of $\Delta\delta/\delta\!=\!0.10$.\par
 Since the uncertainties mentioned in this section do not propagate symmetrically (i.e. see Table \ref{erttt}), their combined effect was simulated as follows. Let us call $|\Delta_{\rm P}|$ the absolute error assigned to a given $P$ quantity, i.e. $P\!=\!E(B-V)$, $\gamma$, $\log g$ and $\delta$. We divide each $|\Delta_P|$ into 4 parts: $\epsilon_P=|\Delta_P|/4$, so that we can assign to each quantity $P$ 9 independent estimates, i.e. $P\equiv$ $P-4\epsilon_P$, $P-3\epsilon_P$,..., $P$,..., $P+3\epsilon_P$, $P+4\epsilon_P$. Then, as we have 4 different parameters to which we assign an error, each of them is taken with one possible simulated value $P\pm{n\epsilon_P}$ ($n$ from 1 to 4); we have $9^4\!=\!6561$ possible combinations that produce as many estimates of $T_{\rm eff}^f$ and $\theta^f$ around the `central' one where all $\Delta_P=0.0$. The frequency of $T_{\rm eff}^f$ and $\theta^f$ values thus obtained have a bell-shaped distribution, not always symmetric. Let us call $Q(\epsilon_{T_{\rm eff}})$ and $Q(\epsilon_{\theta})$ these distributions, where $\epsilon_{T_{\rm eff}}\!=\!T_{\rm eff}-T_{\rm eff}^o$ and $\epsilon_{\theta}\!=\!\theta-\theta^o$ are the displacements from the respective $T_{\rm eff}^o$ and $\theta^o$ values for which all $\Delta_P=0.0$. The errors indicated in Table \ref{prog_stars} are then the mean errors defined as:

\begin{equation}
\label{errm}
\overline{\epsilon} = \int_{-\infty}^{+\infty}\!|\epsilon|Q(\epsilon){\rm d}\epsilon/\int_{-\infty}^{+\infty}\!Q(\epsilon){\rm d}\epsilon
\end{equation}

\noindent that would correspond to $\overline{\epsilon}\!=\!\sqrt{2/\pi}\sigma$ if the distributions $Q(\epsilon)$ are `normal', Gaussian, with dispersion $\sigma$. On average it is $\langle\overline{\epsilon}/T_{\rm eff}^f\rangle\!\simeq\!0.05$, which is of the same order as the error expected for the $T_{\rm eff}(\lambda_1,D)$ values estimated in \S\ref{thecal}, c.f. relation (\ref{er}).

\end{appendix}

\setlongtables
\onecolumn
\setcounter{table}{1}
\begin{longtable}{r|l|rc|clcc|r}
\caption{\label{prog_stars}Program stars, observed and derived parameters.}\\
\noalign{\smallskip}
\hline
\noalign{\smallskip}
 HD & SpT/LC & $\lambda_1$ & $D$ & $E(B-V)$ & $T^f_{\rm eff}\pm\Delta T_{\rm eff}$ & $\theta^f\pm\Delta\theta$ & $\gamma$ & $T_{\rm eff}(\lambda_1, D)$\\
    &        &             & dex & mag      & \ \ \ \ \ \ \ K  & mas & & K \ \ \ \ \\
\hline
\noalign{\smallskip}
\endfirsthead
\caption{continued.}\\
\noalign{\smallskip} 
\hline
\noalign{\smallskip}
 HD & SpT/LC & $\lambda_1$ & $D$ & $E(B-V)$ & $T_{\rm eff}\pm\Delta T_{\rm eff}$ & $\theta\pm\Delta\theta$ & $\gamma$ & $T_{\rm eff}(\lambda_1, D)$\\
\noalign{\smallskip}
\hline
\noalign{\smallskip}
\endhead
\noalign{\smallskip}
\hline
\endfoot
    358 & B5.5V     &  58.0 & 0.288 & 0.023 & 12750$\pm 590$ & 1.016$\pm0.010$ & 1.00  & 14580 \ \ \ \ \\
    886 & B2IV      &  54.0 & 0.155 & 0.007 & 21870$\pm1390$ & 0.444$\pm0.021$ & 1.00  & 22740 \ \ \ \ \\
   2905 & B1Ia      &  31.0 & 0.070 & 0.339 & 22160$\pm1420$ & 0.348$\pm0.021$ & 1.13  & 21330 \ \ \ \ \\
   3360 & B2IV      &  54.0 & 0.161 & 0.031 & 21850$\pm1390$ & 0.307$\pm0.021$ & 1.00  & 22220 \ \ \ \ \\
   3369 & B5V       &  54.0 & 0.259 & 0.046 & 15980$\pm 910$ & 0.305$\pm0.015$ & 1.00  & 16030 \ \ \ \ \\
   4142 & B5.5V     &  59.0 & 0.283 & 0.044 & 16430$\pm 940$ & 0.162$\pm0.015$ & 1.00  & 14720 \ \ \ \ \\
   4727 & B4V       &  58.0 & 0.249 & 0.029 & 16290$\pm 930$ & 0.266$\pm0.015$ & 1.00  & 16460 \ \ \ \ \\
  11241 & B2V       &  65.0 & 0.162 & 0.066 & 22080$\pm1410$ & 0.141$\pm0.021$ & 1.00  & 22420 \ \ \ \ \\
  12767 & B5V       &  55.1 & 0.279 & 0.000 & 13250$\pm 640$ & 0.281$\pm0.011$ & 1.00  & 15000 \ \ \ \ \\
  12953 & A2Ia      &  -1.0 & 0.231 & 0.599 & \ \ 9750$\pm 330$ & 0.591$\pm0.007$ & 1.58 &9940 \ \ \ \ \\
  13267 & B6Ia      &  16.0 & 0.155 & 0.490 & 14380$\pm 760$ & 0.259$\pm0.012$ & 0.76  & 13460 \ \ \ \ \\
  14055 & A1V       &  71.0 & 0.471 & 0.019 & \ \ 9440$\pm 310$ & 0.540$\pm0.007$ & 1.00 &9840 \ \ \ \ \\
  14228 & B6V       &  61.7 & 0.296 & 0.015 & 12470$\pm 560$ & 0.507$\pm0.010$ & 1.00  & 14150 \ \ \ \ \\
  14489 & A0Ib      &  11.0 & 0.310 & 0.420 & \ \ 9840$\pm 340$ & 0.549$\pm0.007$ & 2.04&10650 \ \ \ \ \\
  14818 & B2Ib-Ia   &  31.0 & 0.086 & 0.496 & 18300$\pm1100$ & 0.202$\pm0.018$ & 1.32  & 20260 \ \ \ \ \\
  15130 & A0.5IV    &  52.8 & 0.490 & 0.019 & \ \ 9920$\pm 350$ & 0.346$\pm0.008$ & 1.00&10040 \ \ \ \ \\
  15318 & B9.5V     &  61.9 & 0.440 & 0.020 & 10630$\pm 400$ & 0.421$\pm0.008$ & 1.00  & 10700 \ \ \ \ \\
  16046 & A0IV      &  56.4 & 0.448 & 0.021 & 10910$\pm 420$ & 0.314$\pm0.008$ & 1.00  & 10730 \ \ \ \ \\
  16582 & B2IV      &  52.6 & 0.154 & 0.020 & 23160$\pm1500$ & 0.239$\pm0.022$ & 1.00  & 22610 \ \ \ \ \\
  16908 & B4V       &  54.0 & 0.231 & 0.046 & 17520$\pm1030$ & 0.243$\pm0.017$ & 1.00  & 17430 \ \ \ \ \\
  17081 & B6IV      &  46.0 & 0.311 & 0.016 & 13680$\pm 680$ & 0.344$\pm0.011$ & 1.00  & 14150 \ \ \ \ \\
  17573 & B7V       &  54.0 & 0.339 & 0.014 & 12980$\pm 610$ & 0.476$\pm0.010$ & 1.00  & 12950 \ \ \ \ \\
  18604 & B6V-IV    &  50.4 & 0.301 & 0.033 & 13940$\pm 710$ & 0.286$\pm0.011$ & 1.00  & 14380 \ \ \ \ \\
  19356 & B7V       &  60.0 & 0.324 & 0.047 & 12800$\pm 590$ & 1.033$\pm0.010$ & 1.00  & 13240 \ \ \ \ \\
  20041 & A0Ia      &  10.0 & 0.276 & 0.816 & 10800$\pm 420$ & 0.643$\pm0.008$ & 2.02  & 10980 \ \ \ \ \\
  21291 & B9Ia      &   9.0 & 0.213 & 0.495 & 11420$\pm 470$ & 0.794$\pm0.009$ & 1.28  & 11640 \ \ \ \ \\
  21364 & B7.5V     &  64.8 & 0.337 & 0.044 & 13070$\pm 620$ & 0.473$\pm0.010$ & 1.00  & 12510 \ \ \ \ \\
  21389 & A0Ia      &   4.0 & 0.233 & 0.633 & 11040$\pm 440$ & 0.886$\pm0.009$ & 1.22  & 10860 \ \ \ \ \\
  21447 & A1V       &  72.0 & 0.466 & 0.026 & \ \ 9240$\pm 300$ & 0.337$\pm0.007$ & 1.00 &9890 \ \ \ \ \\
  21790 & B9IV      &  45.8 & 0.408 & 0.017 & 11760$\pm 500$ & 0.317$\pm0.009$ & 1.00  & 11610 \ \ \ \ \\
  21856 & B1IV      &  57.0 & 0.116 & 0.183 & 25370$\pm1670$ & 0.124$\pm0.023$ & 1.00  & 27200 \ \ \ \ \\
  22928 & B5III     &  42.0 & 0.281 & 0.044 & 14890$\pm 820$ & 0.583$\pm0.013$ & 1.00  & 15100 \ \ \ \ \\
  22951 & B1V       &  62.0 & 0.112 & 0.253 & 29330$\pm1980$ & 0.183$\pm0.024$ & 1.00  & 27990 \ \ \ \ \\
  23180 & B1III     &  49.0 & 0.121 & 0.267 & 22840$\pm1470$ & 0.384$\pm0.021$ & 0.97  & 24190 \ \ \ \ \\
  23227 & B4.5III   &  43.4 & 0.247 & 0.019 & 16230$\pm 930$ & 0.215$\pm0.015$ & 1.00  & 16470 \ \ \ \ \\
  23288 & B7V       &  56.0 & 0.329 & 0.117 & 14020$\pm 720$ & 0.224$\pm0.011$ & 1.00  & 13220 \ \ \ \ \\
  23324 & B7V       &  56.0 & 0.334 & 0.061 & 13210$\pm 640$ & 0.201$\pm0.010$ & 1.00  & 13050 \ \ \ \ \\
  23338 & B6IV      &  46.0 & 0.296 & 0.059 & 14180$\pm 740$ & 0.351$\pm0.011$ & 1.00  & 14650 \ \ \ \ \\
  23408 & B6.5III   &  38.5 & 0.313 & 0.129 & 14310$\pm 750$ & 0.464$\pm0.011$ & 1.00  & 13960 \ \ \ \ \\
  23625 & B2V       &  59.0 & 0.175 & 0.294 & 23980$\pm1560$ & 0.114$\pm0.022$ & 1.00  & 21350 \ \ \ \ \\
  23753 & B7IV      &  50.0 & 0.350 & 0.065 & 12680$\pm 580$ & 0.228$\pm0.010$ & 1.00  & 12740 \ \ \ \ \\
  23850 & B7III     &  39.0 & 0.339 & 0.086 & 13020$\pm 620$ & 0.526$\pm0.010$ & 1.00  & 13240 \ \ \ \ \\
  23923 & B9V       &  58.0 & 0.411 & 0.054 & 11640$\pm 490$ & 0.170$\pm0.009$ & 1.00  & 11350 \ \ \ \ \\
  24131 & B1V       &  61.0 & 0.124 & 0.265 & 29130$\pm1960$ & 0.132$\pm0.024$ & 1.00  & 26670 \ \ \ \ \\
  24398 & B1Ib      &  37.0 & 0.085 & 0.328 & 22040$\pm1410$ & 0.659$\pm0.021$ & 0.98  & 22580 \ \ \ \ \\
  24431 & O9.5III   &  54.0 & 0.077 & 0.653 & 27370$\pm1830$ & 0.155$\pm0.023$ & 1.41  & 31380 \ \ \ \ \\
  24760 & B0IV-III  &  54.5 & 0.089 & 0.086 & 27160$\pm1810$ & 0.394$\pm0.023$ & 1.00  & 29810 \ \ \ \ \\
  25204 & B4IV      &  50.1 & 0.243 & 0.050 & 16970$\pm 990$ & 0.447$\pm0.016$ & 1.00  & 16829 \ \ \ \ \\
  25490 & $>$A1V    &  74.0 & 0.496 & 0.007 & \ \ 8990$\pm280$ & 0.605$\pm0.007$ & 1.00 & 9300 \ \ \ \ \\
  27376 & B6V       &  61.5 & 0.303 & 0.012 & 12460$\pm 560$ & 0.522$\pm0.010$ & 1.00  & 13910 \ \ \ \ \\
  27396 & B4IV      &  51.0 & 0.240 & 0.146 & 16720$\pm 970$ & 0.267$\pm0.016$ & 1.00  & 16960 \ \ \ \ \\
  27962 & A1V       &  71.0 & 0.463 & 0.000 & \ \ 8680$\pm 260$ & 0.515$\pm0.008$ & 1.00 &9970 \ \ \ \ \\
  29248 & B2IV      &  51.2 & 0.144 & 0.048 & 23610$\pm1530$ & 0.262$\pm0.022$ & 1.00  & 23120 \ \ \ \ \\
  29305 & B7V       &  61.3 & 0.328 & 0.021 & 12120$\pm 530$ & 0.603$\pm0.010$ & 1.00  & 13030 \ \ \ \ \\
  30211 & B4IV      &  45.0 & 0.248 & 0.024 & 15770$\pm 890$ & 0.346$\pm0.015$ & 1.00  & 16510 \ \ \ \ \\
  30614 & B0.5Ia    &  32.0 & 0.057 & 0.277 & 25340$\pm1670$ & 0.274$\pm0.023$ & 0.95  & 22620 \ \ \ \ \\
  31647 & A1V       &  76.0 & 0.455 & 0.032 & \ \ 9750$\pm 330$ & 0.346$\pm0.007$ & 1.00 &9950 \ \ \ \ \\
  32309 & B9V       &  63.0 & 0.425 & 0.018 & 10800$\pm 420$ & 0.315$\pm0.008$ & 1.00  & 10960 \ \ \ \ \\
  32549 & B9IV      &  45.1 & 0.438 & 0.000 & \ \ 9520$\pm 320$ & 0.392$\pm0.007$ & 1.00&11040 \ \ \ \ \\
  32630 & B4V       &  59.0 & 0.232 & 0.020 & 17940$\pm1070$ & 0.460$\pm0.017$ & 1.00  & 17390 \ \ \ \ \\
  33802 & B6V       &  59.5 & 0.299 & 0.028 & 13280$\pm 640$ & 0.336$\pm0.011$ & 1.00  & 14140 \ \ \ \ \\
  33904 & B7IV      &  49.6 & 0.331 & 0.015 & 12390$\pm 550$ & 0.590$\pm0.010$ & 1.00  & 13390 \ \ \ \ \\
  33949 & B7III     &  41.3 & 0.358 & 0.033 & 12750$\pm 590$ & 0.358$\pm0.010$ & 1.00  & 12710 \ \ \ \ \\
  34085 & B8Ia      &  12.0 & 0.177 & 0.058 & 12130$\pm 530$ & 2.713$\pm0.010$ & 1.18  & 12330 \ \ \ \ \\
  34503 & B6IV      &  44.8 & 0.303 & 0.045 & 14450$\pm 770$ & 0.470$\pm0.012$ & 1.00  & 14440 \ \ \ \ \\
  35468 & B2III     &  49.0 & 0.146 & 0.023 & 21840$\pm1390$ & 0.775$\pm0.021$ & 1.00  & 22530 \ \ \ \ \\
  35497 & B5.5IV    &  44.0 & 0.288 & 0.022 & 13960$\pm 710$ & 1.116$\pm0.011$ & 1.00  & 14910 \ \ \ \ \\
  35600 & A0Ib      &  18.0 & 0.375 & 0.288 & 10870$\pm 420$ & 0.324$\pm0.008$ & 0.69  & 10260 \ \ \ \ \\
  36267 & B5V       &  60.9 & 0.257 & 0.020 & 16020$\pm 910$ & 0.314$\pm0.015$ & 1.00  & 15930 \ \ \ \ \\
  36371 & B5Ia      &  15.0 & 0.129 & 0.471 & 15370$\pm 860$ & 0.476$\pm0.014$ & 1.17  & 14060 \ \ \ \ \\
  36512 & O8-9V     &  68.1 & 0.079 & 0.029 & 32340$\pm2220$ & 0.142$\pm0.025$ & 1.00  & 32060 \ \ \ \ \\
  36822 & B0IV      &  55.0 & 0.095 & 0.111 & 28340$\pm1900$ & 0.198$\pm0.024$ & 1.00  & 29330 \ \ \ \ \\
  37128 & B0Ib      &  39.1 & 0.060 & 0.059 & 24670$\pm1620$ & 0.681$\pm0.022$ & 1.16  & 24440 \ \ \ \ \\
  37468 & O9.5V     &  71.0 & 0.082 & 0.061 & 31270$\pm2130$ & 0.230$\pm0.025$ & 1.00  & 31420 \ \ \ \ \\
  37481 & B2V       &  61.0 & 0.142 & 0.035 & 24350$\pm1590$ & 0.101$\pm0.022$ & 1.00  & 24700 \ \ \ \ \\
  37744 & B2V       &  65.0 & 0.141 & 0.057 & 25690$\pm1700$ & 0.088$\pm0.023$ & 1.00  & 25000 \ \ \ \ \\
  38666 & O8-9V     &  66.5 & 0.072 & 0.017 & 31510$\pm2150$ & 0.111$\pm0.025$ & 1.00  & 33170 \ \ \ \ \\
  38771 & B0II      &  46.0 & 0.081 & 0.057 & 23170$\pm1500$ & 0.620$\pm0.022$ & 1.00  & 26730 \ \ \ \ \\
  39970 & B9Ib-Ia   &  12.0 & 0.261 & 0.487 & 11530$\pm 480$ & 0.349$\pm0.009$ & 1.51  & 11250 \ \ \ \ \\
  40111 & B1Ib      &  38.0 & 0.084 & 0.183 & 24660$\pm1620$ & 0.197$\pm0.022$ & 1.22  & 22950 \ \ \ \ \\
  40183 & A1V       &  69.0 & 0.485 & 0.012 & \ \ 8910$\pm 280$ & 1.506$\pm0.007$ & 1.00 &9680 \ \ \ \ \\
  40312 & B9IV      &  47.0 & 0.448 & 0.000 & \ \ 9890$\pm 340$ & 0.959$\pm0.008$ & 1.00&10850 \ \ \ \ \\
  40589 & B9Ib      &  12.0 & 0.284 & 0.378 & 11660$\pm 490$ & 0.295$\pm0.009$ & 1.33  & 10970 \ \ \ \ \\
  41117 & B2Ia      &  24.0 & 0.050 & 0.494 & 20740$\pm1310$ & 0.371$\pm0.020$ & 1.53  & 20640 \ \ \ \ \\
  41753 & B4V       &  55.0 & 0.229 & 0.032 & 18800$\pm1140$ & 0.251$\pm0.018$ & 1.00  & 17560 \ \ \ \ \\
  42087 & B3Ia      &  25.0 & 0.102 & 0.384 & 16460$\pm 940$ & 0.258$\pm0.015$ & 1.18  & 17440 \ \ \ \ \\
  43112 & B1V       &  58.0 & 0.117 & 0.032 & 27900$\pm1870$ & 0.091$\pm0.024$ & 1.00  & 27190 \ \ \ \ \\
  43384 & B4Ia      &  20.0 & 0.125 & 0.621 & 14780$\pm 800$ & 0.304$\pm0.013$ & 1.38  & 15140 \ \ \ \ \\
  44743 & B1III     &  45.0 & 0.112 & 0.036 & 25320$\pm1670$ & 0.586$\pm0.023$ & 1.00  & 23690 \ \ \ \ \\
  46300 & A1Ib      &  17.0 & 0.421 & 0.083 & \ \ 9800$\pm 340$ & 0.455$\pm0.007$ & 1.00 &9740 \ \ \ \ \\
  46769 & B7III-II  &  30.0 & 0.288 & 0.151 & 13930$\pm 710$ & 0.210$\pm0.011$ & 0.45  & 12990 \ \ \ \ \\
  47105 & A1.5IV    &  61.0 & 0.518 & 0.001 & \ \ 9040$\pm 280$ & 1.435$\pm0.007$ & 1.00 &9280 \ \ \ \ \\
  47240 & B1Ib      &  35.0 & 0.090 & 0.372 & 21540$\pm1370$ & 0.157$\pm0.021$ & 1.33  & 21670 \ \ \ \ \\
  47432 & O8-9Ib    &  43.0 & 0.050 & 0.389 & 25620$\pm1690$ & 0.133$\pm0.023$ & 1.77  & 28550 \ \ \ \ \\
  47670 & B7III     &  33.0 & 0.331 & 0.014 & 12120$\pm 530$ & 0.625$\pm0.010$ & 1.09  & 12760 \ \ \ \ \\
  47964 & B7IV-III  &  42.0 & 0.347 & 0.021 & 12100$\pm 530$ & 0.193$\pm0.009$ & 1.00  & 13040 \ \ \ \ \\
  48434 & B0II      &  44.0 & 0.072 & 0.241 & 25290$\pm1670$ & 0.128$\pm0.023$ & 1.87  & 26770 \ \ \ \ \\
  48977 & B3V       &  60.0 & 0.210 & 0.039 & 19590$\pm1210$ & 0.125$\pm0.019$ & 1.00  & 18820 \ \ \ \ \\
  49567 & B4III     &  37.0 & 0.209 & 0.051 & 17270$\pm1010$ & 0.125$\pm0.016$ & 1.11  & 17110 \ \ \ \ \\
  52089 & B1III     &  45.0 & 0.120 & 0.034 & 22010$\pm1400$ & 0.801$\pm0.021$ & 1.30  & 23240 \ \ \ \ \\
  53138 & B4Ia      &  22.0 & 0.130 & 0.038 & 14920$\pm 820$ & 0.580$\pm0.013$ & 1.00  & 15620 \ \ \ \ \\
  53244 & B6III     &  41.9 & 0.289 & 0.033 & 13690$\pm 690$ & 0.378$\pm0.011$ & 1.00  & 14830 \ \ \ \ \\
  58350 & B5Ia      &  19.0 & 0.152 & 0.012 & 12670$\pm 580$ & 0.882$\pm0.010$ & 1.00  & 14180 \ \ \ \ \\
  66811 & O6-7III   &  55.3 & 0.051 & 0.043 & 37250$\pm2630$ & 0.371$\pm0.027$ & 1.00  & 36630 \ \ \ \ \\
  67797 & B4IV      &  50.9 & 0.251 & 0.028 & 16680$\pm 960$ & 0.280$\pm0.016$ & 1.00  & 16450 \ \ \ \ \\
  68520 & B5IV      &  46.0 & 0.287 & 0.022 & 14090$\pm 730$ & 0.326$\pm0.011$ & 1.00  & 14950 \ \ \ \ \\
  71155 & A1V       &  74.0 & 0.467 & 0.025 & 10060$\pm 360$ & 0.538$\pm0.008$ & 1.00  &  9800 \ \ \ \ \\
  74280 & B2IV-V    &  53.3 & 0.189 & 0.019 & 19410$\pm1200$ & 0.254$\pm0.019$ & 1.00  & 20080 \ \ \ \ \\
  77327 & A1III     &  47.0 & 0.509 & 0.008 & \ \ 9080$\pm 290$ & 0.691$\pm0.007$ & 1.00 &9790 \ \ \ \ \\
  79447 & B3IV      &  50.5 & 0.220 & 0.015 & 17250$\pm1010$ & 0.333$\pm0.016$ & 1.00  & 18010 \ \ \ \ \\
  79469 & A0V       &  63.7 & 0.441 & 0.023 & 10980$\pm 430$ & 0.490$\pm0.009$ & 1.00  & 10630 \ \ \ \ \\
  83754 & B5V       &  59.1 & 0.253 & 0.024 & 16150$\pm 920$ & 0.211$\pm0.015$ & 1.00  & 16220 \ \ \ \ \\
  83944 & B9V       &  66.2 & 0.410 & 0.006 & 11550$\pm 480$ & 0.335$\pm0.009$ & 1.00  & 11150 \ \ \ \ \\
  86440 & B6II      &  25.0 & 0.232 & 0.049 & 13980$\pm 720$ & 0.493$\pm0.011$ & 1.18  & 13650 \ \ \ \ \\
  87737 & A0Ib      &  18.0 & 0.411 & 0.053 & \ \ 9820$\pm 340$ & 0.685$\pm0.007$ & 1.00 &9900 \ \ \ \ \\
  89021 & A1V       &  64.0 & 0.488 & 0.006 & \ \ 8790$\pm 270$ & 0.754$\pm0.007$ & 1.00 &9730 \ \ \ \ \\
  91316 & B1.5Ib    &  34.8 & 0.094 & 0.056 & 19830$\pm1230$ & 0.312$\pm0.020$ & 1.21  & 21370 \ \ \ \ \\
  95418 & $>$A1V    &  70.0 & 0.496 & 0.004 & \ \ 9470$\pm 310$ & 1.112$\pm0.007$ & 1.00 &9480 \ \ \ \ \\
  97633 & $>$A1V    &  67.0 & 0.510 & 0.022 & \ \ 9180$\pm 290$ & 0.772$\pm0.007$ & 1.00 &9280 \ \ \ \ \\
  98664 & B9IV      &  54.4 & 0.438 & 0.023 & 10680$\pm 410$ & 0.466$\pm0.008$ & 1.00  & 10970 \ \ \ \ \\
  98718 & B5V       &  60.1 & 0.261 & 0.014 & 16760$\pm 970$ & 0.346$\pm0.016$ & 1.00  & 15730 \ \ \ \ \\
 100600 & B3V       &  58.0 & 0.218 & 0.033 & 18230$\pm1090$ & 0.129$\pm0.018$ & 1.00  & 18280 \ \ \ \ \\
 100841 & A1III     &  34.6 & 0.518 & 0.020 & \ \ 9880$\pm 340$ & 0.768$\pm0.008$ & 1.00 &9720 \ \ \ \ \\
 100889 & B9V       &  57.2 & 0.435 & 0.021 & 11280$\pm 460$ & 0.334$\pm0.009$ & 1.00  & 10940 \ \ \ \ \\
 106625 & B7IV      &  49.4 & 0.348 & 0.009 & 12360$\pm 550$ & 0.799$\pm0.010$ & 1.00  & 12820 \ \ \ \ \\
 106911 & B4V       &  58.2 & 0.238 & 0.026 & 15330$\pm 850$ & 0.321$\pm0.014$ & 1.00  & 17060 \ \ \ \ \\
 108767 & A0.5V     &  71.0 & 0.448 & 0.020 & 10580$\pm 400$ & 0.777$\pm0.008$ & 1.00  & 10270 \ \ \ \ \\
 109026 & B4V       &  58.9 & 0.227 & 0.027 & 16740$\pm 970$ & 0.357$\pm0.016$ & 1.00  & 17700 \ \ \ \ \\
 111123 & B1IV      &  54.3 & 0.125 & 0.041 & 27030$\pm1800$ & 0.780$\pm0.023$ & 1.00  & 25740 \ \ \ \ \\
 112185 & A1IV      &  58.0 & 0.511 & 0.000 & \ \ 9240$\pm 300$ & 1.504$\pm0.007$ & 1.00 &9530 \ \ \ \ \\
 112413 & B7V       &  59.0 & 0.335 & 0.021 & 11630$\pm 490$ & 0.718$\pm0.009$ & 1.00  & 12890 \ \ \ \ \\
 116656 & A1V       &  67.0 & 0.470 & 0.033 & \ \ 9340$\pm 300$ & 1.361$\pm0.007$ & 1.00 &9980 \ \ \ \ \\
 120315 & B4V       &  61.0 & 0.227 & 0.015 & 17870$\pm1060$ & 0.832$\pm0.017$ & 1.00  & 17710 \ \ \ \ \\
 123299 & A0IV      &  53.0 & 0.482 & 0.025 & 10430$\pm 390$ & 0.571$\pm0.008$ & 1.00  & 10190 \ \ \ \ \\
 129056 & B2IV      &  47.8 & 0.145 & 0.041 & 23100$\pm1490$ & 0.537$\pm0.022$ & 1.00  & 22380 \ \ \ \ \\
 129246 & $>$A1V    &  79.0 & 0.488 & 0.000 & \ \ 8990$\pm 280$ & 0.611$\pm0.007$ & 1.00 &9220 \ \ \ \ \\
 132058 & B2IV      &  52.1 & 0.148 & 0.024 & 24090$\pm1570$ & 0.436$\pm0.022$ & 1.00  & 23000 \ \ \ \ \\
 135742 & B7.5IV    &  44.0 & 0.363 & 0.022 & 12300$\pm 550$ & 0.801$\pm0.010$ & 1.00  & 12500 \ \ \ \ \\
 137422 & $>$A2III  &  35.0 & 0.562 & 0.022 & \ \ 8280$\pm 240$ & 1.022$\pm0.009$ & 0.48 &9110 \ \ \ \ \\
 139006 & A1V       &  69.0 & 0.476 & 0.026 & \ \ 9900$\pm 340$ & 1.175$\pm0.008$ & 1.00 &9820 \ \ \ \ \\
 141003 & A1V       &  69.0 & 0.479 & 0.018 & \ \ 8810$\pm 270$ & 0.682$\pm0.007$ & 1.00 &9770 \ \ \ \ \\
 144217 & B1V       &  65.5 & 0.115 & 0.210 & 30540$\pm2080$ & 0.505$\pm0.024$ & 1.00  & 27750 \ \ \ \ \\
 145389 & B8.5V     &  65.0 & 0.363 & 0.040 & 11700$\pm 490$ & 0.407$\pm0.009$ & 1.00  & 12030 \ \ \ \ \\
 147394 & B5V       &  53.0 & 0.262 & 0.034 & 16350$\pm 940$ & 0.356$\pm0.015$ & 1.00  & 15900 \ \ \ \ \\
 148112 & A0IV      &  57.0 & 0.458 & 0.072 & \ \ 9810$\pm 340$ & 0.417$\pm0.007$ & 1.00&10520 \ \ \ \ \\
 149438 & B0IV      &  58.4 & 0.095 & 0.040 & 31440$\pm2150$& 0.333$\pm0.025$ & 1.00  & 29640 \ \ \ \ \\
 149881 & B0.5III-II&  45.0 & 0.094 & 0.070 & 23420$\pm1520$& 0.063$\pm0.022$ & 1.35  & 24720 \ \ \ \ \\
 155125 & $>$A1V    &  75.0 & 0.478 & 0.000 & \ \ 8620$\pm 260$& 1.205$\pm0.008$ & 1.00 &9570 \ \ \ \ \\
 155763 & B6IV      &  45.0 & 0.314 & 0.026 & 13420$\pm 660$& 0.586$\pm0.011$ & 1.00  & 14080 \ \ \ \ \\
 158094 & B7V       &  53.4 & 0.343 & 0.020 & 12360$\pm 550$ & 0.508$\pm0.010$ & 1.00  & 12840 \ \ \ \ \\
 159975 & B8III     &  35.5 & 0.360 & 0.243 & 12790$\pm 590$ & 0.428$\pm0.010$ & 0.94  & 12470 \ \ \ \ \\
 160578 & B1.5III   &  50.1 & 0.131 & 0.033 & 24720$\pm1620$ & 0.492$\pm0.022$ & 1.12  & 23780 \ \ \ \ \\
 160762 & B3IV      &  51.0 & 0.211 & 0.036 & 19100$\pm1170$ & 0.330$\pm0.019$ & 1.00  & 18590 \ \ \ \ \\
 164353 & B5Ib      &  23.8 & 0.181 & 0.185 & 15420$\pm 860$ & 0.045$\pm0.014$ & 1.11  & 14790 \ \ \ \ \\
 166182 & B2.5III   &  43.0 & 0.174 & 0.074 & 22420$\pm1440$ & 0.232$\pm0.021$ & 1.00  & 19870 \ \ \ \ \\
 169022 & A0III     &  45.5 & 0.498 & 0.024 & \ \ 9520$\pm 320$ & 1.468$\pm0.007$ & 1.00 &9960 \ \ \ \ \\
 172167 & A1V       &  66.0 & 0.489 & 0.000 & \ \ 9470$\pm 310$ & 3.292$\pm0.007$ & 1.00 &9670 \ \ \ \ \\
 173300 & B7IV      &  44.8 & 0.332 & 0.118 & 14990$\pm 830$ & 0.597$\pm0.014$ & 1.00  & 13490 \ \ \ \ \\
 175191 & B2.5V     &  62.1 & 0.187 & 0.000 & 18890$\pm1150$ & 0.711$\pm0.018$ & 1.00  & 20370 \ \ \ \ \\
 176437 & A1III     &  32.0 & 0.493 & 0.040 & 10000$\pm 350$ & 0.740$\pm0.008$ & 1.00&10000 \ \ \ \ \\
 177724 & A1V       &  66.0 & 0.493 & 0.052 & \ \ 9830$\pm 340$ & 0.865$\pm0.007$ & 1.00 &9600 \ \ \ \ \\
 177756 & B9V       &  58.9 & 0.396 & 0.004 & 11780$\pm 500$ & 0.558$\pm0.009$ & 1.00  & 11600 \ \ \ \ \\
 179761 & B8V-IV    &  52.4 & 0.364 & 0.073 & 13060$\pm 620$ & 0.268$\pm0.010$ & 1.00  & 12320 \ \ \ \ \\
 182255 & B5V       &  56.0 & 0.268 & 0.030 & 15300$\pm 850$ & 0.208$\pm0.014$ & 1.00  & 15520 \ \ \ \ \\
 186882 & A0IV      &  50.0 & 0.479 & 0.031 & 10150$\pm 360$ & 0.858$\pm0.008$ & 1.00  & 10280 \ \ \ \ \\
 188209 & B0Ib      &  42.0 & 0.065 & 0.159 & 25260$\pm1670$ & 0.124$\pm0.023$ & 1.33  & 26040 \ \ \ \ \\
 191692 & B9IV      &  52.0 & 0.462 & 0.011 & 10340$\pm 380$ & 0.712$\pm0.008$ & 1.00  & 10570 \ \ \ \ \\
 192425 & A1V       &  77.0 & 0.461 & 0.033 & \ \ 9120$\pm 290$ & 0.370$\pm0.007$ & 1.00 &9810 \ \ \ \ \\
 192907 & A0V       &  59.0 & 0.451 & 0.023 & 10500$\pm 390$ & 0.408$\pm0.008$ & 1.00  & 10570 \ \ \ \ \\
 195556 & B4II      &  33.0 & 0.204 & 0.128 & 17680$\pm1040$ & 0.236$\pm0.017$ & 1.07  & 16320 \ \ \ \ \\
 195810 & B6IV      &  48.0 & 0.303 & 0.032 & 14540$\pm 780$ & 0.370$\pm0.012$ & 1.00  & 14370 \ \ \ \ \\
 196867 & B9IV      &  53.0 & 0.417 & 0.018 & 11220$\pm 450$ & 0.512$\pm0.009$ & 1.00  & 11370 \ \ \ \ \\
 197345 & A2Ia      &   4.0 & 0.366 & 0.056 & \ \ 8720$\pm 260$ & 2.255$\pm0.008$ & 1.70 &8630 \ \ \ \ \\
 198001 & $>$A1V-IV &  64.0 & 0.529 & 0.021 & \ \ 9370$\pm 310$ & 0.612$\pm0.007$ & 1.00 &8980 \ \ \ \ \\
 198478 & B4Ia      &  20.0 & 0.109 & 0.562 & 15390$\pm 860$ & 0.517$\pm0.014$ & 1.21  & 15880 \ \ \ \ \\
 199081 & B4V-IV    &  51.0 & 0.248 & 0.028 & 16180$\pm 920$ & 0.240$\pm0.015$ & 1.00  & 16590 \ \ \ \ \\
 202850 & B9.5Ia    &  10.0 & 0.265 & 0.200 & 11170$\pm 450$ & 0.536$\pm0.009$ & 1.78  & 11110 \ \ \ \ \\
 204172 & B0II-Ib   &  42.0 & 0.073 & 0.142 & 24110$\pm1570$ & 0.110$\pm0.022$ & 1.36  & 25370 \ \ \ \ \\
 205021 & B1V       &  59.0 & 0.120 & 0.037 & 26920$\pm1790$ & 0.312$\pm0.023$ & 1.00  & 26960 \ \ \ \ \\
 206672 & B3III     &  43.0 & 0.208 & 0.089 & 18360$\pm1100$ & 0.241$\pm0.018$ & 1.27  & 18150 \ \ \ \ \\
 207260 & A2Ia      &   6.0 & 0.378 & 0.506 & \ \ 8980$\pm 280$ & 1.010$\pm0.007$ & 1.00 &8790 \ \ \ \ \\
 207330 & B3III     &  44.0 & 0.190 & 0.078 & 17890$\pm1060$ & 0.301$\pm0.017$ & 1.00  & 19270 \ \ \ \ \\
 207971 & B7IV      &  47.1 & 0.338 & 0.011 & 12520$\pm 570$ & 0.649$\pm0.010$ & 1.00  & 13230 \ \ \ \ \\
 209481 & O6-7IV    &  57.0 & 0.062 & 0.348 & 30570$\pm2080$ & 0.149$\pm0.024$ & 1.00  & 34280 \ \ \ \ \\
 209744 & B1III     &  50.0 & 0.120 & 0.363 & 25440$\pm1680$ & 0.112$\pm0.023$ & 1.00  & 24510 \ \ \ \ \\
 209819 & B8V       &  64.3 & 0.348 & 0.020 & 12310$\pm 550$ & 0.376$\pm0.010$ & 1.00  & 12320 \ \ \ \ \\
 209952 & B5V       &  51.7 & 0.282 & 0.013 & 13920$\pm 710$ & 1.084$\pm0.011$ & 1.00  & 14990 \ \ \ \ \\
 209961 & B2IV      &  49.0 & 0.180 & 0.159 & 21100$\pm1330$ & 0.119$\pm0.020$ & 1.00  & 20420 \ \ \ \ \\
 209975 & B1Ia      &  32.0 & 0.063 & 0.331 & 24720$\pm1620$ & 0.222$\pm0.022$ & 0.45  & 22190 \ \ \ \ \\
 210191 & B4III     &  38.8 & 0.202 & 0.040 & 17890$\pm1060$ & 0.140$\pm0.017$ & 1.08  & 17750 \ \ \ \ \\
 210418 & $>$A1V    &  76.0 & 0.480 & 0.033 & \ \ 8840$\pm 270$ & 0.747$\pm0.007$ & 1.00 &9500 \ \ \ \ \\
 212061 & A0V       &  64.5 & 0.458 & 0.022 & 10490$\pm 390$ & 0.527$\pm0.008$ & 1.00  & 10280 \ \ \ \ \\
 212120 & B5V       &  55.0 & 0.282 & 0.050 & 15250$\pm 850$ & 0.288$\pm0.014$ & 1.00  & 14900 \ \ \ \ \\
 212593 & B8.5II    &  21.0 & 0.299 & 0.179 & 11150$\pm 440$ & 0.441$\pm0.009$ & 1.53  & 11370 \ \ \ \ \\
 212883 & B2.5IV-III&  48.0 & 0.183 & 0.085 & 20400$\pm1280$ & 0.101$\pm0.020$ & 1.00  & 20140 \ \ \ \ \\
 212978 & B2III     &  46.0 & 0.176 & 0.081 & 20990$\pm1320$ & 0.113$\pm0.020$ & 1.00  & 20360 \ \ \ \ \\
 213420 & B2III     &  44.0 & 0.167 & 0.131 & 19750$\pm1230$ & 0.267$\pm0.019$ & 1.00  & 20460 \ \ \ \ \\
 213558 & A1.5V     &  75.0 & 0.465 & 0.038 & \ \ 9840$\pm 340$ & 0.590$\pm0.007$ & 1.00 &9800 \ \ \ \ \\
 213976 & B2V       &  64.0 & 0.165 & 0.105 & 22800$\pm1470$ & 0.073$\pm0.021$ & 1.00  & 22170 \ \ \ \ \\
 213998 & B9V       &  58.5 & 0.397 & 0.017 & 11740$\pm 500$ & 0.434$\pm0.009$ & 1.00  & 11590 \ \ \ \ \\
 214240 & B5III     &  43.0 & 0.265 & 0.126 & 16330$\pm 930$ & 0.139$\pm0.015$ & 1.00  & 15750 \ \ \ \ \\
 214652 & B2V       &  60.0 & 0.183 & 0.105 & 22350$\pm1430$ & 0.081$\pm0.021$ & 1.00  & 20730 \ \ \ \ \\
 214680 & O6-7V     &  65.0 & 0.068 & 0.089 & 32380$\pm2220$ & 0.136$\pm0.025$ & 1.00  & 33820 \ \ \ \ \\
 214923 & B9IV      &  50.3 & 0.420 & 0.005 & 11430$\pm 470$ & 0.581$\pm0.009$ & 1.00  & 11360 \ \ \ \ \\
 214993 & B1III     &  49.0 & 0.120 & 0.111 & 24130$\pm1570$ & 0.155$\pm0.022$ & 1.00  & 24260 \ \ \ \ \\
 214994 & A0.5V     &  62.0 & 0.477 & 0.050 & \ \ 9930$\pm 350$ & 0.364$\pm0.008$ & 1.00 &9970 \ \ \ \ \\
 215191 & B2IV      &  52.0 & 0.145 & 0.131 & 22800$\pm1470$ & 0.099$\pm0.021$ & 1.00  & 23230 \ \ \ \ \\
 217101 & B2V       &  63.0 & 0.151 & 0.093 & 22760$\pm1460$ & 0.107$\pm0.021$ & 1.00  & 23700 \ \ \ \ \\
 217811 & B4III     &  45.0 & 0.225 & 0.217 & 19330$\pm1190$ & 0.132$\pm0.019$ & 1.00  & 17420 \ \ \ \ \\
 218045 & A1IV      &  55.0 & 0.500 & 0.021 & \ \ 9850$\pm 340$ & 1.036$\pm0.008$ & 1.00 &9850 \ \ \ \ \\
 218376 & B0.5III   &  46.0 & 0.095 & 0.244 & 27100$\pm1810$ & 0.204$\pm0.023$ & 1.28  & 25190 \ \ \ \ \\
 218407 & B3V       &  60.0 & 0.184 & 0.174 & 21200$\pm1340$ & 0.101$\pm0.020$ & 1.00  & 20650 \ \ \ \ \\
 219688 & B5V       &  51.8 & 0.266 & 0.021 & 15260$\pm 850$ & 0.297$\pm0.014$ & 1.00  & 15730 \ \ \ \ \\
 222173 & B8IV      &  44.0 & 0.371 & 0.038 & 12620$\pm 580$ & 0.369$\pm0.010$ & 1.00  & 12340 \ \ \ \ \\
 222661 & B9.5V     &  66.0 & 0.428 & 0.021 & 10860$\pm 420$ & 0.381$\pm0.008$ & 1.00  & 10820 \ \ \ \ \\
 223640 & B6.5V     &  55.9 & 0.321 & 0.000 & 11740$\pm 500$ & 0.260$\pm0.009$ & 1.00  & 13510 \ \ \ \ \\
 224990 & B5V       &  54.2 & 0.259 & 0.032 & 16100$\pm 910$ & 0.215$\pm0.015$ & 1.00  & 16030 \ \ \ \ \\
\noalign{\smallskip}
\hline
\noalign{\smallskip}
\multicolumn{9}{l}{The parameter $\lambda_1$ is given as $\lambda_1-3700$ \AA .}\\
\multicolumn{9}{l}{The notation $>$SpT/LC means that the spectral type is cooler than the indicated one.}\\
\end{longtable}

\setlongtables
\onecolumn
\setcounter{table}{4}
\begin{longtable}{r|r|rrrrrr}
\kill
\caption{\label{tab_comp}Compilation of effective temperatures determined for the program stars by other authors.}\\
\noalign{\smallskip}
\hline
\noalign{\smallskip}
 HD & $T_{\rm eff}^f$ & $T_{\rm eff}^{(1)}$ & $T_{\rm eff}^{(2)}$ & $T_{\rm eff}^{(3)}$ &
$T_{\rm eff}^{(4)}$ & $T_{\rm eff}^{(5)}$ & $T_{\rm eff}^{(6)}$  \ \ \ \ \\
\noalign{\smallskip}
\hline
\noalign{\smallskip}
\endfirsthead
\caption{continued.}\\
\noalign{\smallskip} 
\hline
\noalign{\smallskip}
 HD & $T_{\rm eff}^f$ & $T_{\rm eff}^{(1)}$ & $T_{\rm eff}^{(2)}$ & $T_{\rm eff}^{(3)}$ & 
$T_{\rm eff}^{(4)}$ & $T_{\rm eff}^{(5)}$ & $T_{\rm eff}^{(6)}$  \ \ \ \ \\
\noalign{\smallskip}
\hline
\noalign{\smallskip}
\endhead
\noalign{\smallskip}
\hline
\endfoot
     358 & 12750 &  --   &  --   & --    &  --   &  --   &  --   \ \ \ \ \\
     886 & 21870 &  --   & 21990 & --    &  --   & 21250 &  --   \ \ \ \ \\
    2905 & 22160 &  --   &  --   & 21600 &  --   & 24000 & 22500 \ \ \ \ \\
    3360 & 21850 &  --   & 22210 & --    &  --   &  --   &  --   \ \ \ \ \\
    3369 & 15980 & 15520 &  --   & --    &  --   &  --   &  --   \ \ \ \ \\
    4142 & 16430 & 16670 &  --   & --    &  --   &  --   &  --   \ \ \ \ \\
    4727 & 16290 & 15850 &  --   & --    &  --   &  --   &  --   \ \ \ \ \\
   11241 & 22080 & 25350 &  --   & --    &  --   &  --   &  --   \ \ \ \ \\
   12767 & 13250 &  --   &  --   & --    &  --   &  --   &  --   \ \ \ \ \\
   12953 &  9750 &  --   &  --   & --    &  --   &  --   &  --   \ \ \ \ \\
   13267 & 14380 &  --   &  --   & --    &  --   & 15000 &  --   \ \ \ \ \\
   14055 &  9440 &  --   &  --   & --    & 9490  &  --   &  --   \ \ \ \ \\
   14228 & 12470 & 12900 &  --   & --    & 13080 &  --   &  --   \ \ \ \ \\
   14489 &  9840 &  --   &  --   & --    &  --   &  --   &  --   \ \ \ \ \\
   14818 & 18300 &  --   &  --   & --    &  --   & 20000 & 18250 \ \ \ \ \\
   15130 &  9920 &  --   &  --   & --    & 9890  &  --   &  --   \ \ \ \ \\
   15318 & 10630 &  --   &  --   & --    & 10600 &  --   &  --   \ \ \ \ \\
   16046 & 10910 &  --   &  --   & --    &  --   &  --   &  --   \ \ \ \ \\
   16582 & 23160 &  --   & 23570 & --    &  --   &  --   &  --   \ \ \ \ \\
   16908 & 17520 & 16850 &  --   & --    & 16940 &  --   &  --   \ \ \ \ \\
   17081 & 13680 & 13210 & 13040 & --    & 13170 & 13100 &  --   \ \ \ \ \\
   17573 & 12980 & 12570 &  --   & --    & 12400 &  --   &  --   \ \ \ \ \\
   18604 & 13940 & 13240 &  --   & --    &  --   &  --   &  --   \ \ \ \ \\
   19356 & 12800 & 14290 &  --   & --    &  --   &  --   &  --   \ \ \ \ \\
   20041 & 10800 &  --   &  --   & --    &  --   &  --   &  --   \ \ \ \ \\
   21291 & 11420 &  --   &  --   & --    &  --   & 11500 &  --   \ \ \ \ \\
   21364 & 13040 & 12650 &  --   & --    & 12540 &  --   &  --   \ \ \ \ \\
   21389 & 11040 &  --   &  --   & --    &  --   &  --   &  --   \ \ \ \ \\
   21447 &  9240 &  --   &  --   & --    &  --   &  --   &  --   \ \ \ \ \\
   21790 & 11760 &  --   &  --   & --    &  --   &  --   &  --   \ \ \ \ \\
   21856 & 25370 & 25880 &  --   & --    &  --   &  --   &  --   \ \ \ \ \\
   22928 & 14890 &  --   & 13940 & --    & 14200 &  --   &  --   \ \ \ \ \\
   22951 & 29330 & 28970 &  --   & --    &  --   &  --   &  --   \ \ \ \ \\
   23180 & 22840 & 22860 &  --   & --    &  --   &  --   &  --   \ \ \ \ \\
   23227 & 16230 &  --   &  --   & --    &  --   &  --   &  --   \ \ \ \ \\
   23288 & 14020 &  --   &  --   & --    &  --   &  --   &  --   \ \ \ \ \\
   23324 & 13220 & 12760 &  --   & --    &  --   &  --   &  --   \ \ \ \ \\
   23338 & 14180 &  --   &  --   & --    &  --   &  --   &  --   \ \ \ \ \\
   23408 & 14310 &  --   &  --   & --    &  --   &  --   &  --   \ \ \ \ \\
   23625 & 23980 & 25530 &  --   & --    &  --   &  --   &  --   \ \ \ \ \\
   23753 & 12680 &  --   &  --   & --    &  --   &  --   &  --   \ \ \ \ \\
   23850 & 13020 &  --   &  --   & --    &  --   &  --   &  --   \ \ \ \ \\
   23923 & 11640 &  --   &  --   & --    &  --   &  --   &  --   \ \ \ \ \\
   24131 & 29130 & 28380 &  --   & --    &  --   &  --   &  --   \ \ \ \ \\
   24398 & 22040 &  --   & 19870 & 19900 &  --   & 23000 &  --   \ \ \ \ \\
   24431 & 27370 &  --   &  --   & --    &  --   &  --   &  --   \ \ \ \ \\
   24760 & 27160 & 25760 &  --   & --    &  --   &  --   &  --   \ \ \ \ \\
   25204 & 16970 &  --   &  --   & --    &  --   &  --   &  --   \ \ \ \ \\
   25490 &  8990 &  --   &  --   & --    & 9260  & 9250  &  --   \ \ \ \ \\
   27376 & 12460 &  --   &  --   & --    &  --   &  --   &  --   \ \ \ \ \\
   27396 & 16720 &  --   & 16800 & --    &  --   &  --   &  --   \ \ \ \ \\
   27962 &  8680 &  --   &  --   & --    &  --   & 9000  &  --   \ \ \ \ \\
   29248 & 23610 &  --   & 23500 & 21600 &  --   &  --   &  --   \ \ \ \ \\
   29305 & 12120 &  --   &  --   & --    &  --   &  --   &  --   \ \ \ \ \\
   30211 & 15770 & 15540 & 14950 & --    & 15630 &  --   &  --   \ \ \ \ \\
   30614 & 25400 &  --   & 25000 & 26300 &  --   & 32500 & 29000 \ \ \ \ \\
   31647 &  9750 &  --   &  --   & --    &  --   &  --   &  --   \ \ \ \ \\
   32309 & 10800 &  --   & 10200 & --    &  --   &  --   &  --   \ \ \ \ \\
   32549 &  9520 &  --   & 10200 & --    &  --   &  --   &  --   \ \ \ \ \\
   32630 & 17940 & 17020 & 17580 & --    & 17050 & 16380 &  --   \ \ \ \ \\
   33802 & 13280 &  --   & 11860 & --    &  --   &  --   &  --   \ \ \ \ \\
   33904 & 12390 &  --   &  --   & --    &  --   &  --   &  --   \ \ \ \ \\
   33949 & 12750 & 11690 &  --   & --    &  --   &  --   &  --   \ \ \ \ \\
   34085 & 12130 &  --   & 11660 & --    &  --   & 13000 &  --   \ \ \ \ \\
   34503 & 14450 & 13510 & 14260 & --    & 14620 &  --   &  --   \ \ \ \ \\
   35468 & 21840 & 21040 & 21760 & 20900 & 21940 &  --   &  --   \ \ \ \ \\
   35497 & 13960 & 13650 & 13820 & --    & 13490 &  --   &  --   \ \ \ \ \\
   35600 & 10870 &  --   &  --   & --    &  --   & 11000 &  --   \ \ \ \ \\
   36267 & 16020 & 17100 &  --   & --    &  --   &  --   &  --   \ \ \ \ \\
   36371 & 15370 &  --   &  --   & --    &  --   & 16500 &  --   \ \ \ \ \\
   36512 & 32340 &  --   & 34350 & --    &  --   &  --   &  --   \ \ \ \ \\
   36822 & 28340 &  --   &  --   & --    & 31120 &  --   &  --   \ \ \ \ \\
   37128 & 24670 &  --   & 24960 & 26400 &  --   & 28500 & 27250 \ \ \ \ \\
   37468 & 31270 &  --   & 31560 & --    &  --   &  --   &  --   \ \ \ \ \\
   37481 & 24350 & 24550 &  --   & --    &  --   &  --   &  --   \ \ \ \ \\
   37744 & 25690 & 24830 &  --   & --    &  --   &  --   &  --   \ \ \ \ \\
   38666 & 31510 &  --   &  --   & --    &  --   &  --   &  --   \ \ \ \ \\
   38771 & 23170 &  --   & 26390 & 23100 &  --   & 27500 & 26250 \ \ \ \ \\
   39970 & 11530 &  --   &  --   & --    &  --   &  --   &  --   \ \ \ \ \\
   40111 & 24660 &  --   &  --   & 21400 &  --   &  --   &  --   \ \ \ \ \\
   40183 &  8910 &  --   &  --   & --    &  --   &  --   &  --   \ \ \ \ \\
   40312 &  9890 &  --   &  --   & --    &  --   &  --   &  --   \ \ \ \ \\
   40589 & 11660 &  --   &  --   & --    &  --   &  --   &  --   \ \ \ \ \\
   41117 & 20740 &  --   & 17460 & --    &  --   & 19500 & 19000 \ \ \ \ \\
   41753 & 18800 & 16970 & 17410 & --    & 17640 &  --   &  --   \ \ \ \ \\
   42087 & 16460 &  --   &  --   & --    &  --   & 20500 & 18000 \ \ \ \ \\
   43112 & 27900 & 28510 &  --   & --    &  --   &  --   &  --   \ \ \ \ \\
   43384 & 14780 &  --   &  --   & --    &  --   &  --   &  --   \ \ \ \ \\
   44743 & 25320 &  --   & 25500 & 24700 &  --   &  --   &  --   \ \ \ \ \\
   46300 &  9800 &  --   &  --   & --    &  --   &  --   &  --   \ \ \ \ \\
   46769 & 13930 & 14520 &  --   & --    &  --   &  --   &  --   \ \ \ \ \\
   47105 &  9040 &  --   & 9260  & --    &  9380 & 9150  &  --   \ \ \ \ \\
   47240 & 21540 & 20610 &  --   & --    &  --   &  --   &  --   \ \ \ \ \\
   47432 & 25620 &  --   &  --   & --    &  --   &  --   &  --   \ \ \ \ \\
   47670 & 12120 & 11910 & 11610 & --    & 11950 &  --   &  --   \ \ \ \ \\
   47964 & 12100 &  --   &  --   & --    &  --   &  --   &  --   \ \ \ \ \\
   48434 & 25290 & 24270 &  --   & --    &  --   &  --   &  --   \ \ \ \ \\
   48977 & 19590 & 19770 &  --   & --    &  --   &  --   &  --   \ \ \ \ \\
   49567 & 17270 & 17100 &  --   & --    &  --   &  --   &  --   \ \ \ \ \\
   52089 & 22010 &  --   & 21830 & 21000 &  --   &  --   &  --   \ \ \ \ \\
   53138 & 14920 &  --   & 14760 & --    &  --   & 18500 & 16000 \ \ \ \ \\
   53244 & 13690 &  --   & 13600 & --    &  --   &  --   &  --   \ \ \ \ \\
   58350 & 12670 &  --   & 13150 & --    &  --   & 16000 & 15000 \ \ \ \ \\
   66811 & 37250 &  --   & 33760 & 35100 &  --   &  --   &  --   \ \ \ \ \\
   67797 & 16680 &  --   & 15460 & --    &  --   &  --   &  --   \ \ \ \ \\
   68520 & 14090 & 14420 &  --   & --    &  --   &  --   &  --   \ \ \ \ \\
   71155 & 10060 &  --   &  --   & --    &  --   &  --   &  --   \ \ \ \ \\
   74280 & 19410 & 18680 & 18790 & --    & 18650 &  --   &  --   \ \ \ \ \\
   77327 &  9080 &  --   &  --   & --    &  --   &  --   &  --   \ \ \ \ \\
   79447 & 17250 & 17220 &  --   & --    &  --   &  --   &  --   \ \ \ \ \\
   79469 & 10980 &  --   &  --   & --    & 10640 &  --   &  --   \ \ \ \ \\
   83754 & 16140 &  --   & 15400 & --    &  --   &  --   &  --   \ \ \ \ \\
   83944 & 11550 & 10740 &  --   & --    &  --   &  --   &  --   \ \ \ \ \\
   86440 & 13980 & 13300 &  --   & --    &  --   &  --   &  --   \ \ \ \ \\
   87737 &  9820 &  --   &  --   & --    &  --   &  --   &  --   \ \ \ \ \\
   89021 &  8790 &  --   &  --   & --    &  9160 & 9000  &  --   \ \ \ \ \\
   91316 & 19830 &  --   &  --   & 20300 &  --   &  --   & 22000  \ \ \ \ \\
   95418 &  9470 &  --   &  9170 & --    &  9650 & 9600  &  --   \ \ \ \ \\
   97633 &  9180 &  --   &  --   & --    &  9130 & 9250  &  --   \ \ \ \ \\
   98664 & 10680 &  --   &  --   & --    &  --   &  --   &  --   \ \ \ \ \\
   98718 & 16760 &  --   &  --   & --    &  --   &  --   &  --   \ \ \ \ \\
  100600 & 18230 & 19320 &  --   & --    &  --   &  --   &  --   \ \ \ \ \\
  100841 &  9880 &  --   & 10470 & --    &  --   &  --   &  --   \ \ \ \ \\
  100889 & 11280 &  --   &  --   & --    & 10810 &  --   &  --   \ \ \ \ \\
  106625 & 12360 &  --   & 12450 & --    &  --   &  --   &  --   \ \ \ \ \\
  106911 & 15330 &  --   &  --   & --    &  --   &  --   &  --   \ \ \ \ \\
  108767 & 10580 &  --   &  --   & --    &  --   &  --   &  --   \ \ \ \ \\
  109026 & 16740 &  --   & 15510 & --    &  --   &  --   &  --   \ \ \ \ \\
  111123 & 27030 &  --   & 27600 & --    &  --   &  --   &  --   \ \ \ \ \\
  112185 &  9240 &  --   &  --   & --    &  --   &  --   &  --   \ \ \ \ \\
  112413 & 11630 &  --   &  --   & --    &  --   &  --   &  --   \ \ \ \ \\
  116656 &  9340 &  --   &  --   & --    &  --   &  --   &  --   \ \ \ \ \\
  120315 & 17870 & 17360 & 16770 & --    & 17100 & 16900 &  --   \ \ \ \ \\
  123299 & 10430 &  --   &  --   & --    &  --   & 10000 &  --   \ \ \ \ \\
  129056 & 23100 & 20320 &  --   & --    &  --   &  --   &  --   \ \ \ \ \\
  129246 &  8990 &  --   &  --   & --    &  --   &  --   &  --   \ \ \ \ \\
  132058 & 24090 &  --   & 23780 & 22800 &  --   &  --   &  --   \ \ \ \ \\
  135742 & 12300 & 12060 &  --   & --    & 12310 & 12130 &  --   \ \ \ \ \\
  137422 &  8280 &  --   &  --   & --    &  --   &  --   &  --   \ \ \ \ \\
  139006 &  9900 &  --   &  --   & --    &  --   &  --   &  --   \ \ \ \ \\
  141003 &  8810 &  --   &  --   & --    &  --   &  --   &  --   \ \ \ \ \\
  144217 & 30540 &  --   &  --   & --    &  --   &  --   &  --   \ \ \ \ \\
  145389 & 11700 &  --   & 10860 & --    &  --   &  --   &  --   \ \ \ \ \\
  147394 & 16350 & 15030 & 15210 & --    & 15050 & 15000 &  --   \ \ \ \ \\
  148112 &  9810 &  --   &  --   & --    &  --   &  --   &  --   \ \ \ \ \\
  149438 & 31440 &  --   &  --   & 31800 & 29920 &  --   &  --   \ \ \ \ \\
  149881 & 23420 & 21130 &  --   & --    & 29280 &  --   &  --   \ \ \ \ \\
  155125 &  8620 &  --   &  --   & --    &  --   &  --   &  --   \ \ \ \ \\
  155763 & 13420 &  --   & 12960 & --    & 13090 &  --   &  --   \ \ \ \ \\
  158094 & 12360 & 11480 & 12010 & --    &  --   &  --   &  --   \ \ \ \ \\
  159975 & 12790 &  --   & 11650 & --    &  --   &  --   &  --   \ \ \ \ \\
  160578 & 24720 &  --   & 25780 & 24400 &  --   &  --   &  --   \ \ \ \ \\
  160762 & 19100 & 17640 & 17810 & --    & 17460 &  --   &  --   \ \ \ \ \\
  164353 & 15420 & 13490 &  --   & --    &  --   & 16500 & 15500 \ \ \ \ \\
  166182 & 22420 & 21380 & 20320 & --    &  --   &  --   &  --   \ \ \ \ \\
  169022 &  9520 &  --   &  9460 & --    &  9540 &  --   &  --   \ \ \ \ \\
  172167 &  9470 &  --   &  9660 & --    &  9620 &  --   &  --   \ \ \ \ \\
  173300 & 14990 & 12390 &  --   & --    & 12110 &  --   &  --   \ \ \ \ \\
  175191 & 18890 &  --   & 18990 & --    &  --   &  --   &  --   \ \ \ \ \\
  176437 & 10000 &  --   &  --   & --    &  9830 & 9550  &  --   \ \ \ \ \\
  177724 &  9830 &  --   &  9480 & --    &  9550 & 9500  &  --   \ \ \ \ \\
  177756 & 11780 & 11560 & 11410 & --    & 11430 &  --   &  --   \ \ \ \ \\
  179761 & 13060 &  --   &  --   & --    &  --   & 13000 &  --   \ \ \ \ \\
  182255 & 15300 &  --   &  --   & --    &  --   &  --   &  --   \ \ \ \ \\
  186882 & 10150 &  --   &  9880 & --    & 10180 &  --   &  --   \ \ \ \ \\
  188209 & 25260 &  --   & 31080 & 27700 &  --   &  --   &  --   \ \ \ \ \\
  191692 & 10340 &  --   &  --   & --    & 10170 &  --   &  --   \ \ \ \ \\
  192425 &  9120 &  --   &  --   & --    &  --   &  --   &  --   \ \ \ \ \\
  192907 & 10500 &  --   & 10650 & --    &  --   & 10250 &  --   \ \ \ \ \\
  195556 & 17680 &  --   & 18140 & --    &  --   &  --   &  --   \ \ \ \ \\
  195810 & 14540 & 13550 & 13610 & --    &  --   &  --   &  --   \ \ \ \ \\
  196867 & 11220 &  --   & 10820 & --    & 10910 &  --   &  --   \ \ \ \ \\
  197345 &  8720 &  --   &  --   & --    &  --   &  --   &  --   \ \ \ \ \\
  198001 &  9370 &  --   &  --   & --    &  9180 &  9200 &  --   \ \ \ \ \\
  198478 & 15390 &  --   & 14270 & --    &  --   & 18000 & 17170 \ \ \ \ \\
  199081 & 16180 & 16560 &  --   & --    &  --   &  --   &  --   \ \ \ \ \\
  202850 & 11170 &  --   &  --   & --    &  --   &  --   & 11000 \ \ \ \ \\
  204172 & 24110 &  --   &  --   & 25200 &  --   & 28500 & 28500 \ \ \ \ \\
  205021 & 26920 &  --   & 25780 & 26200 &  --   &  --   &  --   \ \ \ \ \\
  206672 & 18360 & 17390 & 17890 & --    & 17110 &  --   &  --   \ \ \ \ \\
  207260 &  8980 &  --   &  --   & --    &  --   &  --   &  --   \ \ \ \ \\
  207330 & 17890 & 17100 &  --   & --    &  --   &  --   &  --   \ \ \ \ \\
  207971 & 12520 & 12170 & 11800 & --    & 11950 &  --   &  --   \ \ \ \ \\
  209481 & 30570 &  --   &  --   & --    &  --   &  --   &  --   \ \ \ \ \\
  209744 & 25440 &  --   &  --   & --    &  --   &  --   &  --   \ \ \ \ \\
  209819 & 12310 & 12470 & 11280 & --    &  --   &  --   &  --   \ \ \ \ \\
  209952 & 13920 & 13450 & 13660 & --    & 13930 &  --   &  --   \ \ \ \ \\
  209961 & 21100 & 21430 &  --   & --    &  --   &  --   &  --   \ \ \ \ \\
  209975 & 24720 &  --   &  --   & --    &  --   & 32500 &  --   \ \ \ \ \\
  210191 & 17890 &  --   &  --   & --    &  --   &  --   &  --   \ \ \ \ \\
  210418 &  8840 &  --   &  --   & --    &  --   &  --   &  --   \ \ \ \ \\
  212061 & 10490 &  --   &  --   & --    & 10290 &  --   &  --   \ \ \ \ \\
  212120 & 15250 & 14580 &  --   & --    & 14690 &  --   &  --   \ \ \ \ \\
  212593 & 11150 &  --   &  9930 & --    &  --   &  --   & 11800 \ \ \ \ \\
  212883 & 20400 & 22700 &  --   & --    &  --   &  --   &  --   \ \ \ \ \\
  212978 & 20990 & 20370 &  --   & --    & 20790 &  --   &  --   \ \ \ \ \\
  213420 & 19750 & 19850 & 20840 & --    & 20490 &  --   &  --   \ \ \ \ \\
  213558 &  9840 &  --   &  --   & --    &  9600 &  --   &  --   \ \ \ \ \\
  213976 & 22800 &  --   &  --   & --    &  --   &  --   &  --   \ \ \ \ \\
  213998 & 11740 & 11490 & 11220 & --    & 11530 &  --   &  --   \ \ \ \ \\
  214240 & 16330 & 17380 &  --   & --    &  --   &  --   &  --   \ \ \ \ \\
  214652 & 22350 & 21930 &  --   & --    &  --   &  --   &  --   \ \ \ \ \\
  214680 & 32380 &  --   & 33560 & --    & 33110 & 33000 &  --   \ \ \ \ \\
  214923 & 11430 & 11320 &  --   & --    & 11500 &  --   &  --   \ \ \ \ \\
  214993 & 24130 & 23600 &  --   & --    &  --   &  --   &  --   \ \ \ \ \\
  214994 &  9930 &  --   &  --   & --    &  9560 & 9530  &  --   \ \ \ \ \\
  215191 & 22800 & 22930 &  --   & --    & 24050 &  --   &  --   \ \ \ \ \\
  217101 & 22760 & 23120 &  --   & --    &  --   &  --   &  --   \ \ \ \ \\
  217811 & 19330 & 19100 &  --   & --    &  --   &  --   &  --   \ \ \ \ \\
  218045 &  9850 &  --   &  --   & --    &  9850 &  --   &  --   \ \ \ \ \\
  218376 & 27100 &  --   & 27470 & --    &  --   & 29000 &  --   \ \ \ \ \\
  218407 & 21200 & 23230 &  --   & --    &  --   &  --   &  --   \ \ \ \ \\
  219688 & 15260 &  --   & 15210 & --    & 14910 &  --   &  --   \ \ \ \ \\
  222173 & 12610 & 11850 & 11870 & --    & 11870 &  --   &  --   \ \ \ \ \\
  222661 & 10860 &  --   &  --   & --    &  --   &  --   &  --   \ \ \ \ \\
  223640 & 11740 &  --   &  --   & --    &  --   &  --   &  --   \ \ \ \ \\
  224990 & 16100 &  --   &  --   & --    & 15330 &  --   &  --   \ \ \ \ \\
\noalign{\smallskip}
\hline
\multicolumn{8}{l}{$T_{\rm eff}^{(1)}$ : Gulati et al. (1989),Castelli (1991)}  \ \ \ \ \\
\multicolumn{8}{l}{$T_{\rm eff}^{(2)}$ : Underhill et al. (1979), Code et al. (1976), Malagnini et al. (1986)} \ \ \ \ \\
\multicolumn{8}{l}{$T_{\rm eff}^{(3)}$ : Remie \& Lamers (1981)} \ \ \ \ \\
\multicolumn{8}{l}{$T_{\rm eff}^{(4)}$ : Morossi \& Malagnini (1985), Malagnini et al. (1983), Malagnini \&} \ \ \ \ \\
\multicolumn{8}{l}{Morossi (1990)} \ \ \ \ \\
\multicolumn{8}{l}{$T_{\rm eff}^{(5)}$ : McErlean et al. (1999), \citet{ade02}} \ \ \ \ \\
\multicolumn{8}{l}{$T_{\rm eff}^{(6)}$ : Searle et al. (2008), Markova \& Puls (2008), Crowther et al. (2006)} \ \ \ \ \\
\end{longtable}

\end{document}